%% file: 0_main.tex
\documentclass[manuscript]{acmart}

\usepackage{xcolor}
\usepackage[ruled,vlined]{algorithm2e}
\usepackage{verbatimbox}
\usepackage{makecell}
\usepackage{xspace}
\usepackage{enumitem}
\setlist{nosep}
\usepackage{cleveref}
\usepackage{caption}
\usepackage{subcaption}
\usepackage{placeins}
\usepackage{pifont} 
\usepackage{multirow}
\usepackage{arydshln} 
\usepackage{tabularx}
\usepackage{bm}
\usepackage{adjustbox}
\usepackage[normalem]{ulem}

\newcommand*\rot{\rotatebox{90}}

\usepackage{hyperref}
\hypersetup{bookmarksdepth=3}

\setcitestyle{maxcitenames=1}

\AtBeginDocument{%
  }

\input{1_variables.tex}

\input{3_dlbook_mathcommands.tex}

\setcopyright{acmlicensed}
\copyrightyear{2025}
\acmYear{2025}
\acmDOI{XXXXXXX.XXXXXXX}

\acmJournal{TORS}
\acmVolume{0}
\acmNumber{0}
\acmArticle{0}
\acmMonth{1}

\newif\ifworkinprogress
\workinprogressfalse

\ifworkinprogress

    \newcommand{\mmedits}[1]{#1}

    \newcommand{\cg}[1]{#1}

  \newcommand{\newtext}[1]{#1}    
  \newcommand{\newresults}[1]{#1}
  \newcommand{\reviewTwo}[1]{\textcolor{orange}{#1}}
  
  \newcommand{\reviewRemoved}[1]{\textcolor{gray}{\sout{#1}}}
  \newcommand{\reviewRemovedFootnote}[1]{\footnote{\reviewRemoved{#1}}}
\else

    \newcommand{\cg}[1]{#1}
    
	\newcommand{\mmedits}[1]{#1}
    \newcommand{\newtext}[1]{#1}
    \newcommand{\newresults}[1]{#1}
    \newcommand{\reviewTwo}[1]{#1}
    
    \newcommand{\reviewRemoved}[1]{}
    \newcommand{\reviewRemovedFootnote}[1]{}
\fi

\newlist{questions}{enumerate}{2}
\setlist[questions,1]{label=RQ\arabic*.,ref=RQ\arabic*}
\setlist[questions,2]{label=(\alph*),ref=\thequestionsi(\alph*)}

\begin{document}

\title{Single-Branch Network Architectures to Close the Modality Gap in Multimodal Recommendation}

\author{Christian Ganhör}
\email{christian.ganhoer@jku.at}
\orcid{0000-0003-1850-2626}
\affiliation{%
  \institution{Institute of Computational Perception, Johannes Kepler University Linz}
  \streetaddress{Altenberger Straße 69}
  \city{Linz}
  \country{Austria}
}

\author{Marta Moscati}
\email{marta.moscati@jku.at}
\orcid{0000-0002-5541-4919}
\affiliation{%
  \institution{Institute of Computational Perception, Johannes Kepler University Linz}
  \streetaddress{Altenberger Straße 69}
  \city{Linz}
  \country{Austria}
}

\author{Anna Hausberger}
\email{anna.hausberger@jku.at}
\orcid{0009-0003-5562-681X}
\affiliation{%
  \institution{Institute of Computational Perception, Johannes Kepler University Linz}
  \streetaddress{Altenberger Straße 69}
  \city{Linz}
  \country{Austria}
}

\author{Shah Nawaz}
\email{shah.nawaz@jku.at}
\orcid{0000-0002-7715-4409}
\affiliation{%
  \institution{Institute of Computational Perception, Johannes Kepler University Linz}
  \streetaddress{Altenberger Straße 69}
  \city{Linz}
  \country{Austria}
}

\author{Markus Schedl}
\email{markus.schedl@jku.at}
\orcid{0000-0003-1706-3406}
\affiliation{
  \institution{Institute of Computational Perception, Johannes Kepler University Linz and AI Lab, Linz Institute of Technology}
  \city{Linz}
  \country{Austria}
}

\authornote{Corresponding author.}

\renewcommand{\shortauthors}{Ganhör et al.}

\begin{abstract}
    \input{0_sections/0_abstract}
\end{abstract}



\received{20 February 2007}
\received[revised]{12 March 2009}
\received[accepted]{5 June 2009}

\maketitle

\section{Introduction}
\label{sec:intro}
\input{0_sections/1_intro}

\section{Related Work}
\label{sec:related-work}
\input{0_sections/2_related_work}

\section{Methodology}
\label{sec:methodology}
\input{0_sections/3_methodology}

\section{Experimental Setup}
\label{sec:experimental-setup}
\input{0_sections/4_experiment_setup}

\section{Results}
\label{sec:results}
\input{0_sections/5_results}

\section{Conclusion}
\label{sec:conclusion}
\input{0_sections/6_conclusion}

\begin{acks}
\input{0_sections/7_acknowledgements}
\end{acks}

\bibliographystyle{ACM-Reference-Format}
\bibliography{0_main.bib}

\appendix
\section{Appendix}
\input{0_sections/A_appendix}

\end{document}

%% file: 1_variables.tex
\usepackage{array}
\usepackage{amsmath}
\usepackage{xspace}

\newcommand{\eg}{e.\,g., }
\newcommand{\ie}{i.\,e., }

\newcommand{\ndcg}{\text{NDCG}\xspace}

\newcommand{\precision}{\text{Precision}\xspace}
\newcommand{\ap}{\text{AP}\xspace}
\newcommand{\map}{\text{MAP}\xspace}
\newcommand{\Recall}{\text{Recall}\xspace}

\newcommand{\Coverage}{\text{Coverage}\xspace}

\newcommand{\mrr}{\text{MRR}\xspace}

\newcommand{\euclideanDist}{\text{ED}\xspace}
\newcommand{\cosineSim}{\text{CS}\xspace}

\newcommand{\apName}{AP}
\newcommand{\mapName}{MAP}

\newcommand{\ndcgName}{NDCG}
\newcommand{\dcgName}{DCG}
\newcommand{\idcgName}{IDCG}
\newcommand{\recName}{Recall}
\newcommand{\recallName}{\recName} 
\newcommand{\precName}{Precision}
\newcommand{\precisionName}{\precName} 
\newcommand{\covName}{Coverage}
\newcommand{\coverageName}{\covName} 

\newcommand{\aptName}{APLT}
\newcommand{\arpName}{ARP}

\newcommand{\fscoreName}{F-score}
\newcommand{\plName}{PL}
\newcommand{\rrName}{RR}
\newcommand{\mrrName}{MRR}

\newcommand{\metricAt}[2]{\text{#1}@\ensuremath{#2}\xspace}

\newcommand{\ndcgk}{\metricAt{\ndcgName}{k}}
\newcommand{\dcgk}{\metricAt{\dcgName}{k}}
\newcommand{\idcgk}{\metricAt{\idcgName}{k}}
\newcommand{\preck}{\metricAt{\precisionName}{k}}
\newcommand{\reck}{\metricAt{\recallName}{k}}
\newcommand{\apk}{\metricAt{\apName}{k}}

\newcommand{\covk}{\metricAt{\coverageName}{k}}
\newcommand{\aptk}{\metricAt{\aptName}{k}}
\newcommand{\arpk}{\metricAt{\arpName}{k}}

\newcommand{\fscorek}{\metricAt{\fscoreName}{k}}
\newcommand{\plk}{\metricAt{\plName}{k}}
\newcommand{\rrk}{\metricAt{\rrName}{k}}

\newcommand{\ndcgten}{\metricAt{\ndcgName}{10}}
\newcommand{\precten}{\metricAt{\precisionName}{10}}
\newcommand{\recten}{\metricAt{\recallName}{10}}

\newcommand{\mapten}{\metricAt{\mapName}{10}}
\newcommand{\covten}{\metricAt{\coverageName}{10}}
\newcommand{\aptten}{\metricAt{\aptName}{10}}
\newcommand{\arpten}{\metricAt{\arpName}{10}}

\newcommand{\fscoreten}{\metricAt{\fscoreName}{10}}
\newcommand{\plten}{\metricAt{\plName}{10}}

\newcommand{\mrrten}{\metricAt{\mrrName}{10}}

\newcommand{\topk}{top-$k$\xspace}

\newcommand{\abs}[1]{| #1 |}
\newcommand{\card}[1]{\abs{#1}}
\newcommand{\set}[1]{\{ #1 \}}
\newcommand{\setText}[1]{\{#1\}}

\newcommand{\norm}[1]{\left\lVert#1\right\rVert}
\newcommand{\normFixedSize}[1]{\lVert#1\rVert}

\def\Tabref#1{Table~\ref{#1}}

\newcommand{\mf}{\text{MF}\xspace}
\newcommand{\dmf}{\text{DeepMF}\xspace}
\newcommand{\rand}{\text{Rand}\xspace}
\newcommand{\pop}{\text{Pop}\xspace}

\newcommand{\clcrec}{\text{CLCRec}\xspace}

\newcommand{\mlonem}{\text{ML-1M}\xspace}
\newcommand{\onion}{\text{Onion}\xspace}
\newcommand{\amazonvid}{\text{Amazon}\xspace}

\newcommand{\dropoutnet}{\text{DropoutNet}\xspace}
\newcommand{\dropoutnetBest}{\text{DropoutNet}\textsubscript{\text{best}}\xspace}
\newcommand{\dropoutnetTwo}{\text{DropoutNet}\textsubscript{\text{one}}\xspace}

\newcommand{\sibrarVar}{\texttt{SiBraR}}  
\newcommand{\sibrar}{\sibrarVar\xspace}
\newcommand{\sibrarlong}{a multimodal \textbf{Si}ngle-\textbf{Bra}nch embedding network for \textbf{R}ecommendation\xspace}
\newcommand{\sibrarlongna}{multimodal \textbf{Si}ngle-\textbf{Bra}nch embedding network for \textbf{R}ecommendation\xspace}
\newcommand{\sibrarBest}{\sibrarVar\textsubscript{\text{best}}\xspace}
\newcommand{\sibrarTwo}{\sibrarVar\textsubscript{\text{one}}\xspace}
\newcommand{\sibrarNoReg}{\sibrarVar\textsubscript{\text{S}}\xspace}
\newcommand{\sibrarReg}{\sibrarVar\textsubscript{\text{SC}}\xspace}
\newcommand{\sibrarSamp}{\sibrarNoReg}
\newcommand{\sibrarSampContr}{\sibrarReg}

\newcommand{\sibrarItem}{\texttt{Item-\sibrar}\xspace}
\newcommand{\sibrarUser}{\texttt{User-\sibrar}\xspace}
\newcommand{\sibrarUserItem}{\texttt{User-Item-\sibrar}\xspace}

\newcommand{\sibrarSpec}{\sibrarVar\textsubscript{\text{mod-specific}}\xspace}

\newcommand{\mubrarVar}{\texttt{MuBraR}}  
\newcommand{\mubrar}{\mubrarVar\xspace}
\newcommand{\mubrarItem}{\texttt{Item-\mubrar}\xspace}
\newcommand{\mubrarUser}{\texttt{User-\mubrar}\xspace}

\newcommand{\mubrarVanilla}{\mubrarVar\xspace}

\newcommand{\mubrarlongna}{multimodal \textbf{Mu}lti-\textbf{Bra}nch embedding network for \textbf{R}ecommendation\xspace}

\newcommand{\mubrarBest}{\mubrarVar\textsubscript{\text{best}}\xspace}
\newcommand{\mubrarTwo}{\mubrarVar\textsubscript{\text{one}}\xspace}

\newcommand{\mubrarSamp}{\mubrarVar\textsubscript{\text{S}}\xspace}
\newcommand{\mubrarNoReg}{\mubrarVar\textsubscript{\text{S}}\xspace}
\newcommand{\mubrarReg}{\mubrarVar\textsubscript{\text{SC}}\xspace}
\newcommand{\mubrarSampContr}{\mubrarReg}

\newcommand{\cfbr}{\text{CF}\xspace} 
\newcommand{\cfbrs}{\text{CFs}\xspace} 

\newcommand{\cbrs}{\text{CBRSs}\xspace} 
\newcommand{\hrss}{\text{HRSs}\xspace} 
\newcommand{\hrs}{\text{HRS}\xspace} 

\newcommand{\baserec}{\text{BRS}\xspace} 
\newcommand{\baserecs}{\text{BRSs}\xspace} 

\newcommand{\intraVar}{\text{intra}}
\newcommand{\interVar}{\text{inter}}
\newcommand{\intra}{\intraVar\xspace}
\newcommand{\inter}{\interVar\xspace}

\newcommand{\interCos}{\interVar_\cosineSim}

\newcommand{\interEucl}{\interVar_\euclideanDist}

\newcommand{\repo}{\href{https://github.com/hcai-mms/single-branch-networks}{https://github.com/hcai-mms/single-branch-networks}}
\newcommand{\repoConfigs}{\href{https://github.com/hcai-mms/single-branch-networks}{https://github.com/hcai-mms/single-branch-networks}}

\newcommand{\setUsers}{\ensuremath{\mathcal{U}}}
\newcommand{\setItems}{\ensuremath{\mathcal{I}}}
\newcommand{\setNegs}{\ensuremath{\mathcal{N}}}
\newcommand{\setMods}{\ensuremath{\mathcal{M}}}
\newcommand{\setModsItem}{\ensuremath{\setMods^{item}}}
\newcommand{\setModsUser}{\ensuremath{\setMods^{user}}}
\newcommand{\batch}{\ensuremath{\mathcal{B}}\xspace}

\newcommand{\numMods}{\ensuremath{N_{\text{mod}}}}
\newcommand{\numNegs}{\ensuremath{N_{\text{neg}}}}
\newcommand{\numUsers}{\ensuremath{M}}
\newcommand{\numItems}{\ensuremath{N}}

\newcommand{\matrixInter}{\ensuremath{R}}
\newcommand{\matrixInterElement}{\ensuremath{r}}

\newcommand{\recScore}{\ensuremath{y}}
\newcommand{\recScorePred}{\ensuremath{\hat{\recScore}}}

\newcommand{\loss}{\ensuremath{\mathcal{L}}}
\newcommand{\lossBPR}{\ensuremath{\loss_\text{BPR}}}
\newcommand{\lossInfo}{\ensuremath{\loss_\text{InfoNCE}}}
\newcommand{\lossInfoS}{\ensuremath{\loss_\text{SInfoNCE}}}

\newcommand{\metricCosineSim}{cosine similarity}
\newcommand{\metricCosineSimShort}{\cosineSim}

\newcommand{\metricEuclidean}{Euclidean distance}
\newcommand{\metricEuclideanShort}{\euclideanDist}

\newcommand{\setRel}{\ensuremath{\text{Rel}}} 
\newcommand{\setRet}{\ensuremath{\text{Rec}}} 

\newcommand{\indicatorRelui}{\ensuremath{\mathbf{1}_{\setRel(u_i)}}}

%% file: 3_dlbook_mathcommands.tex
\def\Figref#1{Figure~\ref{#1}}

\def\Secref#1{Section~\ref{#1}}

\def\eqref#1{equation~\ref{#1}}
\def\Eqref#1{Equation~\ref{#1}}

\def\1{\bm{1}}

\def\ve{{\bm{e}}}

\def\vi{{\bm{i}}}

\def\vu{{\bm{u}}}

\DeclareMathAlphabet{\mathsfit}{\encodingdefault}{\sfdefault}{m}{sl}
\SetMathAlphabet{\mathsfit}{bold}{\encodingdefault}{\sfdefault}{bx}{n}

\def\sR{{\mathbb{R}}}



%% file: 0_sections/0_abstract.tex
Traditional recommender systems rely on collaborative filtering (\cfbr), using past user--item interactions to help users discover new items in a vast collection. In \textit{cold start}, \ie when interaction histories of users or items are not available, content-based recommender systems (\cbrs) use side information instead. 
\newtext{Most commonly, user demographics and item descriptions are used for user and item cold start, respectively.} 
Hybrid recommender systems (\hrss) often employ multimodal learning to combine collaborative and user and item side information, which we jointly refer to as \textit{modalities}. Though \hrss can provide recommendations when some modalities are missing, their quality degrades. 
In this work, we utilize \textit{single-branch neural networks} equipped with \textit{weight sharing}, \textit{modality sampling}, and \textit{contrastive loss} to provide accurate recommendations even in missing modality scenarios, including cold start. 
\mmedits{Compared to multi-branch architectures, the weights of the encoding modules are shared for all modalities; in other words, all modalities are encoded using the same neural network. This, together with the contrastive loss, is essential in reducing the modality gap, while the modality sampling is essential in modeling missing modality during training. Simultaneously leveraging these techniques results in more accurate recommendations.} We compare these networks with multi-branch alternatives and conduct extensive experiments on the {MovieLens 1M}, {Music4All-Onion}, and {Amazon Video Games} datasets. Six accuracy-based and four beyond-accuracy-based metrics help assess the recommendation quality for the different training paradigms and their hyperparameters on single- and multi-branch networks in warm-start and missing modality scenarios. We quantitatively and qualitatively study the effects of these different aspects on bridging the modality gap. 
Our results show that single-branch networks provide competitive recommendation quality in warm start, and significantly better performance in missing modality scenarios. \cg{Moreover, our study of modality sampling and contrastive loss on both single- and multi-branch architectures indicates a consistent positive impact on accuracy metrics across all datasets. Overall, the three training paradigms collectively encourage modalities of the same item to be embedded closer together than those of different items, as measured by \metricEuclidean\xspace and \metricCosineSim. This results in embeddings that are less distinguishable and more interchangeable, as indicated by a 7-20\% drop in modality prediction accuracy.} 
\reviewTwo{Our full experimental setup, including training and evaluating code for all algorithms, their hyperparameter configurations, and our result analysis notebooks, is available at \repo.}

%% file: 0_sections/1_intro.tex
Recommender systems (RSs) offer an important tool for users to discover interesting items in the vast and ever-increasing collections of, \eg movies, songs, and shopping items, alleviating users' information overload.
Although many RSs still build on collaborative filtering (\cfbr), using past user--item interactions, scientific developments related to multimedia content analysis enabled content-based recommenders (\cbrs). These RSs leverage a broad set of media types (text, audio, image, video) and are capable of effectively dealing with \textit{cold-start} scenarios~\cite{10.1145/3407190}, \ie when user and/or item interaction histories are unavailable. While \cfbr-based RSs struggle due to their dependence on interaction histories in cold start, the performance of \cbrs might be limited by not using collaborative information in warm start, \ie when interactions are available in sufficient quantities. 
Thus, hybrid RSs (\hrss) overcome the shortcomings of both \cfbrs and \cbrs, by integrating collaborative signals and additional side information of users (\eg age, gender, personal descriptions) and/or items (\eg audio for music RSs; video for movie RSs) to generate recommendations in warm- and cold-start scenarios. For this purpose, state-of-the-art \hrss commonly utilize deep neural networks (DNNs) along with multimodal learning approaches. These networks often contain modality-specific subnetworks (\textit{branches}) for users and items alike, which encode and transform available modalities. Finally, the resulting modality representations per entity (users or items) are aggregated and used in downstream recommendation tasks~\cite{feature_concat,dropoutnet,jeong_2024_mm_rs, chen_2024_mm_video_rec,ren_2024_mm_fusion}. Based on their architecture, we refer to this type of \hrs as multi-branch recommenders, and outline a common structure of them in \Figref{fig:mubrar:vanilla}.
When subsets of the modalities used in training such models are unavailable at inference, the quality of recommendations often declines. This issue is referred to as \textit{missing modality problem}.
To address it, \citet{ganhoer_moscati2024sibrar} propose \sibrarlong (\sibrar, pronounced "zebra"). As outlined in \Figref{fig:sibrar:item}, it leverages only a single shared branch per entity which maps all the modalities of the same entity to a similar region in the shared embedding space. In addition to weight sharing across modalities, \sibrar employs modality sampling and a contrastive loss between multiple modalities to further realize a shared embedding space. 
\cg{This design reduces the dependence on the availability of all modalities.}

While single-branch architectures show promising results in warm- and cold- start scenarios as well as in scenarios of missing modalities, there is a need for further investigating how they work, and to compare them with multi-branch architectures. Moreover, their three main components: weight sharing, modality sampling, and contrastive loss, need to receive additional attention to study which of them, individually and jointly, contribute most to a strong recommendation performance.

In this work, we aim to develop an in-depth understanding of single- and multi-branch architectures for multimodal representation learning for recommendation. Thus, we task ourselves with answering the following research questions:
\begin{questions}
    \item How does \sibrar compare against multi-branch architectures in terms of accuracy and beyond-accuracy performance, in warm-start, missing modality, and cold-start scenarios?\label{rq:1}
    \item How do different training paradigms employed in \sibrar effect its performance?\label{rq:2}
    \item How can we characterize the modality gap for \sibrar and \mubrar?\label{rq:3}
\end{questions}

To achieve our objectives, we introduce \mubrarlongna (\mubrar), serving as a common baseline for multi-branch recommender systems. Moreover, we propose two further variants, \mubrarSamp and \mubrarSampContr, employing modality sampling without and with a contrastive loss, respectively, as used in \sibrar. Doing so, we discover the effects of the individual components on performance and the embedding space. We apply the models on three different recommendation domains (music, movie, and e-commerce) mimicking warm-start as well as user and item cold-start tasks. We evaluate them on six accuracy metrics and four beyond-accuracy metrics. The accuracy is measured with \precName, \recName, \fscoreName, Normalized Discounted Cumulative Gain (\ndcgName), Mean Average Precision (\mapName), and Mean Reciprocal Rank (\mrrName). The beyond-accuracy metrics consist of \covName, Average Popularity of Long Tail Items (\aptName), Average Recommendation Popularity (\arpName), Popularity Lift (\plName) and are used to describe induced model biases. Moreover, we thoroughly evaluate model performances in case of missing modalities. As \sibrar's main intention is to effectively reduce the modality gap, we qualitatively and quantitatively investigate the embedding spaces of single- and multi-branch paradigms. Finally, we take a deep dive into the contrastive loss and analyze how its parameters affect the recommendation quality as well as the embedding space.

We summarize the key contributions of our work as follows: 
\begin{itemize}
    \item We quantitatively compare single- to multi-branch approaches in extensive experiments on three datasets, in warm-start and missing-modality scenarios, including cold start.
    \item We shed light on the factors contributing to strong recommendation performance in single- and multi-branch models.
    \item We qualitatively and quantitatively investigate the embedding spaces of single- and multi-branch approaches, especially with respect to modality gap.
    \item We characterize the importance of the contrastive loss with a detailed analysis.
\end{itemize}

\mmedits{Our analyses show that single-branch recommenders have a clear advantage over multi-branch recommenders in cold start and missing modality, and that for both architectures the use of contrastive loss is paramount for achieving strong results. In our experiments single- and multi-branch architectures proved out to require extensive hyperparameter optimization. Therefore our detailed analysis of the impact of the hyperparameters of the contrastive loss can help practitioners in identifying suitable hyperparameter ranges. In terms of modality gap of embedding spaces, we show that both modality sampling and contrastive loss force modality embeddings of the same items to be closer to each other than to embeddings of other items.}

\cg{The study of single-branch networks for recommendation opens many interesting directions of research. As such, it allows for studying missing modality during training, which the network inherently supports. Further, the impact of different strategies to fuse modalities, \eg by attention, may be of interest in future works.}

The remaining part of this paper is structured as follows: In \Secref{sec:related-work}, we discuss related works on cold-start recommendation and multimodal learning. We outline \sibrar and \mubrar in \Secref{sec:methodology}, followed by descriptions of the experimental setup in \Secref{sec:experimental-setup}. All our experiments' quantitative and qualitative results are presented and discussed in \Secref{sec:results}. Finally, we conclude by considering practical implications of our work, as well as reviewing limitations in \Secref{sec:conclusion}.
\reviewTwo{We also release our full experimental setup, including the training and evaluating pipelines for all algorithms, the hyperparameter configurations, and our result analysis code, at \repo.}

%% file: 0_sections/2_related_work.tex
In this section, we study relevant previous work, which falls within two strands of research: 
\hrss for cold start and multimodal representation learning.

\subsection{Content-Based Cold-Start Recommendation}
    One of the main challenges of RSs~\cite{rshandbook, mrs} is providing relevant recommendations to users with few or no past interactions, or to effectively recommend items that have been interacted with by few or no users; these scenarios are referred to user and item cold start, respectively. We refer the reader to \citet{panda_cold_start_review} for a recent literature review on cold-start recommendation.

    Although there is no unanimous consensus on the number of interactions required to consider a user or an item as ``cold'', in this work, we focus on the \emph{strict} cold-start scenario, \ie when users or items have no interactions.
    To mitigate the low accuracy of recommendations in this scenario, RSs typically complement the user--item interaction data with side information, such as information on the user demographics, or representations of the content of the items. 
    \newtext{Several techniques have been proposed to leverage the additional information. They can be categorized into explicit and implicit methods~\cite{ood_cold_start}.}
    
    \paragraph{\newtext{Explicit techniques.}} \newtext{Methods in this category combine the recommendation loss with additional loss terms. Specifically, several works proposed to additionally minimize the distance of latent representations of side information to latent representations of the interaction profiles. 
    \citet{mrs_content_cold} and \citet{pre_training_cold_start} explicitly model the distance by using the Euclidean distance between the latent representations as a loss term. While the former use fully-connected NNs to compute the latent representations, the latter leverage graph neural networks (GNNs). 
    In a similar way, \citet{cb2cf} and \citet{zhu2019heater} include the Mean Square Error (MSE) between the latent representations
    as an additional loss term to reduce the distance between them.} 
    Another explicit way to reduce the distance between the latent representations is by using contrastive learning. \citet{clcrec} propose the use of a contrastive loss term on the latent embeddings resulting from the deep NNs that encode side information and interaction profiles. \citet{li2024nemcf} propose a GNN-based RSs with a contrastive loss term that uses as positive pairs the items that often co-occur in the interaction profiles of the same users. 
    Other methods to learn latent representations of side information that are effective for recommendation include reconstruction losses. For instance, \citet{uima} propose the use of two autoencoders, one for item content and one for user interaction data, trained to minimize the combination of a recommendation loss term and reconstruction loss terms. Further works explore the use of categorical side information to learn similar latent representations within a category. For instance, in the context of music recommendation, \citet{pulis2023snn} maximize the similarity of audio representations of music tracks of the same music genre. In contrast, in the context of sequential recommendation, \citet{shalaby2022m2trec} complement the recommendation loss of a transformer-based sequential RS with a loss term for predicting item category. Moreover, within the category of explicit methods, \citet{ood_cold_start} assume that the content of items produced at different times evolves continuously over time; they, therefore, introduce a loss term to align the representations of interpolations of items with the interpolation of the item representations. This technique aims to provide effective representations for items whose content is out of distribution compared to the training data, specifically for cold-start items. \mmedits{The advantage of explicit methods is that the required similarity between representations of the interactions and the side information is used directly to optimize the model. This, however, comes at the cost of tuning additional hyperparameters, such as the weight of the similarity loss relative to the recommendation loss.}

    \paragraph{\newtext{Implicit techniques.}} In the category of implicit methods, RSs leverage side information for cold start by relying on training strategies instead of explicit terms in the loss function. The RSs proposed by \citet{feature_concat} and by \citet{dropoutnet} rely on multi-branch architectures that encode interaction information and one or more side information with separate NNs; the resulting representations are concatenated to obtain the representation used to provide recommendations. The difference between those two works is that \citet{dropoutnet} further apply dropout to mimic cold-start during training. Other architectures, such as those proposed by \citet{ihgnn} and by \citet{recruitergcn}, make use of heterogeneous graphs to simultaneously encode information on the interactions and the similarity between side information of different users or items. In a way similar to autoencoder approaches, \citet{mrs_content_cold} propose a NN that takes side information as input and aims at reconstructing the collaborative embedding, while \citet{item_cold_start_nn} propose a NN that takes content as input and outputs the weights of the first layer of the encoder of the collaborative embedding. Finally, as large language models (LLMs) have demonstrated their ability in zero-shot learning, \citet{llm_cikm} and \citet{sanner2023llmrec} proposed the use of pre-trained LLMs to obtain item representations to be leveraged in cold-start scenarios. \mmedits{Compared to explicit methods, implicit methods have the advantage that representations are optimized directly for the recommendation task, without additional loss terms for the similarity. Due to the sparsity and noise in the interaction data, however, this signal might be too weak to enforce the required similarity between representations of the interactions and the side information.}

\subsection{Multimodal Learning}
    Multimodal learning leverages information from multiple modalities, such as audio, image, or text, to improve the performance on various machine learning (ML) tasks, such as classification, ranking, or verification~\cite{multimodal_survey,xu2023multimodal,liu2024multimodal}. 
    \mmedits{Although multimodal tasks differ, the way they are usually addressed is very similar and relies on learning joint representations that either combine, \eg concatenate, multiple modalities, or represent modalities separately, but in a way such that they are close in the embedding space if corresponding to the same entity. }
    Several multimodal methods explored the use of NNs to map multimodal information to joint representations, e.g., multi-branch NNs use separate independent and modality-specific NNs to map each modality to a joint embedding space~\cite{reed2016learning, nagrani2018seeing,nagrani2018learnable,saeed2022fusion,arshad2019aiding}. Recently, multi-branch architectures have also been extended with the use of transformers~\cite{vaswani2017transformers_attention,lu2019vilbert,tan2019lxmert,xu2023multimodal}.
    Such multi-branch methods have achieved remarkable performance using modality-complete data, \ie when all modalities are available for training and evaluation. 
    \cg{However, recent studies have found that such methods lack inherent robustness to missing modalities, leading to a significant drop in performance when certain modalities are missing for the classification task~\cite{ma2022multimodal}. Since recommendation systems often deal with incomplete data (\eg missing user preferences), multi-branch Transformer architectures may not be the most effective choice for such applications.}
    Given the significance of multimodal learning, recent years have seen an increased interest in studies addressing missing modalities for various deep learning tasks~\cite{ma2021smil,zhang2022m3care,lee2023multimodal,lin2023missmodal,wang2023multi,lan2024robust,saeed2024modalityinvariantmultimodallearning}.
    This growing interest has also broadened to multimodal RSs~\cite{wanglrmm2018,malitesta2024we,pomo2025first}, where handling missing modalities is critical to improve performance and robustness.
    Recently, single-branch models~\cite{saeed2023single,tschannen2023clippo}, \ie models that share the same network across multiple modalities, have shown promising results in multimodal learning. Although effective for other multimodal learning tasks, \mmedits{applying single-branch architectures to the domain of recommendation poses the additional challenge of combining dense and sparse data, such as images and user-item interactions. }
    Our work \mmedits{addresses this challenge by adopting \sibrar}, a novel multimodal RS based on a single-branch architecture~\cite{ganhoer_moscati2024sibrar}, and analyzing the effectiveness of single-branch architectures in missing modality scenarios, which are related to cold start in recommendation.

%% file: 0_sections/3_methodology.tex
This section introduces \sibrarlongna (\textit{\sibrar}). We define the notation and mathematically formulate the problem. Then, we describe how \sibrar leverages multimodal information of items and users for recommendation in warm- and cold-start scenarios. Finally, we describe a typical multi-branch architecture that we term \mubrarlongna (\textit{\mubrar}), along with extensions that feature components of \sibrar.

\subsection{Notation}
    Let $\setUsers = \set{u_i}_{i=1}^\numUsers$ be the set of $\numUsers$ users and $\setItems = \set{i_j}_{j=1}^\numItems$ the set of $\numItems$ items. Moreover, let $\matrixInter \in \sR^{\numUsers\times\numItems}$ be a binary matrix of interactions between users and items, where $\matrixInterElement_{i,j}=1$ indicates a \textit{positive} interaction between user~$u_i$ and item~$i_j$. $\matrixInterElement_{i,j}=0$ refers to 
    an \textit{unknown} interaction, as $u_i$ might have consciously chosen not to interact with $i_j$, or just has not encountered it yet. In the rest of this work, and as common in literature~\cite{pan_2008_unknown_negative_sampling,Jannach_2018_implicit}, we refer to items as \textit{negatives} if their interactions are unknown.
    Let $\vu_i \in \sR^\numItems$ be the $i$-th row of~$\matrixInter$, expressing user~$u_i$'s profile, \ie a multi-hot vector encoding what items user~$u_i$ has interacted with. Analogously, let $\vi_j \in \sR^{\numUsers}$ be the $j$-th column of $\matrixInter$, which expresses the profile of item~$i_j$, \ie all users that interacted with $i_j$.\footnote{\newtext{We use bold-face letters (\eg $\vu$ and $\vi$) to denote vectors, while regular letters (\eg $u$ and $i$) for scalars and variables.}} Cold-start users (items) are those with an empty profile, \ie they have no interactions in their profiles.
    We define the set of \textit{modalities} (side information) associated and available for user~$u_i$ (\eg gender and country) and item~$i_j$ (\eg audio and text data) as $\setModsUser_i$ and $\setModsItem_j$, respectively.
    Moreover, we consider also interactions as a modality and therefore include non-empty profiles of users and items in the sets of modalities, \ie $\set{\vu_i} \subseteq \setModsUser_i \iff \normFixedSize{\vu_i} > 0$ and $\set{\vi_j} \subseteq \setModsItem_j \iff \normFixedSize{\vi_j} > 0$, where $\normFixedSize{\cdot}$ is the Euclidean norm.

\subsection{\sibrar}
    We now describe \sibrar, which outputs a recommendation score~$\recScorePred_{i,j}$ for a given user--item pair~$(u_i, i_j)$, by generating an embedding vector~$\ve_i$ for user~$u_i$ and an embedding vector for~$\ve_j$ for item~$i_j$, and by computing their dot product $\recScorePred_{i,j}=\ve_i \cdot \ve_j$. The top~$k$ items of the list of descending $\recScorePred_{il}, 
    l \in \set{1, \ldots, \numItems | \matrixInterElement_{il} = 0 }$ 
    that user~$u_i$ has not interacted with, form the recommendations for~$u_i$. Assuming that different modalities of the same entity (\ie user or item) share a similar set of underlying semantic information, \sibrar is designed to take any available modality as input, and to infer an embedding vector that lies in the same region as other modalities of the same entity in a shared embedding space \cite{ganhoer_moscati2024sibrar}. Moreover, these vectors should be representative of the entity for providing accurate recommendations.
    To achieve this, \sibrar leverages weight-sharing of a deep neural network (DNN)~$g$ to embed different modalities. The network~$g$ is then optimized to provide an accurate representation for entities based on their available modalities taken as input. This encourages the network to map different modalities of the same user or item into the same shared embedding space. For instance, for item~$i_j$, $g(m_1^i) \simeq g(m_2^i) \; \forall m_1^i, m_2^i \in \setModsItem_i$.

    We now describe different variants of \sibrar. First, \sibrarItem leverages multimodal item side information to tackle scenarios of item cold start and missing item modality. On the other hand, \sibrarUser utilizes multimodal user side information to tackle user cold start and missing user modality. Combining both approaches, \sibrarUserItem tackles user and item cold start and missing modality.

\input{resources/figures/architectures/fig-sibrar}
\input{resources/algorithms/item-sibrar.tex}
\subsubsection{\sibrarItem}    
    \Figref{fig:sibrar:item} shows the architecture and training pipeline of \sibrarItem, and Algorithms~\ref{alg:sibrar:train} provides pseudocode for computing the batch loss employed in training. In \sibrarItem, $f_k, \; k \in \set{1, \ldots, \card{\setModsItem}}$ denotes modality-specific functions, each expressed as a single linear layer and a ReLU activation function. They are used to transform different modalities to a common dimension required for the single-branch item embedding network~$g$. 
    If all modalities are already of the same shape, the networks~$f_k$ are unnecessary as $g$ can be designed to take them as input as is. Thus, to ease notation, we omit them in the following.
    The item embedding network~$g$, comprised of a series of linear layers, batch normalizations, and ReLU activation functions, forms a core component of \sibrarItem. In contrast, for the user embedding network~$h$, we consider either an embedding lookup table or an NN architecture similar to DeepMF~\cite{dmf}. For the lookup table, embeddings are randomly initialized and updated during training, while for the NN, the users' profiles are used as inputs to generate user embeddings.

    During training, for a positive user--item pair, where user $u_i$ has interacted with item~$i_j$ ($\matrixInterElement_{i,j} = 1$), a set of $\numNegs$ non-interacted items $\setNegs_{i}$, \ie items the user $u_i$ has not interacted with, are sampled uniformly at random. Moreover, for each pair, $\numMods$ of all the item's modalities~$\setModsItem_j$, are randomly sampled.\footnote{There exist different sampling techniques for negative sampling and modality sampling; we refer the reader to \citet{yang_doesnegativesamplingmatter_2024} and \citet{saeed2024modalityinvariantmultimodallearning}, respectively. 
     \mmedits{Among those, for simplicity we choose sampling uniformly at random, although we are aware that in real-world scenarios modalities are not uniformly available. We leave the application of other modality-sampling strategies for future work.}}
     \cg{The process of modality sampling aims to improve robustness to missing modalities by reducing the dependence on any single input source.}
    Each user embedding is obtained by $\ve_i = h(u_i)$ or $\ve_i = h(\vu_i)$, depending on whether the lookup table or an NN is employed, respectively.
    The single-branch network~$g$ embeds the sampled modalities $\set{m_1^j, \ldots, m_{\numMods}^j} \subseteq \setModsItem_j$ of all positive and negative items alike, sharing weights across items and modalities. Finally, the embedding for item~$i_j$ is obtained by averaging the embeddings of its different modalities:
    \begin{equation}
        \ve_j = \frac{1}{\numMods}\left( g\left(m_1^j\right) + \ldots + g\left(m_{\numMods}^j\right) \right)
    \end{equation}
    
    The recommendation scores are now computed by taking the scalar product between user and item embeddings, $\ve_i$ and $\ve_j$, respectively. Employing Bayesian Personalized Ranking (BPR)~\cite{bpr}, the recommendation loss can be computed based on these scores by
    \begin{equation}
        \lossBPR^{(u_i, i_j)} = -\sum_{k \in \setNegs_{i}}{\ln\sigma\left( \recScorePred_{i,j} - \recScorePred_{ik} \right)},
        \label{eq:bpr}
    \end{equation}
    \newtext{where $\sigma(x) = \frac{1}{1+e^{-x}}$ is the logistic sigmoid function.} 
    
    While the effectiveness of sharing an embedding network to combine modalities in multimodal ML has been shown~\cite{saeed2023single}, we employ a contrastive loss to align embeddings from different modalities further. This loss is optional, and if it is not applied, $\numMods=1$ and only a single modality is sampled per positive user--item pair. When the contrastive loss is applied, $\numMods=2$ modalities per pair are sampled. In addition to the recommendation loss, \sibrarItem adds the symmetric Noise Contrastive Estimation loss~\cite{vandenoord2018symmetricinfonce,radford2021clip} $\lossInfoS$ as an additional loss term. $\lossInfoS$ is computed between pairs of item modality embeddings of the positive and all the sampled negative items $i_j$ and $i_k, \; k \in \setNegs_i$ for user~$u_i$, respectively. It is defined as the symmetric version of the InfoNCE~\cite{radford2021clip} contrastive loss, 
    \begin{equation}
        \lossInfoS=\lossInfo^{1,2}+\lossInfo^{2,1}
        \label{eq:infos}
    \end{equation}
    for the sampled modalities $m_1$ and $m_2$, and where 
    \begin{equation}
        \lossInfo^{1,2} = - \sum_{(u_i, i_j) \in \batch}{\ln \frac{e^{\left(g\left( m_1^j \right) \cdot g\left( m_2^j \right)\right)/\tau}}{\sum_{k \in \set{j} \cup \setNegs_i} e^{\left(g\left( m_1^j \right) \cdot g\left( m_2^k \right)\right)/\tau}}}.
    \end{equation}
    The set \batch is a batch of randomly sampled user--item pairs. Moreover, the parameter $\tau$ controls the expression's contrastive temperature and is considered a hyperparameter in the experiments conducted in \Secref{sec:experimental-setup}. Intuitively, $\lossInfoS$ enforces similarity between different modality embeddings of the same item whereas dissimilarity between modality embeddings of different items.
    The full loss formulation is now given by 
    \begin{equation}
        \loss = \lossBPR + \alpha \lossInfoS,
        \label{eq:total-loss}
    \end{equation}
    where $\alpha$ controls the weight of the contrastive loss term.

    During inference, \sibrarItem utilizes all available item modalities~$\setModsItem_j$ for any item~$i_j$, embedding them with the single-branch network~$g$ and aggregating the resulting embeddings:
    \begin{equation}
        \ve_j = \frac{1}{\card{\setModsItem_j}}\sum_{m^j \in \setModsItem_j}{g(m^j)}
    \end{equation}
    In case of item cold-start, the empty item profile~$\vi_j$ is not considered, \ie $\vi_j \notin \setModsItem_j$, and the resulting embedding is obtained using only item side information. During inference, a single modality, \ie $\card{\setModsItem_j} \ge 1$ is already sufficient to provide recommendations.

\subsubsection{\sibrarUser} \label{sec:methodology:sibrar:user}
    To mitigate the user cold-start and missing modality problems, \ie when a user's profile~$\vu_i$ or any additional side information is missing, \sibrarUser can be employed. Analogous to \sibrarItem, $h$ now produces the item embeddings~$\ve_j$ either from a learnable lookup table or by embedding the item's profile~$\vi_j$ via an NN. The modality-specific user networks $f_k, \; k \in \set{1, \ldots, \card{\setModsUser}}$, simple one-layer NNs, transform individual user modalities to a common space for further usage. Finally, the single-branch user embedding module~$g$, an NN with several linear layers, batch normalizations, and ReLU activation functions, embeds all user modalities into a shared embedding space. By minimizing the full loss (\Eqref{eq:total-loss}), \sibrarUser enforces similarity between modality embeddings of the same user while dissimilarity between modality embeddings of different users. 
    During inference, \sibrarUser generates the embedding for $\vu_i$ by averaging all available user modality embeddings forwarded through the modality-specific networks~$f_k$ and the single-branch~$g$. We again omit the former for simplicity.
    \begin{equation}
        \ve_i = \frac{1}{\card{\setModsUser_i}}\sum_{m^i \in \setModsUser_i}{g(m^i)}
    \end{equation}

\subsubsection{\sibrarUserItem}
    For completeness, by combining \sibrarUser and \sibrarItem the cold-start and missing modality problems can be mitigated for users and items simultaneously. We term this network \sibrarUserItem. We denote the modality-specific entity networks as $f_k^{user}, \; k \in \set{1, \ldots, \card{\setModsUser}}$ and $f_k^{item}, \; k \in \set{1, \ldots, \card{\setModsItem}}$, for users and items, respectively. Moreover, \sibrarUserItem features two single-branch networks $g^{user}$ and $g^{item}$. The training is done by minimizing the loss formulation
    \begin{equation}
        \loss = \lossBPR + \alpha \lossInfoS^{user} + \beta \lossInfoS^{item}
    \end{equation}
    where  $\lossInfoS^{user}$ and $\lossInfoS^{item}$ are the contrastive losses between user and item modalities, respectively. $\alpha$ and $\beta$ are hyperparameters that control the influence of these terms.

\subsection{\mubrar}
    For comparison of \sibrar with standard multimodal NNs found in the literature, we introduce \mubrar. It captures the core concept of having an individual embedding network, typically a DNN, per modality, that many works use~\cite{clcrec,dropoutnet,chen_2024_mm_video_rec,jeong_2024_mm_rs}. 
    \mubrar outputs a recommendation score~$\recScorePred_{i,j}$ for each given user--item pair~$(u_i, i_j)$ based on user and item embeddings~$\ve_i$ and $\ve_u$, and taking their dot product $\recScorePred_{i,j} = \ve_i \cdot \ve_j$. Like \sibrar, the top $k$ items a user has not interacted with are recommended to them. 
    We now describe \mubrarItem, a variant of \mubrar focusing on mitigating the item cold-start and missing modality problem. As \mubrarUser, the user modality-based version, follows accordingly, we will omit its description and refer to \Secref{sec:methodology:sibrar:user} instead.

\input{resources/figures/architectures/fig-mubrar}
\subsubsection{\mubrarItem}
    \Figref{fig:mubrar:vanilla} shows the architecture and training pipeline of \mubrarItem. For user~$u_i$, its embedding~$\ve_i$ is obtained using either an embedding lookup table or embedding the user profile via a DNN. For item~$i_j$, \mubrarItem embeds each available item modality~$m_k^j \in \setModsItem_j$ via a modality-specific network (or branch)~$f_k$, which are DNNs several layers deep. 
    \reviewTwo{As the aim of this study is to provide an in-depth understanding of \sibrar, we accordingly choose to fuse the individual item modality embeddings by averaging:
    \begin{equation}
        \ve_j = \frac{1}{\card{\setModsItem_j}}\sum_{k=1}^{\card{\setModsItem_j}} f_k(m_k^j)
        \label{eq:mubrar:ej-inference}
    \end{equation}
    This aligns more closely with the core concept of \sibrar and thus facilitates an fair comparison between both architectures. Moreover, it allows for seamless handling of missing modalities by excluding their embeddings from the averaging process. Despite these benefits to our study, we also acknowledge the existence of various other strategies for fusing modality embeddings, \eg embedding concatenation (with or without subsequent processing through a NN) as seen in~\cite{dropoutnet, jeong_2024_mm_rs, chen_2024_mm_video_rec}. As such, we select \dropoutnet as an additional baseline to compare \sibrar against (see \Secref{sec:experiments:baselines}).}

    Considering that \sibrar utilizes modality sampling and a contrastive loss for modality embeddings, \newtext{both of which require further investigation regarding their impact on the recommendation task}, we propose two additional variants of \mubrar: (1) \mubrarSamp adds modality sampling and (2) \mubrarReg further extends this idea by adding a contrastive loss term during training. 
    \reviewTwo{Doing so, these two extensions applied on \mubrar (as an architecture without weight-sharing) allows us to gather detailed insights on the utility and effects of \sibrar's three main training paradigms.}
    
    The modified architecture and training pipeline are shown in \Figref{fig:mubrar:item}. In comparison to \sibrarItem (\Figref{fig:sibrar:item}), there is no single-branch network, while the modality-specific layers feature learning capabilities through more linear layers coupled with ReLU activation functions.  
    
    The main and only differences to \sibrarItem are the missing single-branch network and the stronger learning capabilities of the modality-specific layers. Hence, much of the training procedure follows accordingly. For \mubrarSamp, when sampling modalities, $\numMods=1$ and $\ve_j=f_1(m_1),$ where $m_1 \in \setModsItem_j$ is the sampled modality, and $f_1$ is the corresponding modality network. The training loss is simply the BPR loss, as stated in \Eqref{eq:bpr}.
    For \mubrarReg, when utilizing a contrastive loss on the modality representations, we have $\numMods=2$ and 
    \begin{equation}
        \ve_j=\frac{1}{2} \left( f_{1}(m_{1}) + f_{2}(m_{2}) \right)
    \end{equation}
    for the sampled modalities $\set{m_{1}, m_{2}} \subseteq \setModsItem_j$ and their respective networks~$f_{1}$ and $f_{2}$. The training loss is again computed as in \Eqref{eq:total-loss} and \Eqref{eq:infos}, whereas $\lossInfo^{1,2}$ is now given by 
    \begin{equation}
        \lossInfo^{1,2} = - \sum_{(u_i, i_j)\in \batch}{\ln \frac{e^{\left(f_1\left( m_1^j \right) \cdot f_2\left( m_2^j \right)\right)/\tau}}{\sum_{k \in \set{j} \cup \setNegs_i} e^{\left(f_1\left( m_1^j \right) \cdot f_2\left( m_2^k \right)\right)/\tau}}}.
    \end{equation}
    During inference, all available item modalities $\setModsItem_j$ are used to generate an item embedding~$\ve_j$ (\Eqref{eq:mubrar:ej-inference}), and a dot product is calculated with user embeddings~$\ve_i$ to obtain the recommendation score for user--item pair ($u_i$, $i_j$). 

\input{resources/figures/architectures/fig-mubrar-sc}

%% file: resources/figures/architectures/fig-sibrar.tex
\begin{figure}
    \centering
    \includegraphics[width=\columnwidth]{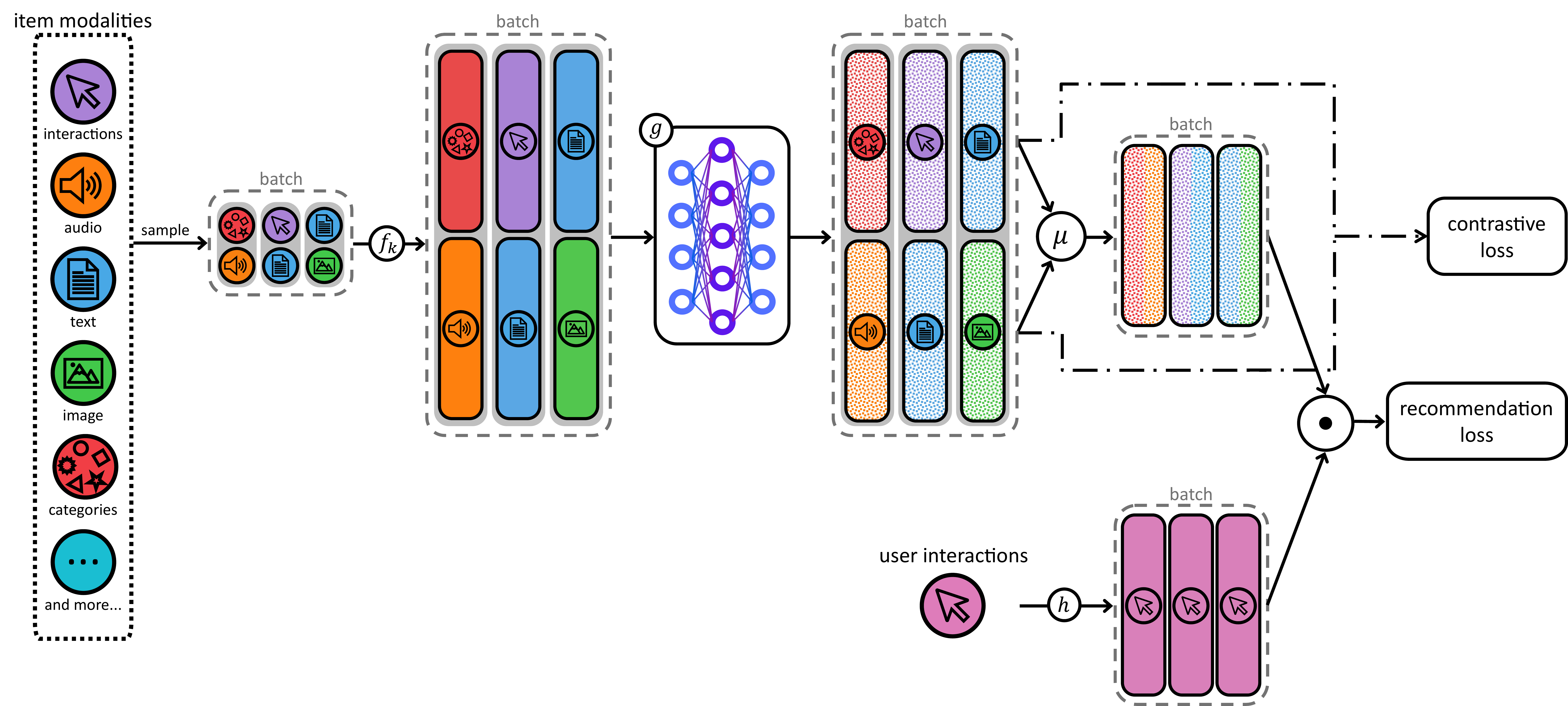}
    \caption{\sibrarItem architecture and its training procedure. \sibrarItem consists of modality-specific item networks $f_k, \; k \in \set{1, \ldots, \numMods}$, the single-branch encoding network $g$ shared across modalities and the user embedding network $h$. For each positive user--item interaction $(u_i, i_j)$ in the training set, two modalities are sampled, independently forwarded through $f_k$ and $g$ before being averaged, indicated by $\mu$, to generate a final representation for item $i_j$. The representation for user $u_i$ is obtained through $h$. Recommendation scores are computed between positive items and the user representations.
    The recommendation loss ($\lossBPR$) considers the difference in scores for positive and negative items. On the other hand, the contrastive loss ($\lossInfoS$) compares the similarity of different modalities of the same item to similarities of modalities to other items in the batch. 
    }
    \label{fig:sibrar:item}
\end{figure}

%% file: resources/figures/architectures/fig-mubrar.tex
\begin{figure}
    \centering
    \includegraphics[width=\columnwidth]{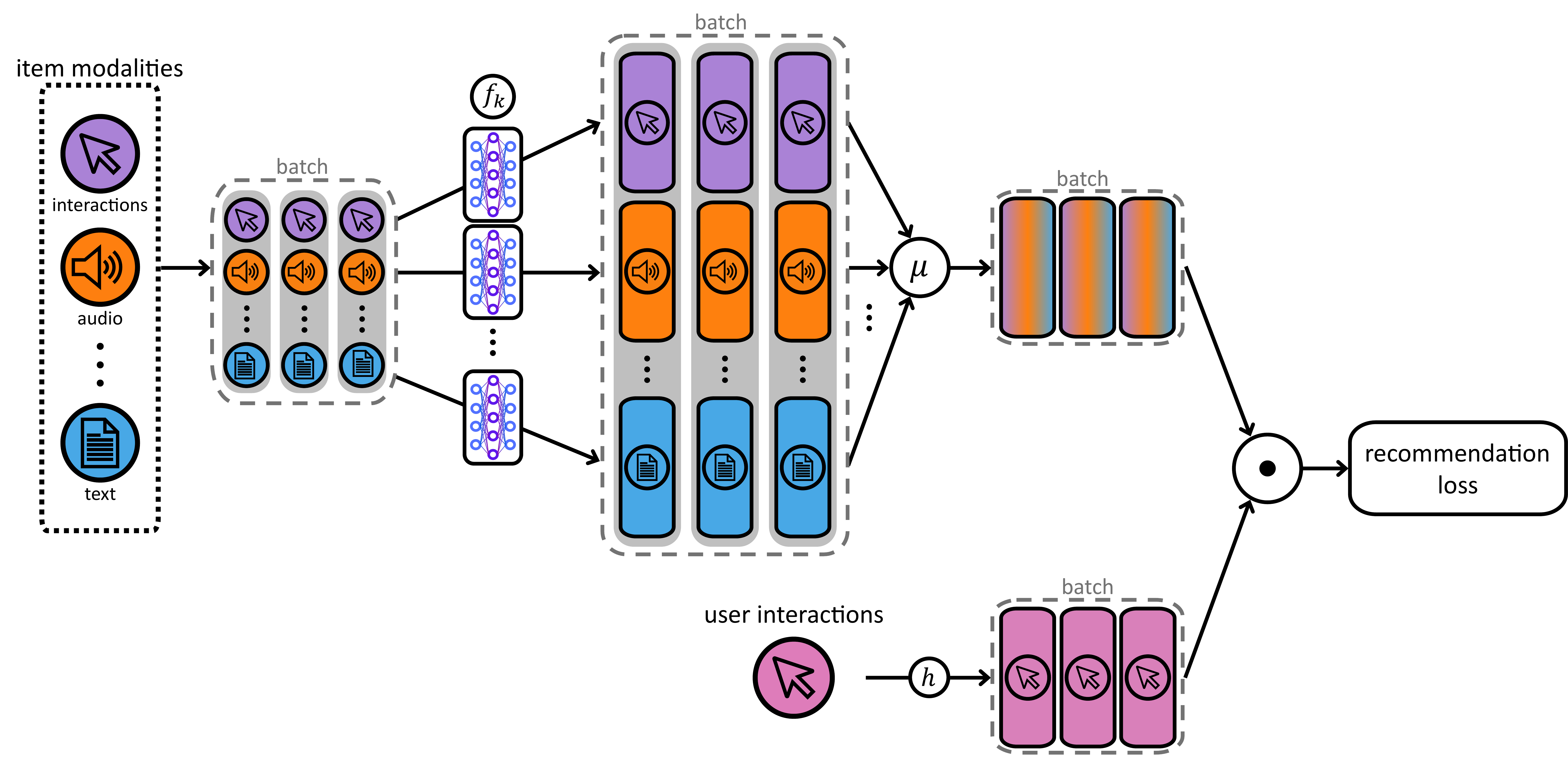}
    \caption{\mubrar model and training procedure. The \mubrar network represents the multi-branch embedding network, and each branch $f_k$ (a DNN), is only used by a single modality. A final item representation is obtained by taking an average over all available modality representations. The representation for a user is obtained through $h$.
    For each user--item interaction pair $(u_i, i_j)$ in the training set, the recommendation score is computed by the dot product between user and item representations. Moreover, the recommendation loss ($\lossBPR$) is computed between the recommendation scores of positive and negative items.
    }
    \label{fig:mubrar:vanilla}
\end{figure}

%% file: resources/figures/architectures/fig-mubrar-sc.tex
\begin{figure}
    \centering
    \includegraphics[width=\columnwidth]{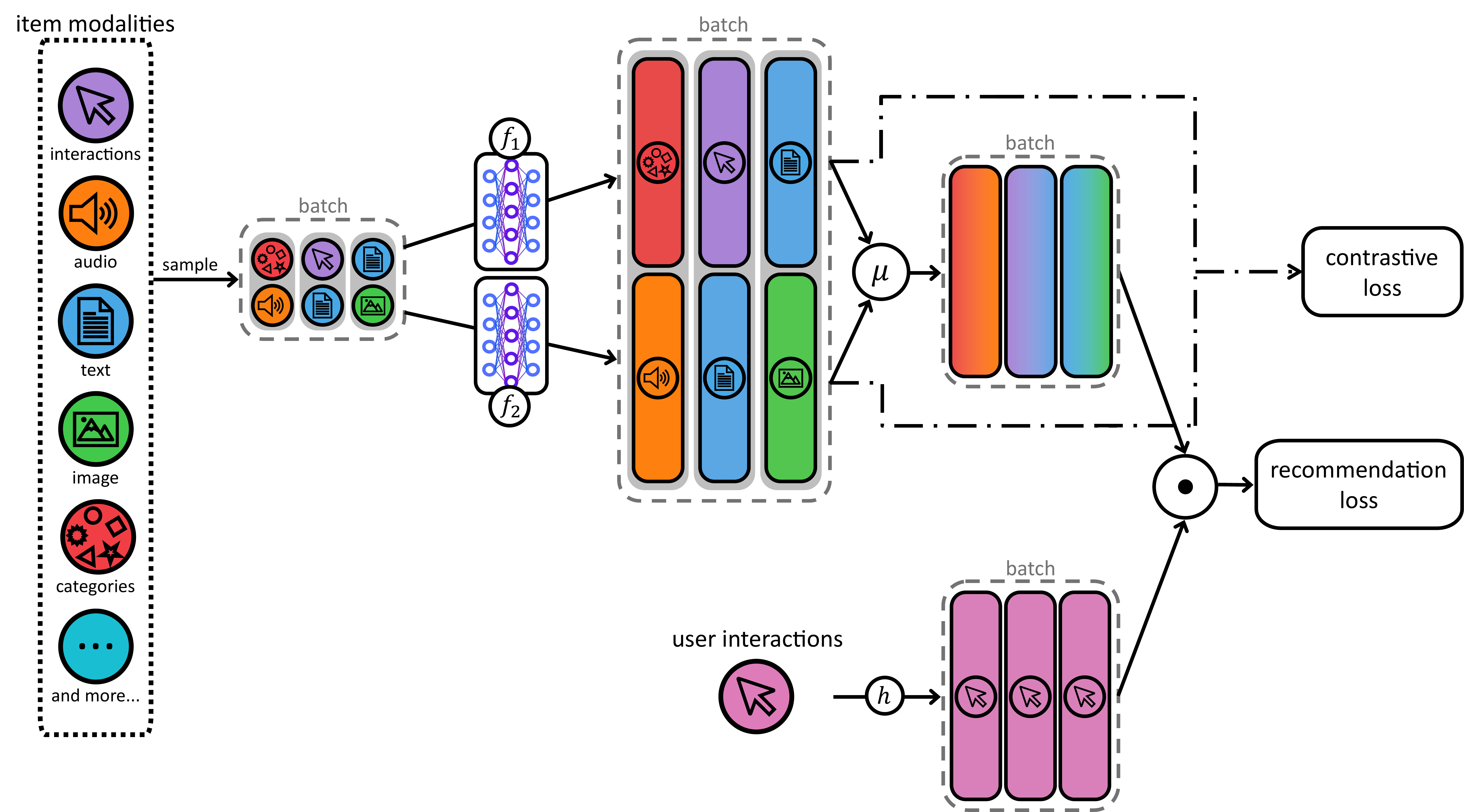}
    \caption{\mubrarSampContr (\mubrarSampContr) architecture and its training procedure. The network consists of modality-specific item networks $f_k, \; k \in \set{1, \ldots, \numMods}$ and the user embedding network $h$. For each positive user--item interaction $(u_i, i_j)$ in the training set, two modalities are sampled (one for \mubrarSamp), independently forwarded through the specific networks $f_k$, before being averaged, indicated by $\mu$, to generate a final representation for item $\vi_j$. The representation $\ve_i$ for user $u_i$ is obtained through $h$. Recommendation scores are computed between positive items and the user representations.
    The recommendation loss ($\lossBPR$) considers the difference in scores for positive and negative items. On the other hand, the contrastive loss ($\lossInfoS$) compares the similarity of different modalities of the same item to similarities of modalities to other items in the batch. 
    \label{fig:mubrar:item}
    }
\end{figure}

%% file: 0_sections/4_experiment_setup.tex
In this section, we describe the experimental setup for our experiments, including the datasets, the baselines used for comparison, the evaluation protocol, the training procedure, and hyperparameter tuning. Moreover, we will publicly release the code used to carry out our experiments.\footnote{\repo} 

\subsection{Datasets}
    \label{sec:dataset}
    \input{resources/tables/tab_dataset_size}
    We use the same dataset selection, preprocessing, and data filtering procedure as described in~\citet{ganhoer_moscati2024sibrar}. 
    We conduct our experiments in three different domains: music, movie, and e-commerce. We carefully selected the datasets by considering their popularity for benchmarking RSs, their number of content features, and the multimodality of available side information.
    
    This work concentrates on the implicit feedback scenario. We therefore construct a binary user--item interaction matrix $\matrixInter$: $\matrixInter_{ij}=1$ if user $u_i$ interacted with item $i_j$ and $R_{ij}=0$ otherwise. 
    
    Table~\ref{tab:dataset-overview} outlines the traits of the datasets after pre-processing (further explained below). The table also gives insights into the modalities and dimensionalities of the side information used in our experiments.

    \subsubsection{Music4All-Onion} (\onion)
        \footnote{\url{https://zenodo.org/records/6609677}}~\cite{music4all_onion} 
        This large-scale multimodal music dataset provides three types of information: 1) user--item interactions, 2) side information on users, and 3) side information on items.
        As side information on users, we select their gender and country. The dataset provides a wide range of item features, including several representations linked to the music tracks' audio signals, lyrics, and YouTube video clips. From these, we select those pertaining to the audio signal in terms of i-vectors~\cite{eghbal2015vectors} and the visual representations of the YouTube video clips extracted with a pre-trained instance of ResNet~\cite{he2016resnet} for our experiments.
        
        As in~\cite{ganhoer_moscati2024sibrar}, we extend the available representations for audio and lyrics by applying more recent NN architectures for music information retrieval and natural language processing to the provided data. 
        We select the NNs \texttt{MusicNN}\footnote{\url{https://github.com/jordipons/musicnn/tree/master}}~\cite{pons2018musicnn1, pons2019musicnn2} and \texttt{Jukebox}\footnote{\url{https://github.com/openai/jukebox/}}~\cite{openai2020jukebox} to extract representations of the $30s$-audio-snippets. After pre-processing the lyrics as described in~\cite{music4all_onion}, we encode the text using  \texttt{all-mpnet-v2}\footnote{\url{https://huggingface.co/sentence-transformers/all-mpnet-base-v2}} pre-trained instance of the SentenceTransformer model~\cite{reimers2019sentencebert} provided by HuggingFace~\cite{wolf2020huggingface}.
                
        We only consider users of age between 10 and 80 and users and items with all side information available. We further follow~\cite{ganhoer_moscati2024sibrar} by creating a subset of listening events over the year 2018 for our experiments to follow a standard practice in the music RS domain~\cite{protomf, gender_fair}. 

        Finally, we transform the interactions to binary implicit feedback, similar to~\cite{protomf,ganhoer_adv_mult_vae, melchiorre2021genderfair}. 
        \newtext{We use a threshold of 2 on the user--item interaction counts, \ie a user must have listened to an item at least twice, and apply \textit{5-core filtering} for users and items, which only keeps users with at least 5 interactions and items that have been interacted with by at least 5 users.}

    \subsubsection{MovieLens 1M} (\mlonem)
        \footnote{\url{https://grouplens.org/datasets/movielens/1m/}}~\cite{movielens} 

        This dataset provides user ratings of movies gathered from the MovieLens website.\footnote{\url{https://movielens.org/}} The dataset consists of user--item interactions, side information on users (age, gender, and occupation) and on items (movie genre). 
        
        As in~\citet{ganhoer_moscati2024sibrar}, we enhance the dataset by crawling Wikipedia for movie plots and further encode these texts with the \texttt{all-mpnet-v2}\footnote{\url{https://huggingface.co/sentence-transformers/all-mpnet-base-v2}} pre-trained instance of the SentenceTransformer model~\cite{reimers2019sentencebert} provided by HuggingFace~\cite{wolf2020huggingface}.
        
        We filter the data to only include users between 10 and 80 years old and those users and items for which all features are available in the dataset. We transform the rating information into binary implicit feedback by keeping only the interaction information of ratings $>=3$. 
        Finally, we apply 5-core filtering on users and items alike.

    \subsubsection{Amazon Video Games} (\amazonvid)
        \footnote{\url{https://amazon-reviews-2023.github.io/index.html}}~\cite{hou2024amazonreviews}
        Amazon Reviews'23 is a large-scale dataset which provides Amazon reviews dated from May 1996 to September 2023. This dataset is an expanded version of the well-established Amazon Reviews'18 dataset~\cite{ni2019amazonreviews}.
        
        \amazonvid provides various item features, including the title, description, price, and at least one product image link per item. In this work, we chose to take the category \textit{Video Games} as a subset of the \amazonvid dataset for our experiments. Following~\cite{ganhoer_moscati2024sibrar}, we decided on this subset by carefully considering multiple factors, such that the resulting subset consists of a large enough set of users, items, and user--item interactions to replicate a real-world scenario, and provides unique features considering our other selected datasets. We filter the reviews to be dated between January 2016 and December 2019. Moreover, we keep only items with available titles, descriptions, and at least one high-resolution image per product. We consider the ratings as implicit feedback with a threshold of 1 and apply 5-core filtering. 
        As in~\cite{ganhoer_moscati2024sibrar}, we use \texttt{all-mpnet-v2}\footnote{\url{https://huggingface.co/sentence-transformers/all-mpnet-base-v2}} pre-trained instance of the SentenceTransformer model~\cite{reimers2019sentencebert} provided by HuggingFace~\cite{wolf2020huggingface} to encode the texts (product title and description) of this dataset. 
        After downloading the first high-resolution image of the product, we preprocess it following the same approach as~\citet{saeed2023single}. First, the image is rescaled such that the smaller edge has a size of $256$. The resulting image is then cropped at the center to a square of $224$ pixels. Each channel is then normalized to the mean and standard deviation of ImageNet~\cite{deng2009imagenet}. We then compute the encoding of the final image with the pre-trained instance of \texttt{ResNet-101}~\cite{he2016resnet} provided by TorchVision\footnote{\url{https://pytorch.org/vision/main/models/generated/torchvision.models.resnet101.html}}~\cite{marcel2010torchvision}.
        
\input{resources/figures/fig-split-random}
\input{resources/figures/fig-split-cold-start}
\subsection{\newtext{Data Scenarios}}\label{sec:data-split}     
    \newtext{To assess the performance of the RSs in different scenarios, we split the datasets in three different ways, each featuring a train, validation, and test set.} While models are trained on the train set, hyperparameter optimization is done on the validation set. The test set is only used to report the final results, estimating a model's performance on unseen data.

    \paragraph{Warm start (random split).} We randomly select 80\%, 10\%, and 10\% of each user's interactions for training, validation, and testing, respectively. \Figref{fig:exp:random} illustrates this scenario.

    \paragraph{User cold start.} We randomly split 80\%, 10\%, and 10\% of all the users in the dataset into disjoint sets of train, validation, and test users, respectively.
    RSs are trained on all user interactions in the train set while ignoring those in validation and test sets. During validation (testing), only validation (test) users are considered, while train and test (validation) users are disregarded. Since test users have not been seen during model training and validation/model selection, this split is suitable for assessing the performance of RSs in user cold-start scenarios. \Figref{fig:exp:cold-start:user} visually explains this user cold-start scenario.
    
    \paragraph{Item cold start.} We randomly split 80\%, 10\%, and 10\% of all items in the dataset into disjoint sets of train, validation, and test items, respectively.
    RSs are trained on all item interactions in the train set while ignoring those in validation and test sets. During validation (testing), only validation (test) items are considered, while train and test (validation) items are disregarded. Since test items have not been seen during model training and validation/model selection, this split is suitable for assessing the performance of RSs in item cold-start scenarios. \Figref{fig:exp:cold-start:item} visually explains this item cold-start scenario.
    
\subsection{Baselines}\label{sec:experiments:baselines}
    We compare \sibrar with two \hrss for cold-start, \clcrec~\cite{clcrec} and \dropoutnet~\cite{dropoutnet}, which are often used as baselines addressing cold-start scenarios, leveraging CF signals and multimodal side information. Moreover, they employ both explicit (\clcrec) and implicit (\dropoutnet) approaches for multimodal learning (see \Secref{sec:related-work}).  
    Additionally, we include two \cfbr models, matrix factorization (\mf) with BPR~\cite{bpr} and Deep Matrix Factorization (\dmf)~\cite{dmf}. We select the former based on its simplicity and yet effectiveness in recommendations tasks by factorizing user and item latent representations. \dmf features a two-tower DNN with user and item interactions as inputs.  
    To compare different single- and multi-branch multimodal networks, we additionally consider the variants described in \Secref{sec:methodology}. We further employ two naive baselines recommender systems (\baserecs):  
    \rand, which, for each user, randomly samples items to recommend, and \pop, which recommends to all users the same most popular items -- as described by the number of individual user interactions in the training set.
    
\subsection{Metrics}
\label{sec:experiments:metrics}
We evaluate the performance of our algorithms by accuracy and beyond-accuracy metrics. While the former estimates whether users would consume recommended items, the latter measures aspects beyond consumption, which can, among others, indicate possible biases the algorithms might exhibit. Moreover, we measure the \textit{modality gap}~\cite{liang_2022_modality_gap,humer2023amumo} by two metrics between the modalities of an item and between different items.

Let $\setRel(u_i) \subset \setItems$ be the list of relevant items for user~$u_i$. When evaluating and testing an algorithm, these are the items in the respective validation or test split.
$\setRet(u_i) \subset \setItems, k=\card{\setRet(u_i)}$ is the ordered list of \topk~items recommended to the user, and $\setRet(u_i)_{j}$ selects the $j$-th~item of this list. $\indicatorRelui(i_j)$ indicates that item~$i_j$ is present in $\setRel(u_i)$, more formally 
\begin{equation*}
    {\indicatorRelui}(i_j) := 
    \begin{cases}
        1 & \text{if $i_j \in \setRel(u_i)$}, \\
        0 & \text{if $i_j \notin \setRel(u_i)$}.
    \end{cases}
\end{equation*}
In the following, we define the metrics on a user level, noting that only averages over all users are reported later on. 
For accuracy and beyond-accuracy metrics, where applicable, we test the best algorithms for significant improvement over the other algorithms using multiple paired t-tests, with Bonferroni correction to account for the multiple comparisons. We consider the improvement statistically significant if $p < 0.05$.

\subsubsection{Accuracy metrics}
\label{sec:experiments:metrics:accuracy}
    We first describe \ndcg, our primary measure of recommendation quality, followed by additional metrics. 

    \textit{Normalized Discounted Cumulative Gain} (\ndcg) evaluates the ranking quality of an RS. In short, it weighs good recommendations, \ie ones that a user likes, stronger the earlier in the list of recommended items they occur. When computed on \topk recommendations, \ndcgk is given by 
    \begin{equation*}
        \ndcgk(u_i) = \frac{\dcgk(u_i)}{\idcgk(u_i)},
    \end{equation*}
    where \dcgk is the Discounted Cumulative Gain (DCG)
    \begin{equation*}
        \dcgk(u_i) = \sum_{j = 1}^{k}\frac{{\indicatorRelui}(\setRet(u_i)_{j})}{\log_2(j+1)}
    \end{equation*}
    and \idcgk is the Ideal Discounted Cumulative Gain (IDCG), when all recommended items are relevant, \ie $\idcgk(u_i) = \sum_{j = 1}^{k}\frac{1}{\log_2(j+1)}$ if there are $k$ relevant items for user $u_i$.

    \textit{\Recall and \precision} for user~$u_i$ only consider the presence of a relevant item in the recommendations, ignoring the order of items in the list:  
    \begin{equation*}
        \reck(u_i) = \frac{\card{\setRel(u_i) \cap \setRet(u_i)}}{\card{\setRel(u_i)}} \qquad
        \text{and} \qquad
        \preck(u_i) = \frac{\card{\setRel(u_i) \cap \setRet(u_i)}}{\card{\setRet(u_i)}}.
    \end{equation*} 

    \textit{\fscorek} applies a harmonic mean on \reck and \preck to represent both metrics at once: 
    \begin{equation*}
        \fscorek(u_i) = 2\frac{\preck(u_i) \cdot \reck(u_i)}{\preck(u_i) + \reck(u_i)}.
    \end{equation*}
    
    \textit{Average Precision} (\ap), also known as \textit{Mean Average Precision} (\map) when averaged over all users, considers the order of recommended items by evaluating \preck at multiple levels of $k$. 
    \begin{equation*}
        \apk(u_i) = \frac{\sum_{j=1}^{k}{\indicatorRelui}(\setRet(u_i)_{j}) \cdot \metricAt{\precisionName}{j}(u_i)}{\card{\setRel(u_i)}}
    \end{equation*}

    \textit{Reciprocal Rank} (\rrName)~\cite{trec8qa}, also known as \textit{Mean Reciprocal Rank} (\mrr) when averaged over all users, assumes that users are satisfied after having encountered the first relevant item recommended at a high rank.
    \begin{equation*}
        \rrk(u_i) = \frac{1}{min_k\set{k ; \setRet(u_i)_{k} \in \setRel(u_i)}}
    \end{equation*}

\subsubsection{Beyond-accuracy metrics}
    A wide variety of metrics exist beyond the common accuracy-based model performance indicators above. In this work, we focus on four metrics that can shed light on the popularity bias of RSs, which describes how RSs might recommend already popular items more often while scarcely or never recommending low popular items. This is problematic due to its adverse effects on many different stakeholders, \eg users may not find items of interest or content providers may miss sales opportunities if their items are not recommended. 
    For a survey on popularity bias, we refer the reader to~\cite{klimashevskaia_survey_pop_bias}.
    
    \textit{\Coverage} indicates what proportion of items of the whole catalog are recommended during evaluation. 
    \begin{equation*}
        \covk = \frac{\card{\bigcup_{u_i \in \setUsers}{\setRet(u_i)}}}{\card{\setItems}}
    \end{equation*}
    
    \textit{Average recommendation popularity} (\arpName)~\cite{klimashevskaia_2022_mitigating_pop_bias} investigates the average popularity of items recommended to the users. While the popularity~$\phi(i_j)$ of an item~$i_j$ can be measured in different ways, we use $\phi(i_j)=\frac{\card{\setUsers_j}}{\card{\setUsers}}$ with $\setUsers_j$ indicating training set users that interacted with the item~\cite{muellner_2024_impact_diff_privacy}.
    \begin{equation*}
        \arpk(u_i) = \frac{1}{\card{\setRet(u_i)}}\sum_{i_j \in \setRet(u_i)}\phi(i_j)
    \end{equation*}

    \textit{Average Percentage of Long Tail Items} (\aptName)~\cite{klimashevskaia_2022_mitigating_pop_bias, abdollahpouri_2019_managing_pop_bias} assesses the proportion of long tail items~$\Gamma$, \ie set of items with the least popularity in the recommendations. We consider 80\% of the least popular items as the long tail. 
    \begin{equation*}
        \aptk(u_i) = \frac{\card{\setRet(u_i) \cap \Gamma}}{\card{\setRet(u_i)}}
    \end{equation*}

    \textit{Popularity lift} (\plName)~\cite{abdollahpouri_2020_connection_popularity} measures how RSs distort the popularity bias of users inherent in their histories. 
    \begin{equation*}
        \plk(u_i) = \frac{\text{PB}_q(u_i) - \text{PB}_p(u_i)}{\text{PB}_p(u_i)},
    \end{equation*}
    where $\text{PB}_p(u_i) = \frac{1}{\card{\setRet(u_i)}}\sum_{i_j \in \setRet(u_i)}\phi(i_j)$ and $\text{PB}_q(u_i) = \frac{1}{\card{\setRel(u_i)}}\sum_{i_j \in \setRel(u_i)}\phi(i_j)$ measure the average item popularity of user $u_i$'s history and recommendations, respectively. 
    
    \subsubsection{Describing the embedding space}
    \label{sec:experiments:metrics:embedding-space}
        We provide quantitative measures to describe the modality gap to gain insights on the shared embedding space of multimodal models. This gap is linked to a model's performance on downstream tasks~\cite{liang_2022_modality_gap}.
        
        \textit{Euclidean distance} (\euclideanDist) and \textit{cosine similarity} (\cosineSim) measure the distance and similarity between two $d$-dimensional vectors $a \in \mathrm{R}^d$ and $b \in \mathrm{R}^d$, respectively. 
        \begin{equation*}
            \euclideanDist(a, b) = \norm{a - b} 
            \qquad \text{and} \qquad 
            \cosineSim(a, b) = \frac{a \cdot b}{\norm{a}\norm{b}},
        \end{equation*}
        where $\norm{\cdot}$ is the Frobenius norm.
        
        \textit{Inter-item metrics} measure the average difference in the modality embeddings of the same type between different items. Let $\setMods$ be the set of $\numMods$ unique modalities of all $\numItems$ items.
        We define $e_j^{m_k} \in \mathrm{R}^d$ to be the embedding for the modality $m_k \in \setMods$ of item $i_j \in \setItems$. The inter-item metric is then given by

        \begin{equation*}
            \inter_f = \frac{1}{\numMods}\frac{2}{\numItems(\numItems-1)}\sum_{k=1}^{\numMods}\sum_{i=1}^{\numItems}\sum_{j=1}^{i-1}{f(e_i^{m_k}, e_j^{m_k})},
        \end{equation*}
        
        where $f(\cdot, \cdot)$ is either \euclideanDist or \cosineSim. Based on the symmetry of $f$, we only compute it once per pair. Moreover, we neglect measuring embeddings to themselves. A low $\interEucl$ expresses close proximity of the same modalities of different items while a low $\interCos$ indicates strong dissimilarity between the modalities.

        Analogously, \textit{intra-item metrics}, which quantify the average difference between modalities of the same item, are defined as
        \begin{equation*}
            \intra_f = \frac{1}{\numItems}\frac{2}{\numMods(\numMods-1)}\sum_{j=1}^{\numItems}\sum_{k=1}^{\numMods}\sum_{l=1}^{k-1}{f(e_j^{m_k}, e_j^{m_l})}.
        \end{equation*}
    
\subsection{Hyperparameter Tuning} \label{sec:hyperopt}
    We perform an extensive hyperparameter search on \sibrar and all other algorithms reported in \Secref{sec:experiments:baselines}. Where applicable, we tune their learning rates, weight decays, dropout rates, model dimensions, and number of layers. For explicit \hrss, \ie those with additional loss terms, we search over possible loss parameters, including contrastive weights and temperatures for \sibrar and \mubrar.
    Moreover, for \hrss, we treat training modalities as another hyperparameter. For \dropoutnet, \sibrar, and \mubrar, which are not restricted by the number of modalities, we try out all possible combinations of various sizes (1 up to \numMods). To make a fair comparison with \clcrec, which supports only one modality besides interactions, we additionally tune all \hrss on only sets consisting of interaction and another modality. We denote the best-performing \hrss in this setting by \dropoutnetTwo, \sibrarTwo and \mubrarTwo, while their best-performing counterparts leveraging any number of modalities as \dropoutnetBest, \sibrarBest and \mubrarBest, respectively.
    All algorithms are trained for up to 50 training epochs. For algorithms requiring negative samples, we sample $10$ for each positive user-item pair. For computing the metrics (\Secref{sec:experiments:metrics}), we use the cut-off threshold $k=10$.
    After each training cycle, the algorithms are evaluated on the validation set and the \ndcgten is computed and used as validation metric for hyperparameter selection. We perform early-stopping by continuously storing the best model over the epochs and stopping the training when no further \ndcgten improvements can be observed for five consecutive epochs. 
    Only the models achieving the highest \ndcgten on the validation set are evaluated on the test set, ensuring no information flow from testing to model training.
    We exclusively use data present in the test sets when reporting the outcomes of our experiments (\Secref{sec:results}).
    We utilize the Weights and Biases (W\&B) platform~\cite{wandb}\footnote{\url{https://wandb.ai/}} for monitoring and the Bayesian optimization of W\&B Sweeps for tuning the hyperparameters. 
    For an overview of all experiments performed and ranges of hyperparameters employed, we refer the reader to \Tabref{tab:hyperparams} in the appendix as well as the the configuration files available in our repository.\footnote{\repoConfigs}

%% file: resources/tables/tab_dataset_size.tex
\begin{center}
\begin{table*}
    \scalebox{0.93}{
    \begin{tabular}{l|rrrr||cccc|cccc}
         &  &  &  &  &\multicolumn{4}{c|}{User Features} & \multicolumn{4}{c}{Item Features} \\
            & \# users & \# items & \# inter.  & sparsity & \rot{Age (\textbf{d})} & \rot{Gender (\textbf{c})} & \rot{Country (\textbf{c})} & \rot{Occupation (\textbf{c})} & \rot{Text (\textbf{v})} & \rot{Audio (\textbf{v})} & \rot{Image (\textbf{v})} & \rot{Genre (\textbf{m})} \\ 
        \midrule
        \onion~\cite{music4all_onion} & 5{,}192    & 13{,}610    & 407{,}653         & 0.99423  &--&2&152&--&768&50; 100; 4{,}800&4{,}096&853\\
        \mlonem~\cite{movielens} & 5{,}816    & 3{,}299    & 813{,}792         & 0.95759 &7&2&--&21&768&--&--&18\\
        \amazonvid~\cite{hou2024amazonreviews} & 11{,}454& 4{,}177&87{,}098&0.99818&--&--&--&--&768; 768&--&2{,}048&--\\
    \end{tabular}}
    \vspace{0.6em}
    \caption{This table provides the summary of datasets. Left side: Characteristics after pre-processing. Right side: Representation type (\textbf{c}ategorical, \textbf{d}iscrete, \textbf{v}ector, \textbf{m}ultilabel), dimensionality (for \textbf{v}) or number of options (for \textbf{c}, \textbf{d}, and \textbf{m}) of each content feature available. Dimensionalities separated by a colon indicate that more than one feature is available; features that are not available or have been neglected are denoted by~-- (see Section~\ref{sec:dataset}).}
    \label{tab:dataset-overview}
\end{table*}
\end{center}

%% file: resources/figures/fig-split-random.tex
\begin{figure}
    \centering
    \begin{tabular}{ll} 
        \toprule
        users   & item interactions           \\
        \midrule
        user $1$ & \textcolor{violet}{3, 1, 32, 8, 18, 10,} \textcolor{teal}{7, 37,} \textcolor{purple}{10, 21}   \\
        user $2$ & \textcolor{violet}{2, 25, 24, 31,} \textcolor{teal}{19,} \textcolor{purple}{7} \\ 
        user $3$ & \textcolor{violet}{1, 39, 15, 9, 19, 31,} \textcolor{teal}{22, 3,} \textcolor{purple}{2, 6}    \\
        $\cdots$    &        \multicolumn{1}{c}{$\cdots$}                \\
        user $n$ & \textcolor{violet}{38, 9, 11, 28, 22, 10, 27,} \textcolor{teal}{3, 23,} \textcolor{purple}{33, 27} \\
        \bottomrule
    \end{tabular}
    \caption{Outline on a warm-start dataset scenario. The item interactions of each user in the dataset are split randomly into \textcolor{violet}{train}, \textcolor{teal}{validation}, and \textcolor{purple}{test} set, as indicated by the different colors. Each user and item is present in the train set, and may be in the validation and test sets.
    }
    \label{fig:exp:random}
\end{figure}

%% file: resources/figures/fig-split-cold-start.tex
\begin{figure}
    \centering
    \begin{subfigure}[b]{0.35\columnwidth}
        \centering
        \includegraphics[width=\textwidth]{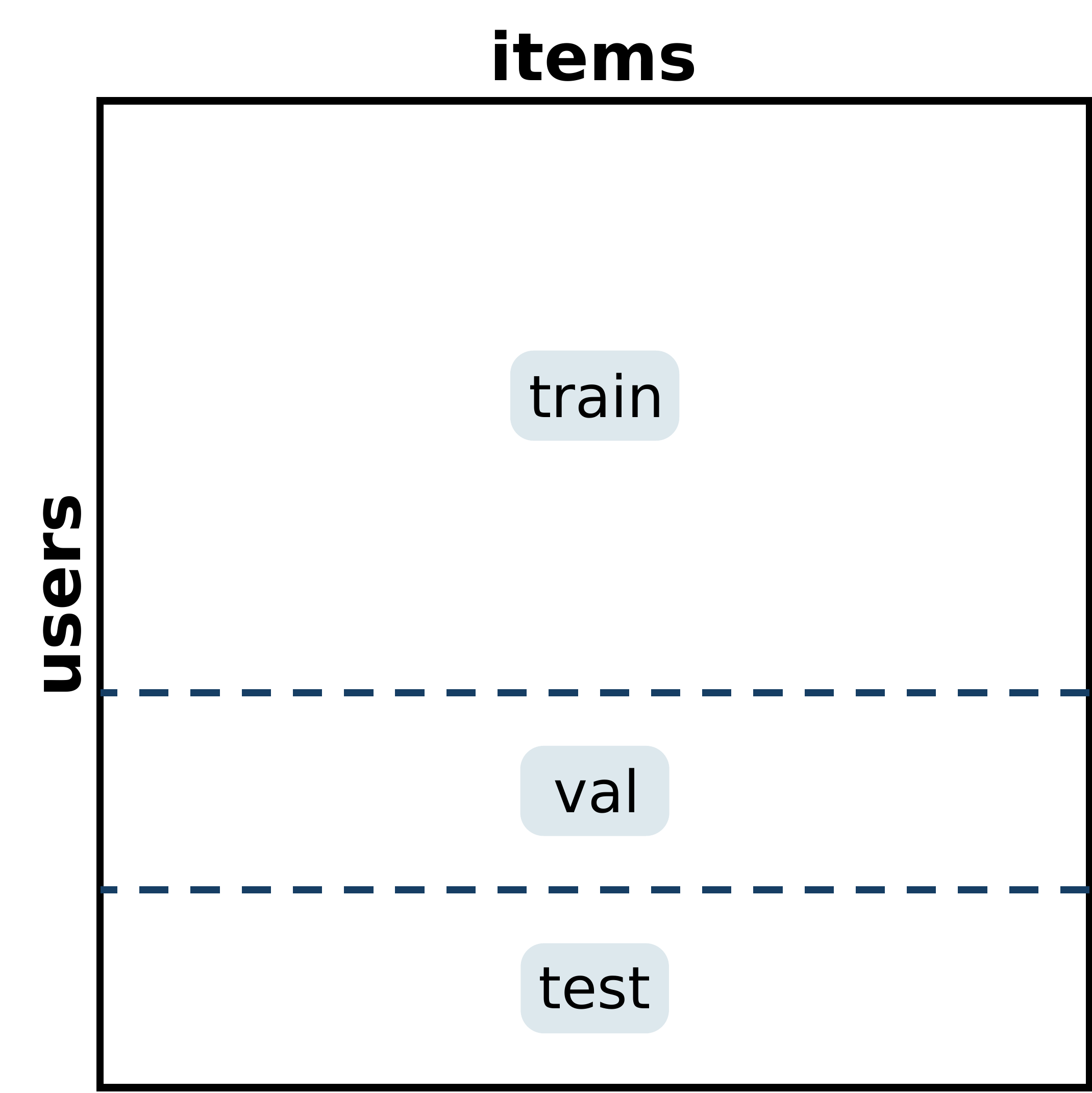}
        \caption{user cold-start}
        \label{fig:exp:cold-start:user}
    \end{subfigure}
    \hspace{1.5em}
    \begin{subfigure}[b]{0.35\columnwidth}
        \centering
        \includegraphics[width=\textwidth]{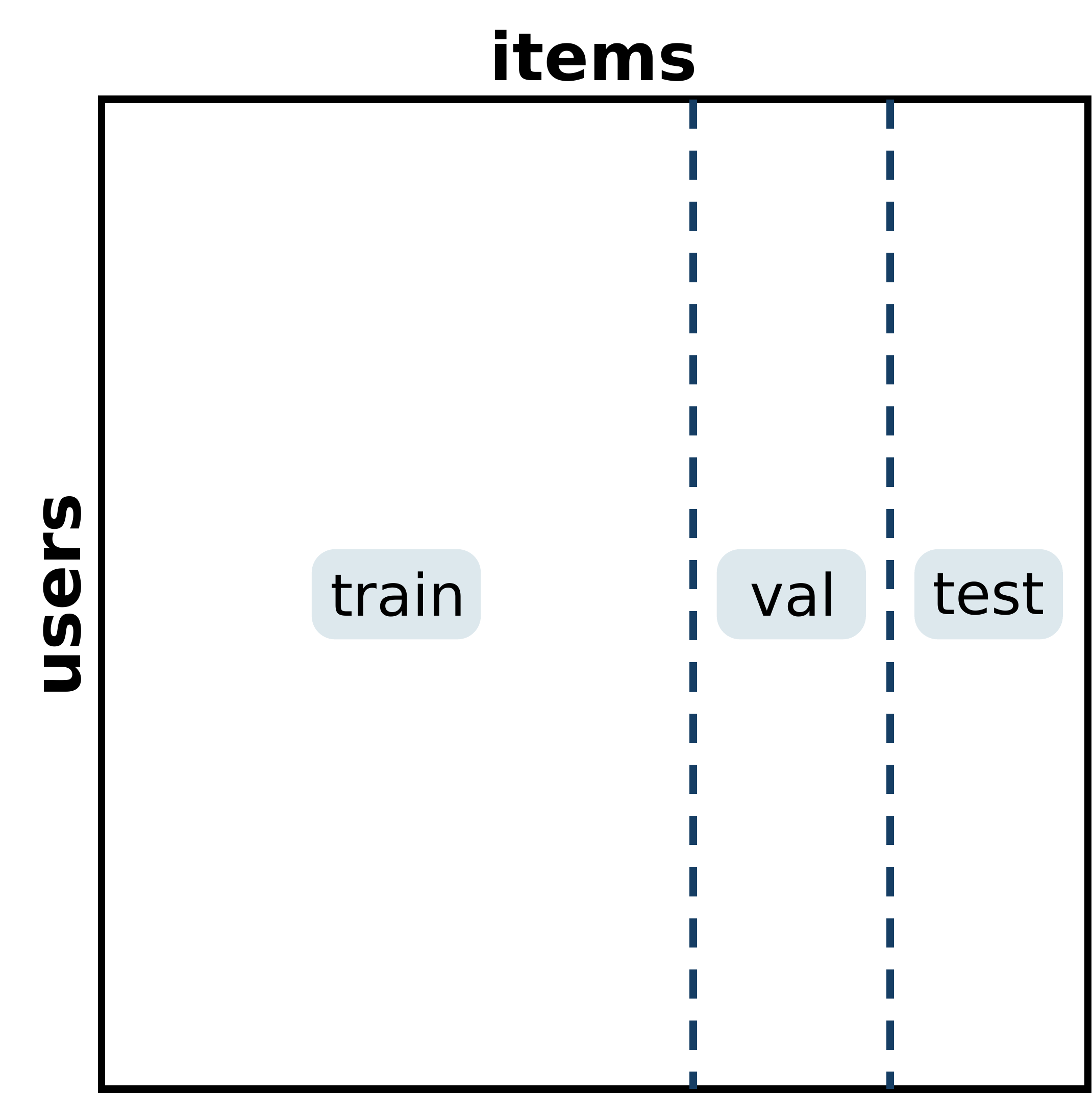}
        \caption{item cold-start}
        \label{fig:exp:cold-start:item}
    \end{subfigure}
    \vspace{-0.5em}
    \caption{Outline on user and item cold-start scenario. The whole square represents all users and items of a dataset, which can be thought of as the interaction matrix containing user--item interaction pairs. It is split into train, validation, and test sets, as indicated by the different subregions labeled \textit{train}, \textit{val}, and \textit{test}, respectively.  
    In (a), users are split randomly such that no user exists in more than one set. Analogously, in (b), no item exists in more than one set.
    }
    \label{fig:exp:cold-start}
\end{figure}

%% file: 0_sections/5_results.tex
In this section, we first compare the performance of \sibrar to standard \cfbrs and \hrss, an then analyze different single- and multi-branch variants and their components. Additionally, we study how contrastive loss parameters affect single-branch networks, and conclude by investigating the modality gap of multimodal RSs. 

\subsection{Baseline comparison}
    The primary aim of this section is to answer \ref{rq:1} (see \Secref{sec:intro}). In the following, we compare \sibrar with basic, collaborative filtering, and hybrid recommenders, including multi-branch variants.
    \input{resources/tables/tab_results_ndcg}
    \subsubsection{Accuracy}\label{sec:results:baseline:accuracy}
        \Tabref{tab:results-ndcg} shows the \ndcgten test scores of all algorithms on different datasets, in warm-start as well as user and item cold-start scenarios. Considering the lack of user side information of the \amazonvid dataset, for this dataset algorithms have only been trained on warm and item cold start, while they have been trained on all scenarios for \mlonem and \onion. 
        Each row indicates the \ndcgten test results of a specific algorithm (see \Secref{sec:experiments:baselines}), which are further grouped by the horizontal solid lines into basic (\baserec), collaborative filtering-based (\cfbr), and hybrid (\hrs) recommender systems. For brevity, we jointly report the best variant of \sibrarItem and \sibrarUser as \sibrar, and of \mubrarItem and \mubrarUser as \mubrar.
        Moreover, for algorithms that allow more than one modality in addition to interactions, \ie \dropoutnet, \mubrar, and \sibrar, for a fair comparison with \clcrec, we list the models leveraging only a single side-modality with the suffix \textit{one}, \eg \sibrarTwo, and those that report the best performance for any number of modalities with the suffix \textit{best}, \eg \sibrarBest. The table further indicates this by the dashed line separating these \hrss. 
        We refer the reader to the appendix for results on the other accuracy metrics mentioned in \Secref{sec:experiments:metrics:accuracy}.

        \paragraph{Warm start.}
            In warm start, \ie where user and item interaction histories are also leveraged for recommendation, \mf, \clcrec, \mubrar, and \sibrar  are the four top-performing algorithms. The order varies with the datasets: \clcrec performs best on \mlonem, \mubrar on \onion, and \mf on \amazonvid. Considering only algorithms with at most one modality besides interactions, \clcrec achieves a top performance on \mlonem, \sibrar on \onion, and \mf on \amazonvid.
        
        \paragraph{Cold start.}
            The weakness of \baserecs and \cfbrs is evident when considering cold-start scenarios, where \hrss outperform them. While a more considerable improvement over \cfbr techniques is observable for item than for user cold-start, we attribute this to the more extensive and informative set of item modalities compared to user modalities (see \Secref{sec:dataset}). As only the demographics of cold-start users are available, the difference between \hrss and \cfbrs is less distinct. Across all three datasets and two cold-start scenarios, \sibrar outperforms all other algorithms. The difference is even more substantial when the model can leverage any number of modalities (\sibrarBest). This indicates the efficiency of \sibrar in leveraging multimodal information available during inference while not being dependent on interactions.

    \input{resources/tables/tab_results_beyond_accuracy}
    \subsubsection{Beyond-accuracy}\label{sec:results:baseline:beyond}
        \Tabref{tab:results-beyond-acc} reports the beyond-accuracy results of all algorithms on warm-start datasets. The rows again refer to the individual \baserecs, \cfbrs, and \hrss. For each dataset, we report \arpten, \aptten, and \plten test scores as columns.\footnote{We refer the reader to the appendix for a table on \covten.}
        The arrows next to the metric names indicate which values hint at a lower popularity bias of the recommendations. A lower value for \arpten ($\downarrow$) means that less popular items are recommended. For \aptten ($\uparrow$), a higher number refers to more tail items being recommended. \plten ($\rightarrow$) measures the discrepancy between the popularity of items in users' histories and in the recommended lists. A value close to $0$ indicates that RSs match users' interests of item popularity. We highlight the best result per metric in bold, and significant improvements with $\dagger$.

        Unsurprisingly, \rand recommends significantly more unpopular tail items than any other algorithm (lowest \arpten; highest \aptten) and more unpopular ones than the items in the users' histories (most negative \plten). In contrast, \pop recommends the most popular items (highest \arpten; lowest \aptten), even more popular items than what users like (most positive \plten). \newtext{Besides \baserecs, the extent to which algorithms exhibit biases in comparison to other algorithms depends on the dataset and metric considered: 
        While \cfbrs are less biased than \hrss on \mlonem, they show a greater bias toward popular items over \mubrar and \sibrar on \onion. For \amazonvid, all \hrss except for \dropoutnet show less bias than \mf, while often more than \dmf.}
        
        Comparing \sibrar and \mubrar, \arpten (\aptten) is lower (higher) for both \mubrar models (\mubrarTwo and \mubrarBest) on \mlonem and \amazonvid, while higher (lower) for \onion. Furthermore, \sibrar displays a more substantial shift toward popular items for \mlonem and \amazonvid, while a weaker one for \onion. 

        Moreover, the beyond-accuracy metrics indicate that \sibrar is one of the least popularity-biased recommenders on modality-rich datasets like \onion, and significantly less biased than others when the bias is measured by \plten. 

    \input{resources/figures/missing_modalities/fig-missing-modality-random-onion-best}
    \input{resources/figures/missing_modalities/fig-missing-modality-random-onion-diff-best}
    \subsubsection{Missing modality}\label{sec:results:baseline:mm}
        \Figref{fig:mm:random:onion} shows the \ndcgten results of \sibrarBest~(a) and \mubrarBest~(b) models on the warm-start \onion dataset. Both models are trained on five modalities (interactions, genres, image, audio, and text) and evaluated on different subsets of these modalities while their weights are frozen.\footnote{We choose the \onion dataset for a fine-grained analysis considering its richer set of modalities compared to the other datasets.}$^{,}$\footnote{We refer the reader to the appendix for plots on the other variants and datasets.} The bar plot in each subfigure presents the test scores for specific subsets of modalities. In addition, the bars are sorted by increasing score and horizontal dashed lines enable the comparison to the best warm-start \cfbrs and \hrss. The middle plot shows which subsets of the modalities are used for evaluation. Each row corresponds to a different modality, a colored box indicates the presence of the modality during evaluation, while a missing box indicates its absence. Finally, the lower plot summarizes the number of modalities used per run.
        
        \paragraph{\sibrarBest} 
            \Figref{fig:mm:onion:sibrar} shows the results for \sibrarBest. The plot indicates that genres are the least informative as a stand-alone modality. Moreover, text and audio are comparable in the resulting \ndcgten, slightly worse than image. 
            By utilizing any modality except for genres, \sibrarBest achieves better results than \dropoutnet. Leveraging any three modality combination or any combination containing interactions, \sibrarBest outperforms \mf, \dmf, and \clcrec. Only \mubrarBest achieves better results in warm start when all modalities are used. 
            The utility of using interactions in recommender systems is underlined by the fact that the single-branch network only achieves peak \ndcgten results when using interactions as part of the input. Moreover, interactions alone already lead to a stronger performance than for almost all two-pair modalities.
            \newtext{In cold start, \ie when no interactions are available and only the modalities \setText{\textit{genres, image, audio, text}} are used as input, \sibrarBest displays its strength by maintaining its advantage over \mf, \dmf, \clcrec, and \dropoutnet while only suffering a slight performance decline.}\footnote{\newtext{This is observable in \Figref{fig:mm:random:onion} by considering the column for which all but the block for interaction are colored.}}
        
        \paragraph{\mubrarBest}
            \Figref{fig:mm:onion:mubrar} shows the results for \mubrarBest, for which genres again reach the lowest \ndcgten when used alone, followed by audio, text, and image. Using interactions or any combination of two modalities, \mubrarBest achieves a higher \ndcgten than \dropoutnet. When any set of at least three modalities (except for \setText{\textit{genres, audio, text}}) is available, or for interactions and any additional modality, \mubrarBest outperforms \dmf, \clcrec, and \mf. Finally, as long as only text is missing, \mubrarBest performs better than warm-start \sibrarBest.

        \paragraph{Comparison}
            To better understand differences between single- and multi-branch networks, we directly compare the \ndcgten results of \sibrarBest and \mubrarBest for specific sets of modalities. \Figref{fig:mm:random:onion:diff:best} shows the percentage difference in \ndcgten ($\%\Delta \ndcgten$) between both algorithms, where we sort the bars by the number of modalities used at test time. A positive and negative value indicate that \sibrarBest performs better and worse than \mubrarBest, respectively. Furthermore, an asterisk $*$ indicates a statistically significant difference of results.
            Especially when only small subsets of all modalities are available, the single-branch network shows its merits by achieving significant performance gains in comparison to its multi-branch counterpart, which is the case in 16 out of 31 modality subsets. In detail, the \ndcgten scores improve by up to 40\%.
            When more modalities become available, the difference in recommendation quality decreases, and the models perform comparably. Finally, \mubrarBest is significantly better than \sibrarBest in three cases, in each of which both algorithms utilize \setText{\textit{interactions, genres, image}} in addition to one or more other modalities as input, \eg when all modalities are available. 

    \paragraph{Takeaway on RQ1.}
        To investigate how \sibrar compares to other algorithms on recommendation performance (\ref{rq:1}), we studied the results on different data scenarios. 
        Our evaluation across accuracy and beyond-accuracy metrics revealed that \sibrar exhibits comparable performance to other \hrss and specifically to \mubrar in warm- and cold-start scenarios. Notably, single-branch recommenders have an advantage over multi-branch recommenders in settings with missing modalities in which \sibrar outperforms its competitors. 
        Overall, the findings demonstrate the effectiveness of single-branch recommenders in leveraging any available data, \mmedits{which we attribute to the effectiveness of weight-sharing in combining multimodal data~\cite{tschannen2023clippo}. Moreover, as modality sampling serves as a way to model missing modality during training, it may improving cold-start performances.}
        
\input{resources/tables/tab_results_sibrar_mubrar_ndcg}
\subsection{Training paradigms}
    \newtext{In this section, we aim to gain insights on modality sampling and contrastive loss by analyzing their effects on single- and multi-branch architectures, therefore addressing~\ref{rq:2}.}

    \subsubsection{Accuracy}
        \Tabref{tab:results-ndcg:sibrar-mubrar} shows the impact of the sampling (S) and the contrastive loss (C) on single- and multi-branch networks in terms of \ndcgten.\footnote{We again refer the reader to the appendix for tables on all other accuracy metrics.} \sibrarSampContr and \sibrarSamp describe a single-branch network with and without contrastive loss, respectively. \newtext{Both versions are outlined in \Figref{fig:sibrar:item}. Further, \mubrarVanilla (see \Figref{fig:mubrar:vanilla}) corresponds to a vanilla multi-branch network. Finally, \mubrarSamp features additional modality sampling and \mubrarSampContr features both, modality sampling and a contrastive loss (see \Figref{fig:mubrar:item}).}
        We refer the reader to \Secref{sec:methodology} for more detail on the algorithms.
        In line with \Tabref{tab:results-ndcg}, we evaluate all variants on the warm- and cold-start datasets, select the runs performing best on the validation set, and report their \ndcgten results on the test set in \Tabref{tab:results-ndcg:sibrar-mubrar}. 
        In warm start, the best-performing variant strongly depends on the dataset: For \mlonem, \mubrarVanilla has a slight advantage over \mubrarSampContr and \sibrarSampContr. For \onion, \mubrarSampContr is the best-performing model, followed by \sibrarSampContr and \mubrarSamp. Finally, \sibrarSampContr is significantly better than the other variants for \amazonvid, followed by \sibrarSamp and \mubrarSampContr. 
        The strength of \sibrar becomes apparent in user and item cold-start scenarios. Here, for each recommendation domain (movie, music, products), \sibrarSampContr outperforms the \mubrar variants. 
        Comparing different variants in each model class, employing modality sampling in \mubrar increases its \ndcgten five out of eight times across datasets and scenarios. Additionally, using a contrastive loss further improves the results in seven out of eight cases (all except for user cold start on \mlonem), featuring a performance improvement of \mubrarSampContr over \mubrar six out of eight times.
        Similarly, for single-branch networks, which use modality sampling by default, including a contrastive loss increases \ndcgten scores of \sibrarSampContr over \sibrarSamp six out of eight times.
        Overall, the results indicate that modality sampling and a contrastive loss can strongly improve the accuracy of single- and multi-branch architectures across all datasets, for warm and cold start alike.
        
    \input{resources/tables/tab_results_beyond_accuracy_sibrar_mubrar}
    \subsubsection{Beyond-accuracy}
        Analogously to \Secref{sec:results:baseline:beyond} and \Tabref{tab:results-beyond-acc}, we evaluate different variants of single- and multi-branch networks on all warm-start datasets for popularity bias. \Tabref{tab:results-beyond-acc:sibrar-mubrar} reports the results for \arpten, \aptten and \plten.
        Studying different multi-branch variants, \mubrar demonstrates the lowest popularity biases for all metrics on \mlonem and \onion. For \amazonvid, it shows comparable results to \mubrarSamp on \arpten and \plten, while better results on \aptten. On the other hand, \mubrarSampContr exhibits a lower popularity bias than \mubrarSamp on \mlonem, while a higher one on \onion and \amazonvid. 
        In contrast, \sibrarSampContr is more biased than \sibrarSamp on \mlonem, while less biased on \onion and \amazonvid. 
        Due to the disparity between popularity bias and recommendation quality, we leave it up to practitioners to trade-off which of the variants to employ in practice.
    
    \input{resources/figures/missing_modalities/fig-missing-modality-random-onion-diff-samp-contr.tex}
    \input{resources/figures/missing_modalities/fig-missing-modality-random-onion-diff-spec.tex}
    \input{resources/tables/tab_missing_modalities_number_random_onion}
    \input{resources/tables/tab_missing_modalities_number_cs_onion}
    \subsubsection{Missing modality}\label{sec:results:sib-mubr:mm}
        \Figref{fig:mm:random:onion:diff:samp-contr} compares the percentage difference in \ndcgten between the best single- and multi-branch networks trained on five modalities (interactions, genres, image, audio, and text) with and without modality sampling (S) and contrastive loss (C) on warm-start \onion. 
        
        \paragraph{\mubrar}
            Specifically, \Figref{fig:mm:random:onion:diff:samp-contr:mubrar} shows the difference between \mubrarSampContr and \mubrar. Here, we sort the bars by the number of modalities used at test time. \newtext{We use a logarithmic scale for the y-axis to accommodate the large performance disparity between \mubrar and \mubrarSampContr in missing modality. }Considering the significant improvements for all modality combinations, the results indicate a strong positive influence of both sampling and contrastive loss on the multi-branch network. Moreover, the \ndcgten improvement tends to be more pronounced when many modalities are unavailable (left side of the figure). 
            
        \paragraph{\sibrar}
            Analogously, \Figref{fig:mm:random:onion:diff:samp-contr:sibrar} summarizes the differences in performance for \sibrar. Again, utilizing a contrastive loss leads to significant improvements in \ndcg. Its benefits are especially useful when modalities are scarce. Overall, the results indicate a positive influence of sampling and contrastive loss on single- and multi-branch recommenders.
            
        \paragraph{Training in missing modality.}
            \newtext{Next, considering that only a subset of the modalities may be available at inference, we compare the recommendation quality of \sibrar trained on all available modalities to instances trained only on those modalities that are also available during inference. Note that training on a smaller set reduces the computational resources required for training and is thus an important aspect for practitioners. 
            Hence, we compare \sibrarBest, the best single-branch network trained on five modalities of warm-start \onion (see \Secref{sec:results:baseline:accuracy}) to newly initialized \sibrar instances trained on specific subsets of modalities, while using the same hyperparameters as \sibrarBest. We collectively refer to these networks as \sibrarSpec.}
            \Figref{fig:mm:random:onion:diff:mod-specific} reports the percentage change in \ndcgten of \sibrarBest over \sibrarSpec per subset of modalities. While the performance in modality-rich scenarios is comparable\footnote{The slight discrepancy is due to randomness in negative and modality sampling.}, \sibrar performs significantly better than specific \sibrarSpec instances across almost all modalities subsets. 
            This result illustrates that training single-branch networks on all available modalities may be beneficial for inference, even if many modalities are unavailable \newtext{and the training is more resource intensive.}

        \paragraph{\newtext{Conclusion on missing modality.}}
            \Tabref{tab:missing-modality:number-of-mods:random:onion} summarizes all previous observations by reporting the average \ndcgten scores of single- and multi-branch variants on warm-start \onion of \mubrar, all trained on five modalities and evaluated on all possible subsets of these modalities.
            While each row represents a different variant, each column corresponds to the number of modalities used during evaluation. The reported values are an average of the results over all possible combinations of a specific number of modalities.
            The table shows the merits of modality sampling, contrastive loss, and of the single-branch architecture. Across all numbers of modalities, \mubrarSampContr is, on average, better than \mubrarSamp, which in turn is, on average, better than \mubrar. Similarly, \sibrarSampContr is better than \sibrarSamp. Moreover, \sibrarSampContr is, on average, substantially better than \mubrarSampContr when only few modalities are available and comparable when at most two modalities are missing. Finally, \sibrarSpec only achieves similar results to \sibrarSampContr in warm-start.
            Conducting a similar analysis on item cold-start \onion, \Tabref{tab:missing-modality:number-of-mods:cs:onion} indicates similar results: \sibrarSampContr shows a solid improvement for low modality counts, and this improvement vanishes with an increasing number of modalities. 
            We refer the reader to the appendix for similar analyses on \mlonem and \amazonvid.

    \input{resources/figures/contrastive_loss/fig-contrastive-loss-gridsearch}
    \input{resources/figures/contrastive_loss/fig-contrastive-loss-ratio}
    \subsubsection{Contrastive loss}
        This section aims to understand the impact and effect of the contrastive loss as well as its two parameters on \sibrar: (1), the weight~$\alpha$, which balances recommendation and contrastive loss terms (\Eqref{eq:total-loss}), and (2), the temperature~$\tau$ controlling the parametrization of the \lossInfoS\xspace loss (\Eqref{eq:infos}). 

        To this end, we perform a grid search over various combinations of contrastive weight~$\alpha$ and temperature~$\tau$ on warm-start \onion, selecting the hyperparameters of \sibrarBest as the basis for this experiment. Specifically, fixing all other hyperparameters, we only vary $\alpha$ and $\tau$, train a new model from scratch, and assess its accuracy and modality gap on the test set.  
        
        \Figref{fig:contrastive:gridsearch:onion} reports the \ndcgten, intra-item cosine similarity and intra-item Euclidean distance results (see \Secref{sec:experiments:metrics:embedding-space}) of different weight--temperature combinations as heatmaps. The left plot correspond to \ndcgten, and the middle and right plot to the intra-item cosine similarity and Euclidean distance, respectively.
        Moreover, the weight $\alpha$ and temperature $\tau$ are plotted logarithmically on the x- and y-axis in each heatmap. The colored bar to the right of each plot outlines the value ranges of the resulting metrics.  

        \paragraph{\ndcgten.}
            Considering the \ndcgten (left-most plot in \Figref{fig:contrastive:gridsearch:onion}), a subregion of well-performing experiments spans diagonally across the heatmap, from low temperatures and low weights to high temperatures and high weights.
            In contrast, experiments with very low weights and high temperatures achieve suboptimal results, similar to ones with high weights and low temperatures.
            To further understand the relationship between temperature and weight, \Figref{fig:contrastive:ratio:onion} reports the ratio $\alpha / \tau$ of the tested weights and temperatures on the X-axis and the \ndcgten result of each run n the y-axis. 
            The horizontal dashed line corresponds to the average performance of models with weight $\alpha=0$ (\ie no contrastive loss). 
            The results indicate the presence of an optimal proportion for the contrastive loss parameters to improve the performance of single-branch networks -- for \onion between $10^{-3}$ and $10^{-1}$.\footnote{This proportion likely also depends on the model class and task-specific losses, which the contrastive loss has to be weighed against.}    
            Using this insight for training, fixing a reasonable value for either weight or temperature and scanning over a broad range for the other parameter might simplify the hyperparameter search. 

        \paragraph{Intra-Item cosine similarity.}
            Analogous to the previous paragraph, the middle plot in \Figref{fig:contrastive:gridsearch:onion} shows the intra-item cosine similarities on warm-start \onion. We again observe a diagonal-spanning subregion with very high similarity values. Off-diagonal combinations, or when the contrastive loss has no influence ($\alpha=0$), exhibit the lowest similarity values. 
            Finally, the diagonal high-similarity subregion roughly aligns with the previous one on \ndcgten, hinting at a positive relationship between intra-item modality similarity and model accuracy. 

        \paragraph{Intra-Item Euclidean distance.}
            The right plot of \Figref{fig:contrastive:gridsearch:onion} demonstrates the direct effect of contrastive weight and temperature on the shared embedding space of \sibrarSampContr. The plot indicates that with decreasing temperature and increasing weight, different modalities of the same items are further pushed together.

        \paragraph{Takeaway on RQ2.}
            \newtext{In the above, we studied the effects of the different training paradigms employed in \sibrar~(\ref{rq:2}). The results suggest that while modality sampling alone may be harmful for model performance in warm start, its merits are evident in missing modality scenarios. \mmedits{ This is to be expected, since modality-sampling can be considered as a way to model missing modality during training. }
            Employing a contrastive loss alongside the modality sampling benefits single- and multi-branch networks alike; both exhibit significantly better results across warm-start and missing-modality scenarios. \mmedits{The improved performance can be explained by the fact that the contrastive loss enhances the similarity between the representations of each modality and that of the collaborative signal, rendering each of them more effective for recommendation.}
            Finally, the optimal choice of the contrastive loss parameters (\ie weight and temperature) is essential for training \sibrar. While combinations of low weights and high temperatures may lead to suboptimal performances in terms of accuracy, high weights might harm the recommendation objective of \sibrar.} \mmedits{The potential negative impact of the contrastive loss on recommendation is to be attributed to the fact that both high weights and low temperatures lead to larger components of the contrastive loss in the overall loss function; these high values could therefore render the signals from the recommendation loss negligible during training.}
            
\input{resources/figures/projections/fig-embedding-space-onion-sibrar}
\input{resources/figures/projections/fig-embedding-space-onion-all-others}
\subsection{Modality gap}
    In this section, we investigate the ability of single- and multi-branch networks to reduce the modality gap by mapping different modalities of the same items into similar regions in the embedding space.\footnote{\cg{We note that studying the modality gap is limited to multimodal algorithms with shared embedding spaces. Thus, as \dropoutnet concatenates it's modalities, we do not study it. Moreover, as our main interest lies in understanding the effects of weight sharing, modality sampling and contrastive loss, we refer the reader to the appendix for results on \clcrec.}}
    This corresponds to answering \ref{rq:3}.
    
    \subsubsection{Qualitative analysis}
        For our analysis, we consider single- and multi-branch variants trained on five item modalities (interactions, genres, audio, image, and text) of warm-start \onion. \Figref{fig:projection:onion:sibrar} shows the input (left) and output (right) embedding space of the single-branch network $g$ of \sibrarSampContr. To visualize the higher-dimensional embeddings in 2D, we consider only the first $10$ principal components obtained through principal component analysis (PCA). We do so to reduce the computational cost for t-Distributed Stochastic Neighbor Embedding (t-SNE)~\cite{hinton2002tnse}, which we apply afterward to further reduce the dimensionality to $2$. For PCA and t-SNE, we use the implementations and default parameters provided by the \textit{scikit-learn}~\cite{sklearn_api} library.\footnote{\url{https://scikit-learn.org/}}
        To enhance perceptibility, we only show the projections of a subset of all items by sampling 500 items uniformly at random.\footnote{See \Figref{fig:projection:all} in the appendix for the projections of \mlonem and \amazonvid.}
        
        \newtext{The left plot in \Figref{fig:projection:onion:sibrar} shows that before applying the single-branch network, all modalities cover separate regions in the input space.
        In contrast, in the output of the single-branch network (right), the individual regions overlap. The worsened separability indicates that the embeddings of different modalities of \sibrarSampContr become more similar, which, in turn, the downstream recommendation task can profit from.}

        \newtext{
        Similarly, \Figref{fig:projection:onion:mubrar-sibrars} shows projections of the embedding spaces of all single- and multi-branch variants. 
        While different modalities embedded by \mubrar form individual subspaces in their joint embedding space, these spaces start overlapping for \mubrarSamp, which indicates a positive influence of modality sampling to reduce the modality gap in multi-branch networks. Finally, the projections of \mubrarSampContr suggest that employing a contrastive loss benefits \mubrar by encouraging its independent branches to map different modalities to the same region in the embedding space. This observation is further supported by comparing the projections of \sibrarSamp and \sibrarSampContr. In \sibrarSampContr, the modalities exhibit reduced separability in the embedding space, indicating more similar representations.}
    
    \input{resources/tables/tab_intra_inter_comparison_cosine}
    \subsubsection{Quantitative analysis}
        To verify the above observations, we quantitatively evaluate the individual embedding spaces of the different \sibrar and \mubrar variants on all warm-start datasets. 

        \paragraph{Cosine similarity.}
            \Tabref{tab:intra-inter-comparison:cosine} reports the average intra- and inter-item modality cosine similarities (see \Secref{sec:experiments:metrics:embedding-space}).\footnote{\newtext{We refer the reader to the appendix for results on \metricEuclidean.}} Each row corresponds to the best model of a specific variant. The columns indicate the dataset and whether they have been computed between different modalities of the same item (intra) or between the same modalities of different items (inter). The cosine similarity is bound by $-1$ and $1$, where values close to -1 and 1 indicate high and low similarity, respectively. 
            \newtext{We indicate the algorithm with the highest average similarity between modalities in bold.}
            Finally, the reported values are an average over the measurements between all possible modality pairs for the respective metrics. 
            As the interpretation for a single result is difficult, we focus on comparing \textit{intra} and \textit{inter} results to draw insights. 
        
            For \mubrar, inter-item cosine similarities are higher than intra-item similarities for all datasets. For \mubrarSamp, which includes sampling, the iter-item cosine similarity is lower than the intra-item similarity for \mlonem, while lower for \onion and \amazonvid. For \mubrarSampContr, which is further equipped with a contrastive loss, and for single-branch networks \sibrarSamp and \sibrarSampContr, the intra-item cosine similarities are consistently lower than the inter-item similarity for all datasets. Moreover, the intra-item values continuously increase when employing sampling and a contrastive loss alongside \mubrar and \sibrar. Except for the \sibrar models on \onion, an increase in intra-item cosine similarity coincides with a decrease in inter-item cosine similarity. Finally, for \mlonem and \onion, \sibrarSampContr achieves a substantially higher intra-item similarity than all other variants and a comparable similarity for \amazonvid.
            Our results indicate that by utilizing modality sampling and a contrastive loss, the single-branch forces multimodal models to push modality representations of the same items closer together than representations of different items. 
            
        \input{resources/tables/tab_modality_prediction}
        \paragraph{Modality prediction.}
            We now study whether the aforementioned shared embedding spaces express the interchangeability of different modalities. To this purpose, we study whether it is possible to determine to which modality type a given embedding vector belongs to.             
            \Tabref{tab:modality-prediction} reports the average accuracy of random forest classifiers\footnote{We use the implementation of scikit-learn~\cite{sklearn_api} with default parameters (\url{https://scikit-learn.org/1.5/modules/generated/sklearn.ensemble.RandomForestClassifier.html}).} trained on this classification task. Each row in the table corresponds to a different single- and multi-branch variant. Moreover, each column reports the results for a specific warm-start dataset. 
            We use the embeddings of the datasets' test splits and further split them for our classification experiment into train and test items with ratios 80:20 by sampling them uniformly at random. Next, we fit a random forest classifier on each train set and evaluate it on the test set. For stronger validity of the results, we repeat this process of dataset splitting and classifier training with $20$ different seeds and report the average accuracy of all classifiers. We also report the expected accuracy of a random predictor. \newtext{Bold values indicate the the algorithms with the least separable embeddings, while $\dagger$ that the classifiers perform significantly worse than on other methods.} 
            For \mubrar and \mubrarSamp, the classifiers are able to perfectly discern the modality types of different embeddings ($\text{accuracy}=1$), indicating low similarity between the embeddings. The classification task is slightly more difficult for \mubrarSampContr, for which classifiers are not able to predict the modalities perfectly on \onion and \amazonvid. 
            Finally, the results of \sibrarSamp and \sibrarSampContr highlight similarities in the modality embeddings in the embedding space, making discerning the modality classes more challenging. \newtext{Although the classifiers trained on \sibrarSampContr embeddings exhibit statistically significant lower accuracies, their performance remains substantially above random.} This suggests persistent semantic differences between modality types, and together with previous insights from \Secref{sec:results:baseline:mm} and \Secref{sec:results:sib-mubr:mm}, the benefits of utilizing all available modalities for generating recommendations.

        \paragraph{Takeaway on RQ3.}
            \newtext{In the section, we investigated the ability of single- and multi-branch networks and their training paradigms to reduce the modality gap between different modalities of the same items (\ref{rq:3}). Our qualitative and quantitative analysis indicates that modality sampling, contrastive loss, and the single-branch network have positive effects on reducing the modality gap by encouraging the models to encode different modalities of the same item more closely together while modalities of different items further apart. Moreover, applying these paradigms increases the difficulty in determining the modality for a given embedding, further suggesting an increase in modality similarity. Overall, the results suggest that reducing the modality gap is a potential factor for increased model performance in missing modality scenarios.}

%% file: resources/tables/tab_results_ndcg.tex
\begin{center}
\begin{table*}
    \centering
    \begin{adjustbox}{max width=\textwidth}
    \begin{tabular}{ll|rrrcrrrcrr}
          & & \multicolumn{3}{c}{Warm-Start} & & \multicolumn{3}{c}{Item Cold-Start} & & \multicolumn{2}{c}{User Cold-Start} \\
          & & \mlonem & \onion & \amazonvid & & \mlonem & \onion & \amazonvid & & \mlonem & \onion \\
          
          \cmidrule{1-5}
          \cmidrule{7-9}
          \cmidrule{11-12}
          \multirow{2}{*}{\baserec} & \rand & 0.0045 & 0.0009 & 0.0012 & \bfseries  & 0.0497 & 0.0074 & 0.0118 & \bfseries  & 0.0445 & 0.0074 \\
          & \pop & 0.1268 & 0.0182 & 0.0255 & \bfseries  & 0.0087 & 0.0063 & 0.0083 & \bfseries  & 0.4583 & 0.0994 \\
         \cmidrule{1-5}
         \cmidrule{7-9}
         \cmidrule{11-12}
         \multirow{2}{*}{\cfbr} & \mf & 0.2589 & 0.1438 & \bfseries 0.0818$\dagger$ & \bfseries  & 0.0300 & 0.0055 & 0.0125 & \bfseries  & 0.4613 & 0.0920 \\
          & \dmf & 0.1165 & 0.1297 & 0.0556 & \bfseries  & 0.0087 & 0.0063 & 0.0083 & \bfseries  & 0.4124 & 0.0575 \\
         \cmidrule{1-5}
         \cmidrule{7-9}
         \cmidrule{11-12}
         \multirow{7}{*}{\hrs} & \clcrec & \bfseries 0.2680$\dagger$ & 0.1378 & 0.0569 & \bfseries  & 0.1762 & 0.1503 & 0.1348 & \bfseries  & 0.4616 & 0.0987 \\
          & \newresults{\dropoutnetTwo} & \newresults{0.2510} & \newresults{0.1078} & \newresults{0.0413} & \newresults{\bfseries } & \newresults{0.2000} & \newresults{0.1221} & \newresults{0.1033} & \newresults{\bfseries } & \newresults{0.4582} & \newresults{0.0947} \\
          & \mubrarTwo & 0.2603 & 0.1471 & 0.0535 & \bfseries  & 0.2751 & 0.2289 & 0.1262 & \bfseries  & 0.4553 & 0.0946 \\
          & \sibrarTwo & 0.2593 & 0.1505 & 0.0728 & \bfseries  & 0.2592 & 0.2183 & \bfseries 0.1479$\dagger$ & \bfseries  & \bfseries 0.4673 & 0.0918 \\
         \cdashline{2-5}[3.0pt/3.0pt]
         \cdashline{7-9}[3.0pt/3.0pt]
         \cdashline{11-12}[3.0pt/3.0pt]
          & \newresults{\dropoutnetBest} & \newresults{0.2510} & \newresults{0.1078} & \newresults{0.0455} & \newresults{\bfseries } & \newresults{0.2000} & \newresults{0.1689} & \newresults{0.1033} & \newresults{\bfseries } & \newresults{0.4582} & \newresults{0.1086} \\
          & \mubrarBest & 0.2603 & \bfseries 0.1762 & 0.0535 & \bfseries  & 0.2949 & 0.2276 & 0.1262 & \bfseries  & 0.4501 & 0.0951 \\
          & \sibrarBest & 0.2581 & 0.1717 & 0.0728 & \bfseries  & \bfseries 0.2994 & \bfseries 0.2297 & \bfseries 0.1479$\dagger$ & \bfseries  & 0.4659 & \bfseries 0.1125 \\

\cmidrule{1-5}
\cmidrule{7-9}
\cmidrule{11-12}
    \end{tabular}
    \end{adjustbox}
    \vspace{0.6em}
    \caption{\ndcgten results for different \baserecs, \cfbrs, and \hrss. Each row indicates the results of the evaluated algorithm, and the columns blocks indicate the data scenario (warm start and user/item cold start) for the datasets (\mlonem, \onion, and \amazonvid). 
    Bold indicates the best-performing RS, $\dagger$ indicates significant improvement over all others (paired $t$-tests with $p<0.05$ considering Bonferroni correction).
    The results of the different scenarios are not comparable due to the different numbers of users and items in the test splits.
    }
    \label{tab:results-ndcg}
\end{table*}
\end{center}

%% file: resources/tables/tab_results_beyond_accuracy.tex
\begin{center}
\begin{table*}
    \centering
    \begin{adjustbox}{max width=\textwidth}
    \begin{tabular}{ll|rrr|rrr|rrr}
            & & \multicolumn{3}{c|}{\mlonem} & \multicolumn{3}{c|}{\onion} & \multicolumn{3}{c}{\amazonvid} \\
            & & \arpName$\downarrow$ & \aptName$\uparrow$ & \plName$\rightarrow$ & \arpName$\downarrow$ & \aptName$\uparrow$ & \plName$\rightarrow$ & \arpName$\downarrow$ & \aptName$\uparrow$ & \plName$\rightarrow$\\

            \midrule
            \multirow{2}{*}{\baserec} & \rand & \bfseries 0.0307$\dagger$ & \bfseries 0.8206$\dagger$ & -0.7257 & \bfseries 0.0045$\dagger$ & \bfseries 0.8024$\dagger$ & -0.5694 & \bfseries 0.0013$\dagger$ & \bfseries 0.8005$\dagger$ & -0.5016 \\
            & \pop & 0.3273 & 0.0000 & 1.9425 & 0.0759 & 0.0000 & 6.2455 & 0.0249 & 0.0000 & 8.7718 \\
           \midrule
           \multirow{2}{*}{\cfbr} & \mf & 0.1690 & 0.1031 & 0.4158 & 0.0213 & 0.1641 & 0.6614 & 0.0100 & 0.1345 & 1.5492 \\
            & \dmf & 0.1174 & 0.0679 & \bfseries -0.0295$\dagger$ & 0.0199 & 0.1266 & 0.5353 & 0.0053 & 0.1976 & \bfseries 0.2144$\dagger$ \\
           \midrule
           \multirow{7}{*}{\hrs} & \clcrec & 0.1698 & 0.1049 & 0.4359 & 0.0227 & 0.1377 & 0.8065 & 0.0069 & 0.2241 & 0.9262 \\
            & \newresults{\dropoutnetTwo} & \newresults{0.1825} & \newresults{0.0845} & \newresults{0.5395} & \newresults{0.0210} & \newresults{0.2143} & \newresults{0.6489} & \newresults{0.0099} & \newresults{0.1441} & \newresults{1.8784} \\
            & \mubrarTwo & 0.1742 & 0.0926 & 0.4584 & 0.0167 & 0.2385 & 0.3137 & 0.0081 & 0.2159 & 0.8433 \\
            & \sibrarTwo & 0.1911 & 0.0846 & 0.6082 & 0.0155 & 0.3126 & \bfseries 0.1942$\dagger$ & 0.0085 & 0.1522 & 1.0296 \\
           \cdashline{2-11}[3.0pt/3.0pt]
            & \newresults{\dropoutnetBest} & \newresults{0.1825} & \newresults{0.0845} & \newresults{0.5395} & \newresults{0.0210} & \newresults{0.2143} & \newresults{0.6489} & \newresults{0.0126} & \newresults{0.1178} & \newresults{2.4068} \\
            & \mubrarBest & 0.1742 & 0.0926 & 0.4584 & 0.0203 & 0.1871 & 0.5870 & 0.0081 & 0.2159 & 0.8433 \\
            & \sibrarBest & 0.1834 & 0.0831 & 0.5438 & 0.0169 & 0.2731 & 0.3064 & 0.0085 & 0.1522 & 1.0296 \\
            
            \midrule
    \end{tabular}
    \end{adjustbox}
    \vspace{0.6em}
    \caption{Beyond-accuracy results (\aptten, \arpten, and \plten) for different \baserecs, \cfbrs, and \hrss on three warm-start datasets (\mlonem, \onion, and \amazonvid). $\uparrow$, $\downarrow$, and $\rightarrow$ next to the metric names indicate the optimal values of individual metrics for a low popularity bias are high, low, and close to zero, respectively. Bold indicates the best performing RS, $\dagger$ indicates significant improvement over all others (paired $t$-tests with $p<0.05$ considering Bonferroni correction).
    }
    \label{tab:results-beyond-acc}
\end{table*}
\end{center}

%% file: resources/figures/missing_modalities/fig-missing-modality-random-onion-best.tex
\begin{figure*}
    \centering
    \begin{subfigure}[b]{0.95\columnwidth}
        \centering
        \includegraphics[width=\textwidth]{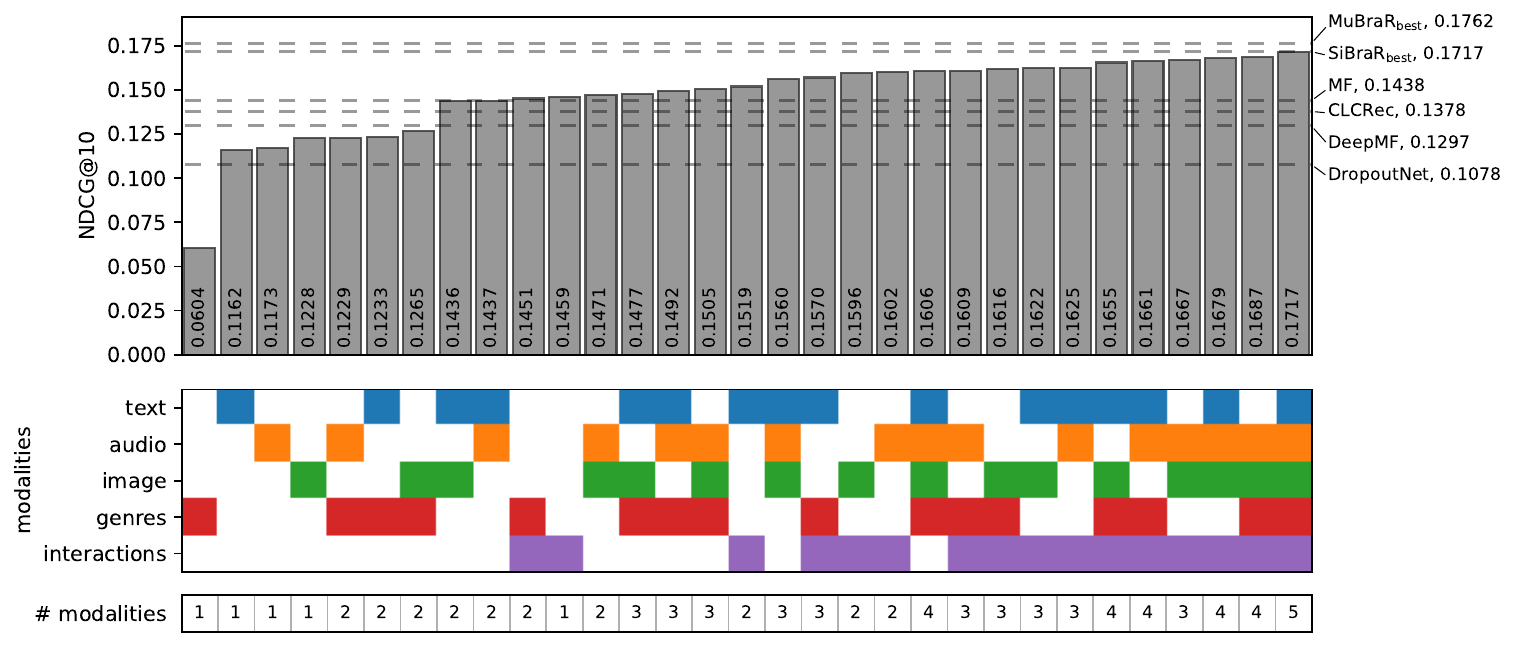}
        \caption{\sibrarBest}
        \label{fig:mm:onion:sibrar}
    \end{subfigure}
    \hspace{1.5em}
    \begin{subfigure}[b]{0.95\columnwidth}
        \centering
        \includegraphics[width=\textwidth]{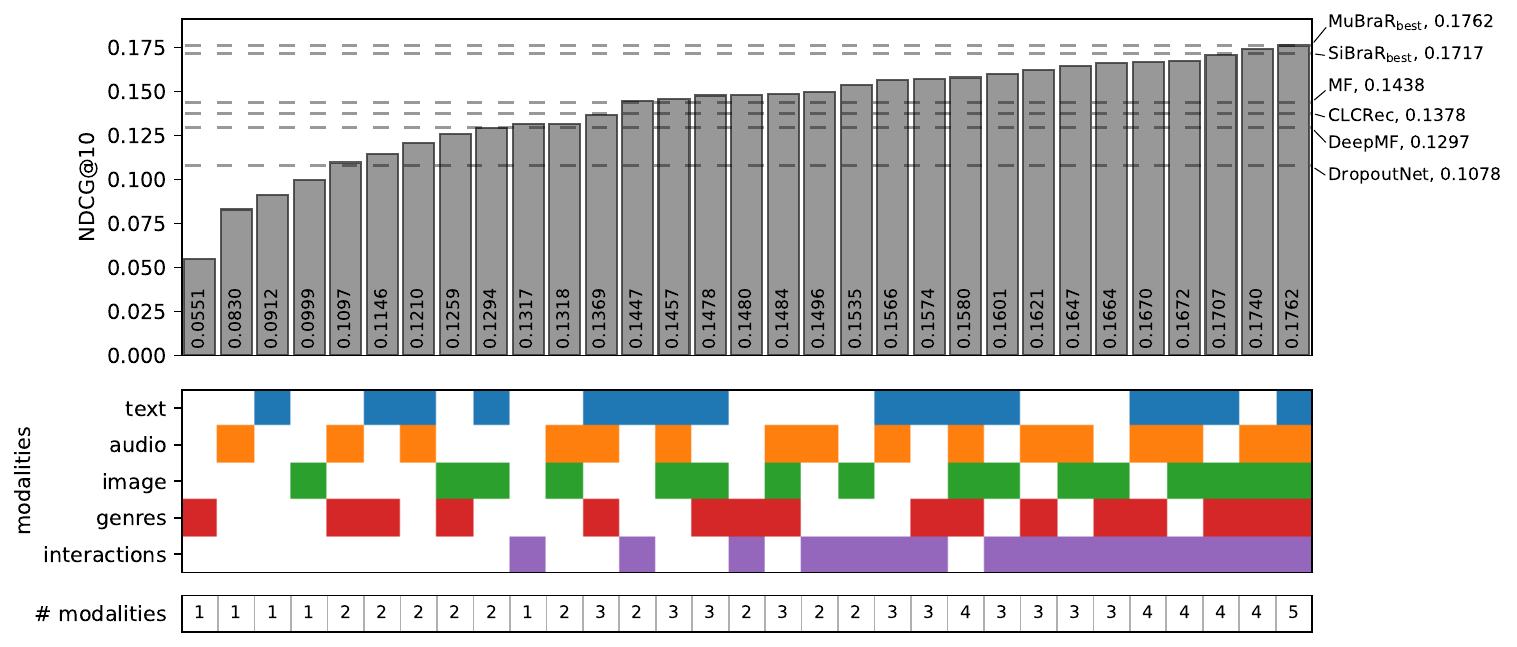}
        \caption{\mubrarBest}
        \label{fig:mm:onion:mubrar}
    \end{subfigure}    
    \caption{\ndcgten results of (a)~\sibrarBest and (b)~\mubrarBest trained on warm-start \onion with interactions, genres, image, audio, and text modalities. The bars correspond to evaluations on the test set with varying evaluation modalities. If a modality is used, its block is filled with the corresponding color in the central plot. The bottom plot counts the number of modalities per bar. The gray dashed horizontal lines in the bar plot show the \ndcgten of the best \cfbrs and \hrss in warm-start \onion.
    }
    \label{fig:mm:random:onion}
\end{figure*}

%% file: resources/figures/missing_modalities/fig-missing-modality-random-onion-diff-best.tex
\begin{figure*}
    \centering    
    \includegraphics[width=0.9\textwidth]{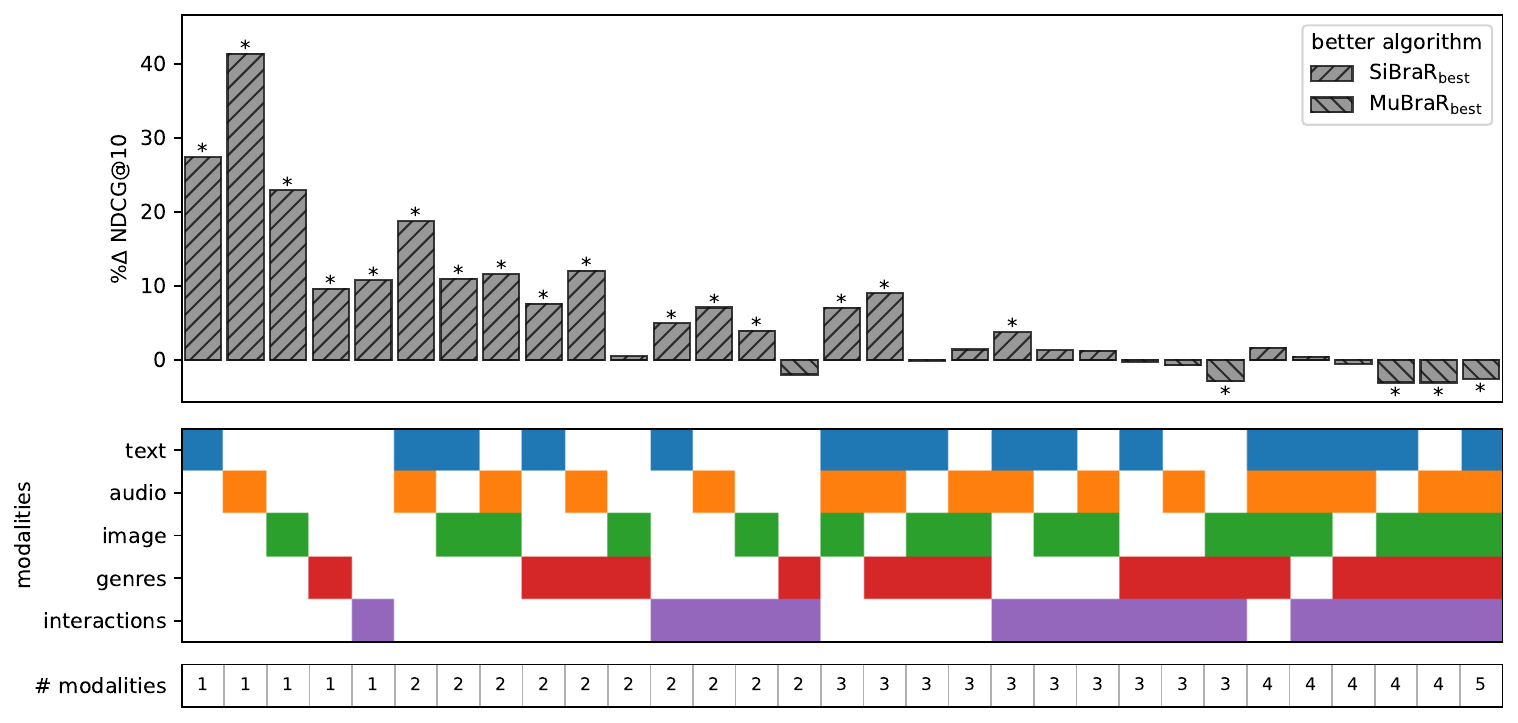}    
    \caption{\ndcgten percent difference between \sibrarBest and \mubrarBest, both trained on warm-start \onion with interactions, genres, image, audio, and text modalities. Positive values indicate that \sibrarBest performs better than \mubrarBest on a specific set of modalities, while negative values indicate a poorer performance. An asterisk '$*$' indicates a significant difference of results per subset of modalities (paired $t$-tests with $p < 0.05$).
    }
    \label{fig:mm:random:onion:diff:best}
\end{figure*}

%% file: resources/tables/tab_results_sibrar_mubrar_ndcg.tex

\begin{center}
\begin{table*}
    \centering
    \begin{adjustbox}{max width=\textwidth}
    \begin{tabular}{l|rrrcrrrcrr}
          & \multicolumn{3}{c}{Warm-Start} & & \multicolumn{3}{c}{Item Cold-Start} & & \multicolumn{2}{c}{User Cold-Start} \\
          & \mlonem & \onion & \amazonvid & & \mlonem & \onion & \amazonvid & & \mlonem & \onion \\
    
          \cmidrule{1-4}
          \cmidrule{6-8}
          \cmidrule{10-11}
          \mubrarVanilla & \bfseries 0.2610 & 0.1259 & 0.0459 & \bfseries  & 0.2349 & 0.2041 & 0.1225 & \bfseries  & 0.4579 & 0.0886 \\
          \mubrarSamp & 0.2427 & 0.1586 & 0.0508 & \bfseries  & 0.2620 & 0.2257 & 0.1078 & \bfseries  & 0.4549 & 0.0902 \\
          \mubrarSampContr & 0.2603 & \bfseries 0.1762 & 0.0535 & \bfseries  & 0.2949 & 0.2276 & 0.1262 & \bfseries  & 0.4501 & 0.0951 \\
          \cdashline{1-4}[3.0pt/3.0pt]
          \cdashline{6-8}[3.0pt/3.0pt]
          \cdashline{10-11}[3.0pt/3.0pt]
          \sibrarSamp & 0.2568 & 0.1550 & 0.0615 & \bfseries  & \bfseries 0.3046$\dagger$ & 0.2195 & 0.0921 & \bfseries  & 0.4639 & \bfseries 0.1125 \\
          \sibrarSampContr & 0.2581 & 0.1717 & \bfseries 0.0728$\dagger$ & \bfseries  & 0.2994 & \bfseries 0.2297 & \bfseries 0.1479$\dagger$ & \bfseries  & \bfseries 0.4659 & 0.1094 \\

          \cmidrule{1-4}
          \cmidrule{6-8}
          \cmidrule{10-11}
    \end{tabular}
    \end{adjustbox}
    \vspace{0.6em}
    \caption{\ndcgten results for the best-performing \sibrar and \mubrar variants on different warm- and user and item cold-start datasets (\mlonem, \onion, and \amazonvid). Bold indicates the best-performing RSs, $\dagger$ indicates significant improvement over all other variants (paired $t$-tests with $p<0.05$ and Bonferroni correction).
    The results of the different data splits are not comparable due to the different numbers of users and items in the test splits.
    }
    \label{tab:results-ndcg:sibrar-mubrar}
\end{table*}
\end{center}

%% file: resources/tables/tab_results_beyond_accuracy_sibrar_mubrar.tex
\begin{center}
\begin{table*}
    \centering
    \begin{adjustbox}{max width=\textwidth}
    \begin{tabular}{l|rrr|rrr|rrr}
        & \multicolumn{3}{c|}{\mlonem} & \multicolumn{3}{c|}{\onion} & \multicolumn{3}{c}{\amazonvid} \\
        & \arpName$\downarrow$ & \aptName$\uparrow$ & \plName$\rightarrow$ & \arpName$\downarrow$ & \aptName$\uparrow$ & \plName$\rightarrow$ & \arpName$\downarrow$ & \aptName$\uparrow$ & \plName$\rightarrow$\\

        \midrule
        \mubrarVanilla & \bfseries 0.1701$\dagger$ & 0.1017 & \bfseries 0.4240$\dagger$ & \bfseries 0.0142$\dagger$ & \bfseries 0.3520$\dagger$ & \bfseries 0.0999$\dagger$ & 0.0065 & \bfseries 0.3061$\dagger$ & 0.5160 \\
        \mubrarSamp & 0.1840 & 0.0659 & 0.5523 & 0.0193 & 0.2131 & 0.5138 & \bfseries 0.0065 & 0.2459 & \bfseries 0.4874$\dagger$ \\
        \mubrarSampContr & 0.1742 & 0.0926 & 0.4584 & 0.0203 & 0.1871 & 0.5870 & 0.0081 & 0.2159 & 0.8433 \\
        \cdashline{0-9}[3.0pt/3.0pt]
        \sibrarSamp & 0.1718 & \bfseries 0.1175$\dagger$ & 0.4434 & 0.0179 & 0.2390 & 0.4084 & 0.0087 & 0.1218 & 1.0685 \\
        \sibrarSampContr & 0.1834 & 0.0831 & 0.5438 & 0.0169 & 0.2731 & 0.3064 & 0.0085 & 0.1522 & 1.0296 \\
        \midrule
    \end{tabular}
    \end{adjustbox}
    \vspace{0.6em}
    \caption{Beyond-accuracy results (\aptten, \arpten, and \plten) for different \mubrar and \sibrar variants on three warm-start datasets (\mlonem, \onion, and \amazonvid). $\uparrow$, $\downarrow$, and $\rightarrow$ next to the metrics indicate the optimal values of individual metrics for a low popularity bias are high, low, and close to zero, respectively. Bold indicates the best performing RS, $\dagger$ indicates significant improvement over all others (paired $t$-tests with $p<0.05$ considering Bonferroni correction).
    }
    \label{tab:results-beyond-acc:sibrar-mubrar}
\end{table*}
\end{center}

%% file: resources/figures/missing_modalities/fig-missing-modality-random-onion-diff-samp-contr.tex
\begin{figure*}
    \centering
    \begin{subfigure}[b]{0.90\columnwidth}
        \centering
        \includegraphics[width=\textwidth]{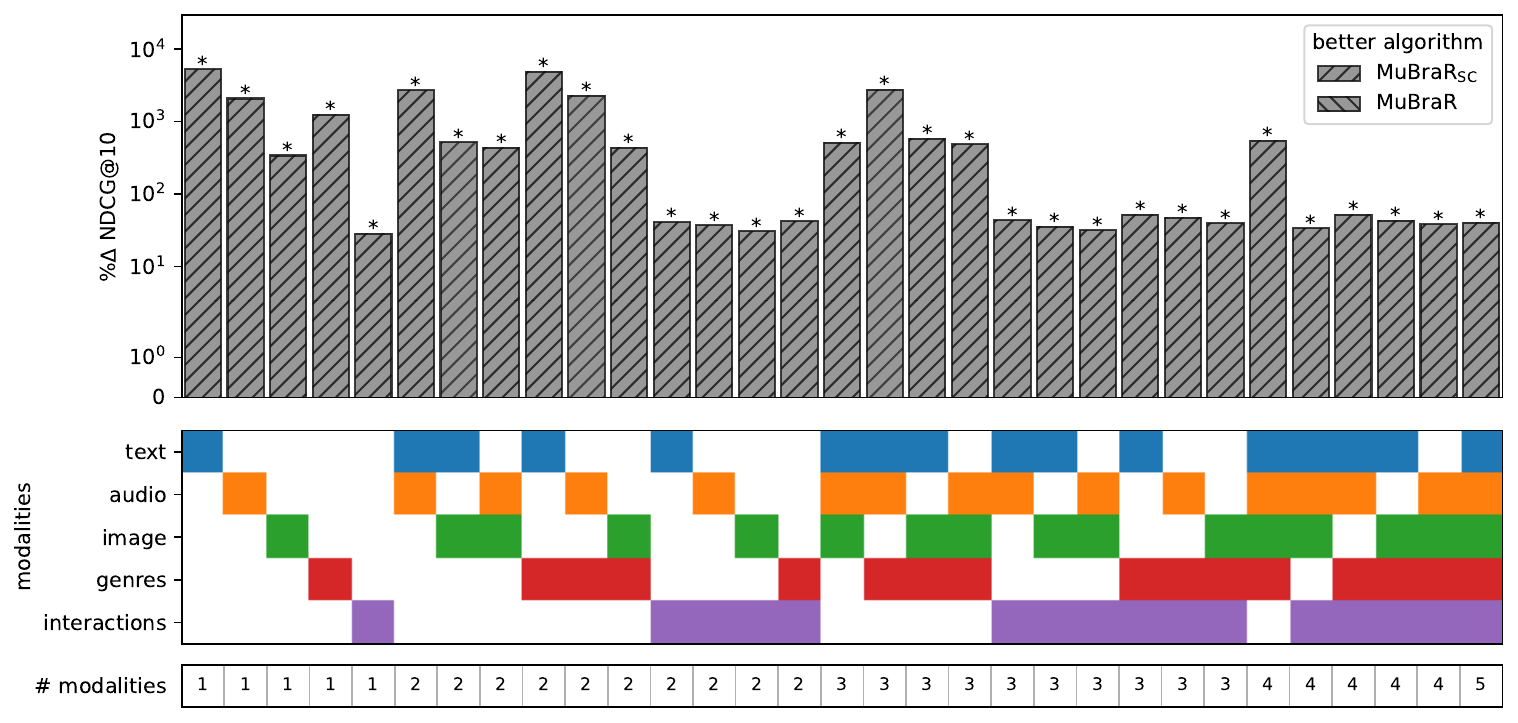}
        \caption{\mubrarSampContr vs. \mubrar}
        \label{fig:mm:random:onion:diff:samp-contr:mubrar}
    \end{subfigure}
    \vspace{1.5em}
    \begin{subfigure}[b]{0.90\columnwidth}
        \centering
        \includegraphics[width=\textwidth]{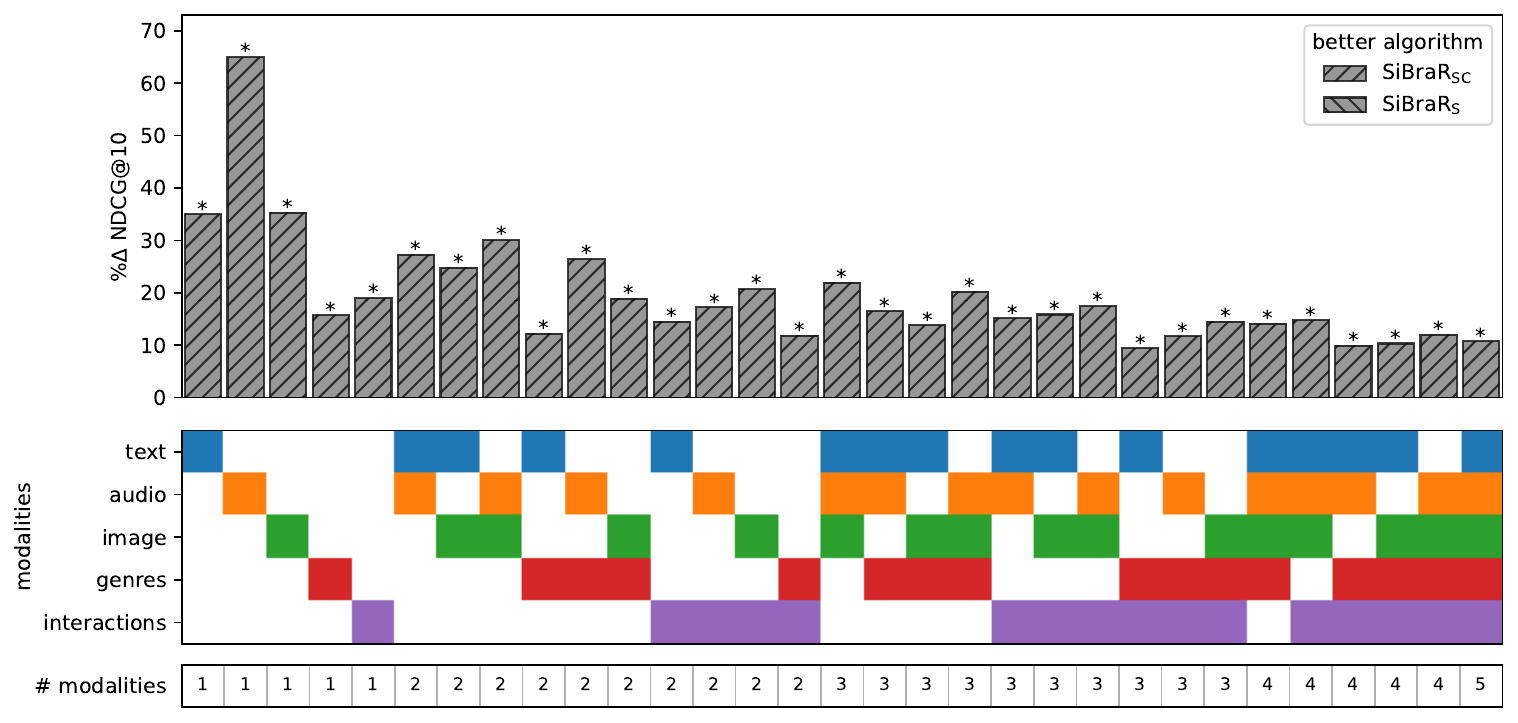}
        \caption{\sibrarSampContr vs. \sibrarSamp}
        \label{fig:mm:random:onion:diff:samp-contr:sibrar}
    \end{subfigure}
    \vspace{-2em}
    \caption{\ndcgten comparison for single- and multi-branch architectures when trained on all modalities of warm-start \onion, with and without utilizing modality sampling (S) and a contrastive loss (C). (a)~shows the difference between \mubrar and \mubrarSampContr, and (b)~the difference between \sibrarSamp and \sibrarSampContr. The colored squares in the middle plot indicate the modalities used during evaluation. Moreover, the bar plot shows the percent difference of \ndcgten, where positive values indicate that a contrastive loss is beneficial for recommendation quality, and worse if negative. \newtext{We use a logarithmic scale for the y-axis of~(a) to ease interpretation.}
    }
    \label{fig:mm:random:onion:diff:samp-contr}
\end{figure*}

%% file: resources/figures/missing_modalities/fig-missing-modality-random-onion-diff-spec.tex
\begin{figure*}
    \centering
    \includegraphics[width=0.9\textwidth]{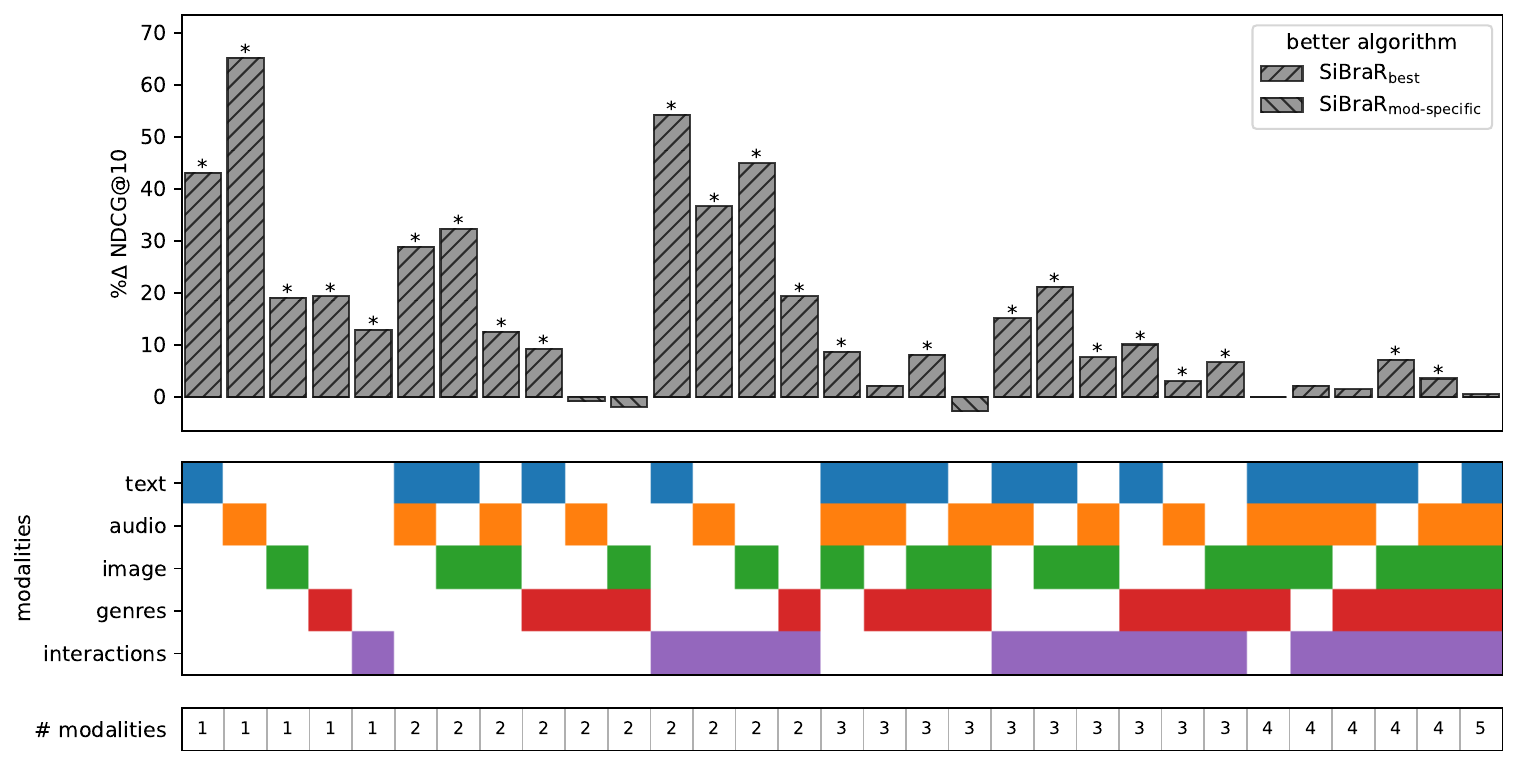}
    \caption{\ndcgten comparison between \sibrarBest trained on all modalities of warm-start \onion, and multiple instances of the same model (and hyperparameters) trained on specific subsets of modalities (\sibrarSpec). The modalities are indicated by the colored squares in the middle plot. The bar plot shows the percent difference of \ndcg, where positive values indicate that \sibrarBest performs better than \sibrarSpec on the given set of modalities and worse if negative.
    }
    \label{fig:mm:random:onion:diff:mod-specific}
\end{figure*}

%% file: resources/tables/tab_missing_modalities_number_random_onion.tex
\begin{table}
    \begin{tabular}{lrrrrr}
         & \multicolumn{5}{c}{\# item modalities} \\
         & 1 & 2 & 3 & 4 & 5 \\
         \midrule
        \mubrarVanilla & 0.0264 & 0.0504 & 0.0751 & 0.0995 & 0.1239 \\
        \mubrarSamp & 0.0572 & 0.0999 & 0.1272 & 0.1457 & 0.1586 \\
        \mubrarSampContr & 0.0921 & 0.1328 & 0.1546 & \textbf{0.1674} & \textbf{0.1762} \\
        \hdashline
        \sibrarSamp & 0.0846 & 0.1186 & 0.1363 & 0.1478 & 0.1550 \\
        \sibrarSampContr & \textbf{0.1124} & \textbf{0.1424} & \textbf{0.1574} & 0.1658 & 0.1717 \\
        \hdashline
        \sibrarSpec & 0.0876 & 0.1187 & 0.1450 & 0.1601 & 0.1697 \\
        \bottomrule
    \end{tabular}
    \vspace{0.6em}
    \caption{\ndcgten results of different \sibrar and \mubrar variants trained on warm-start \onion with the same set of item modalities (interactions, genres, image, audio, and text) and evaluated on subsets of those. \ndcgten values of different subsets with the same number of modalities are averaged. Bold indicates the best-performing algorithm per the number of modalities used. 
    }
    \label{tab:missing-modality:number-of-mods:random:onion}
\end{table}

%% file: resources/tables/tab_missing_modalities_number_cs_onion.tex
\begin{table}
    \begin{tabular}{lrrrr}
         & \multicolumn{4}{c}{\# item modalities} \\
         & 1 & 2 & 3 & 4 \\
         \midrule
        \mubrarVanilla & 0.0748 & 0.1247 & 0.1674 & 0.2041 \\ 
        \mubrarNoReg & 0.1069 & 0.1620 & 0.1996 & 0.2257 \\ 
        \mubrarReg & 0.1194 & 0.1676 & 0.2023 & 0.2276 \\ 
        \hdashline
        \sibrarNoReg & 0.1217 & 0.1640 & 0.1959 & 0.2195 \\ 
        \sibrarReg & \bfseries 0.1338 & \bfseries 0.1762 & \bfseries 0.2078 & \bfseries 0.2297 \\
        \bottomrule
    \end{tabular}
    \vspace{0.6em}
    \caption{\ndcgten results of different \sibrar and \mubrar variants trained on cold-start \onion with the same set of item modalities (interactions, genres, image, audio, and text) and evaluated on subsets of those (excluding interactions). \ndcgten values of different subsets with the same number of modalities are averaged. Bold indicates the best-performing algorithm per the number of modalities used. 
    }
    \label{tab:missing-modality:number-of-mods:cs:onion}
\end{table}

%% file: resources/figures/contrastive_loss/fig-contrastive-loss-gridsearch.tex
\begin{figure}
    \centering    
    \includegraphics[width=\textwidth]{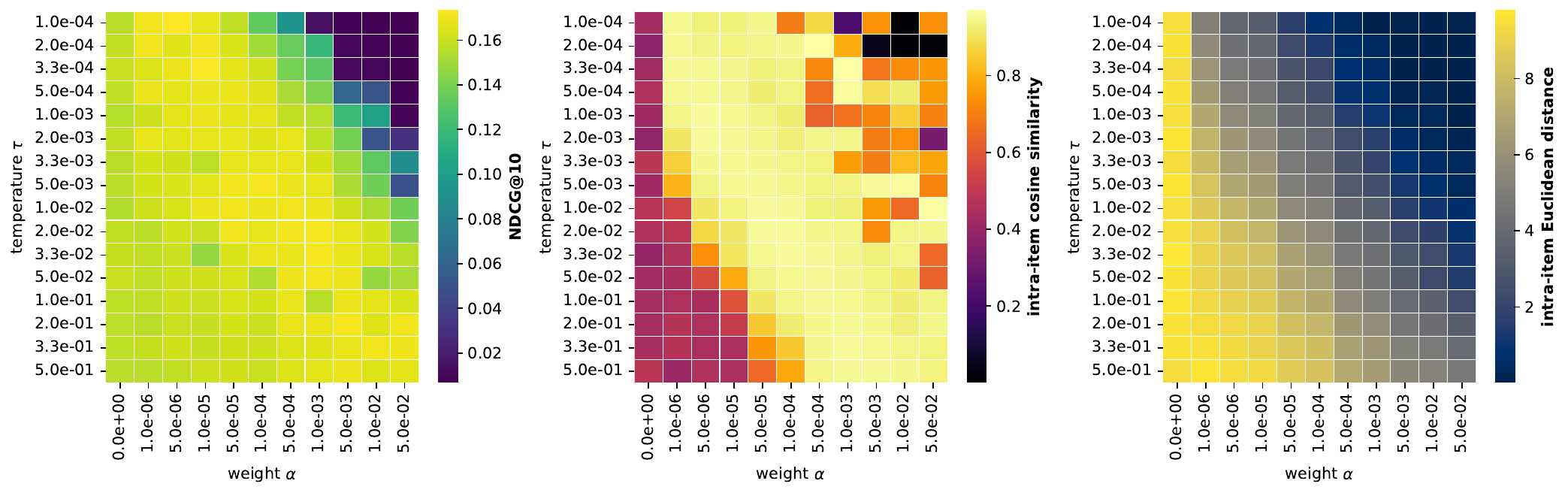}    
    \caption{\ndcgten, intra-item cosine similarity and Euclidean distance results of \sibrarBest on different combinations of contrastive loss temperature~$\tau$ and weight~$\alpha$ on warm-start \onion, while all other hyperparameters are fixed. The left plot shows the \ndcgten results at test time, and the center and right plots show the intra-item consine similarity and Euclidean distance, respectively. When contrastive loss $\alpha=0$, \ie when the loss does not affect training, the minor performance differences are due to randomness in the training pipeline.
    }
    \label{fig:contrastive:gridsearch:onion}
\end{figure}

%% file: resources/figures/contrastive_loss/fig-contrastive-loss-ratio.tex
\begin{figure}
    \centering
    
    \includegraphics[width=0.37\textwidth]{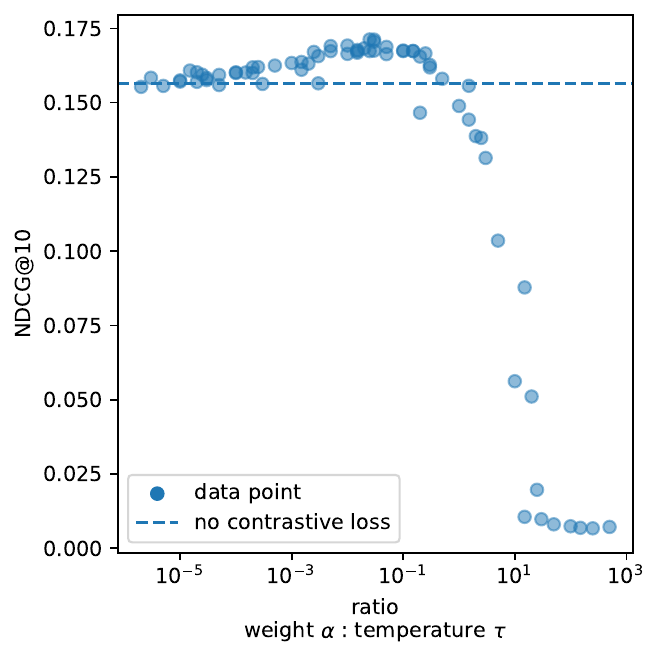}
    
    \caption{\ndcgten results of \sibrarBest on different combinations of contrastive loss temperature~$\tau$ and weight~$\alpha$ on warm-start \onion, while all other hyperparameters are fixed. Each point indicates the performance of a run with a certain ratio ${\alpha}/{\tau}$ and the resulting \ndcgten. The horizontal dashed line shows the average performance for runs where the contrastive loss has no effect on training, \ie when $\alpha=0$.
    }
    \label{fig:contrastive:ratio:onion}
\end{figure}

%% file: resources/figures/projections/fig-embedding-space-onion-sibrar.tex
\begin{figure}
    \centering
    \includegraphics[width=0.70\textwidth]{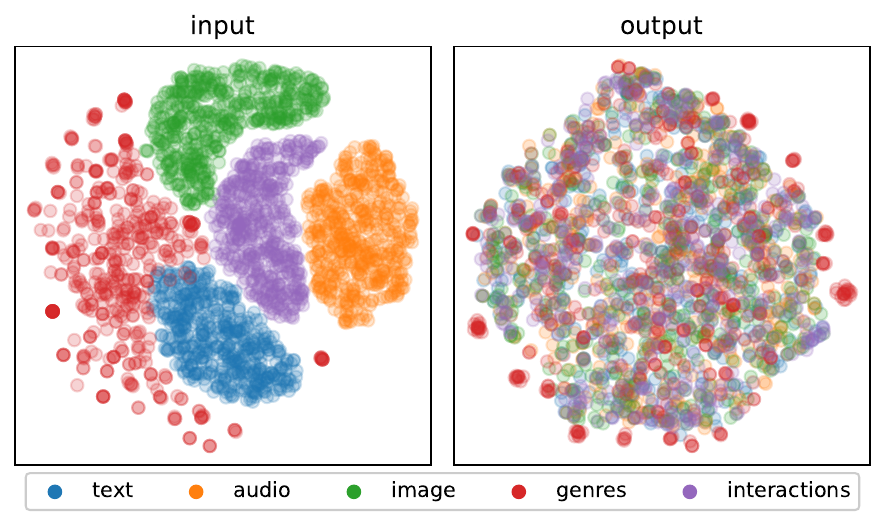}
    \caption{T-SNE projected embeddings before and after the single-branch network of the best performing \sibrarSampContr on warm-start \onion. While different modalities can be differentiated at input~(left), modalities overlap substantially in the shared embedding space after passing the modalities through the network~(right).}
    \label{fig:projection:onion:sibrar}
\end{figure}

%% file: resources/figures/projections/fig-embedding-space-onion-all-others.tex
\begin{figure}
    \centering
    
    \includegraphics[width=\textwidth]{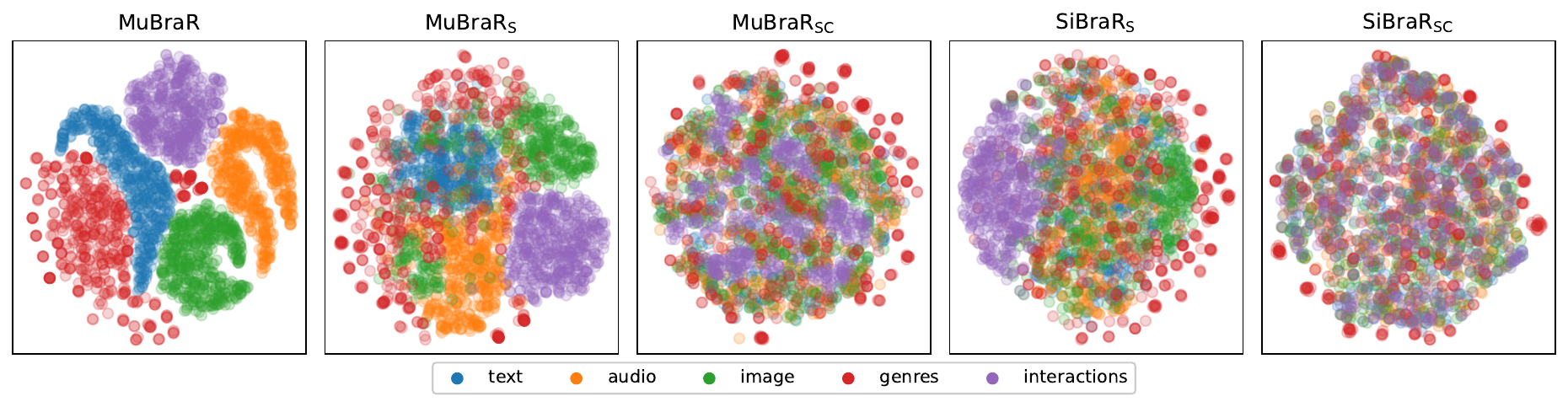}

    \caption{T-SNE projected embeddings of the modality-specific embeddings of the best single- and multi-branch variants
    trained on warm-start \onion. While the different modalities can easily be differentiated for \mubrarVanilla (1st), due to modality sampling (\mubrarSamp, 2nd) and the contrastive loss (\mubrarSampContr, 3rd), the different modalities are less separable. \newtext{For \sibrarSamp (4th from left), the regions of the different modalities overlap, though not as much as for \sibrarSampContr (rightmost).}
    }
    \label{fig:projection:onion:mubrar-sibrars}
\end{figure}

%% file: resources/tables/tab_intra_inter_comparison_cosine.tex
\begin{table}[]
    \begin{adjustbox}{max width=\textwidth}
    \begin{tabular}{l|rr|rr|rr}
        & \multicolumn{2}{c|}{\mlonem} & \multicolumn{2}{c|}{\onion} & \multicolumn{2}{c}{\amazonvid} \\
        & intra & inter & intra & inter & intra & inter\\
        \midrule
\mubrarVanilla 		& 0.136 & 0.489 & 0.032 & 0.486 & 0.063 & \bfseries 0.772 \\
\mubrarSamp 		& 0.308 & 0.333 & 0.288 & 0.282 & 0.707 & 0.423 \\
\mubrarSampContr 	& 0.456 & 0.272 & 0.498 & 0.329 & \bfseries 0.899 & 0.405 \\
\hdashline
\sibrarSamp 		& 0.718 & \bfseries 0.668 & 0.478 & 0.353 & 0.715 & 0.327 \\
\sibrarSampContr	& \bfseries 0.845 & 0.615 & \bfseries 0.929 & \bfseries 0.854 & 0.895 & 0.323 \\
        \bottomrule
    \end{tabular}
    \end{adjustbox}
    \vspace{0.6em}
    \caption{\newtext{Average inter- and intra-item \metricCosineSim\xspace for \sibrar and \mubrar variants. For each warm-start dataset, all variants are trained and evaluated on the same set of modalities: (interactions, genres, text) for \mlonem, (interactions, genres, audio, image, text) for \onion, and (interactions, text) for \amazonvid. Bold indicates the algorithm where modality embeddings exhibit the highest average similarity within each group.}
    }
    \label{tab:intra-inter-comparison:cosine}
\end{table}

%% file: resources/tables/tab_modality_prediction.tex
\begin{table}[]
    \begin{tabular}{lrrr}
        & \mlonem & \onion & \amazonvid \\
        \midrule
    \mubrarVanilla & 1.0000 & 1.0000 & 1.0000 \\
    \mubrarNoReg & 1.0000 & 1.0000 & 1.0000 \\
    \mubrarReg & 1.0000 & 0.9925 & 0.8853 \\
    \hdashline
    \sibrarNoReg & 0.9768 & 0.9298 & 0.9811 \\
    \sibrarReg & \bfseries 0.8983$\dagger$ & \bfseries 0.7771$\dagger$ & \bfseries 0.6823$\dagger$ \\
    \hdashline
    Random & 0.3333 & 0.2000 & 0.5000 \\
        \bottomrule
    \end{tabular}
    \vspace{0.6em}
    \caption{Accuracy of random forest classifiers predicting the modality class of modality embeddings output by \mubrar and \sibrar variants trained on three warm-start datasets (\mlonem, \onion, and \amazonvid). Each result reports the average performance of 20 classifiers trained on different train--test splits and seeds. 80\% of each warm-start test set is used for fitting the classifiers, and 20\% for their evaluation. \newtext{Bold indicates the variant with the least separable embedding space, $\dagger$ indicates significant worse classifier accuracies in comparison to all other methods (paired $t$-tests with $p<0.05$ considering Bonferroni correction). The last row shows the expected accuracy for random guessing.}
    }
    \label{tab:modality-prediction}
\end{table}

%% file: 0_sections/6_conclusion.tex
This work formalizes multimodal single- and multi-branch recommenders, \sibrar and \mubrar, respectively. 
With extensive experiments, we quantitatively compare these approaches on three datasets on warm-start, missing modality, and user and item cold-start scenarios. 
Besides accuracy metrics, we investigate different popularity biases these models exhibit. We break down the factors contributing to strong recommendations employed in \sibrar: weight-sharing, modality sampling, and contrastive loss. 
As the results show, single- and multi-branch recommenders perform comparably well in warm-start, while the former have a clear advantage in cold start and missing modality \cg{with an, on average, more than 10\% increase in \ndcgten for low numbers of available modalities}. Utilizing a contrastive loss is paramount for achieving strong results, even when parts of the training data are missing at inference. We perform a widespread analysis on contrastive loss parameters to shed additional light on their importance. \cg{The study reveals that the right choice of parameters may considerably improve the recommendation performance of a model.} 
Finally, we qualitatively and quantitatively examine the modality gap of embedding spaces of single- and multi-branch approaches. 
Specifically, we show that modality sampling and contrastive loss affect the distance between embeddings of the same item and of different items; both force modality embeddings of the same items to be closer to each other than to embeddings of other items. \cg{Classification, recommendation and other downstream tasks utilizing the embeddings may thus be less affected by the underlying modality source.}

\subsection{Practical considerations}
As outlined in this work, single- and multi-branch networks allow for generating recommendations by utilizing any available combination of modalities used during training. While results in missing-modality scenarios are worse than in warm start, we show that single-branch networks significantly outperform their competitors in the former case.  
Moreover, users and items with missing modalities are often dropped from the datasets before training.
In such cases, the presented modality sampling and contrastive loss naturally allow training on missing-modality data.
To reduce computational resources at inference, \sibrar enables practitioners to precompute embeddings of static modalities, \eg audio or lyrics in the music domain, while (re)computing the embeddings for previously unavailable or updated modalities, \eg interactions. Furthermore, practitioners can trade off recommendation quality and computational costs by selecting only a subset of all modalities for recommendation and utilizing left-out modalities only to ensure a minimum recommendation quality.

\subsection{Limitations \mmedits{and Future Directions}}
\mmedits{The use of single- and multi-branch losses for recommendation poses several challenges and limitations. Compared to other models, both single- and multi-branch recommenders come at the cost of a large number of hyperparameters. We experienced that these models are prone to substantial drops in recommendation quality when the hyperparameters are not optimized; hence tuning single- and multi-branch architectures might require large computational costs. We therefore believe that our analysis on the impact of the contrastive-loss hyperparameters will be helpful in simplifying the tuning phase in future applications.} There are still many open research questions pertaining to single-branch recommenders, which are too extensive for this work to cover. For example, we only sample pairs of modalities when applying the constrastive loss; the effect of sampling multiple pairs at once is not studied. \mmedits{We only considered random sampling of multimodal data for all items during training; this might not be reflective of real-world scenarios in which not all modalities are equally available for all items. \reviewTwo{Further, we only considered averaging as a modality fusion strategy for both single- and multi-branch networks, despite many different techniques, \eg attention-based fusion, emerging in the domain of multimodal learning.} \reviewRemoved{We did not consider architectures based on attention-based fusion techniques, although these have been emerging in the domain of multimodal learning.}} Moreover, while \sibrar and \mubrar models may feature similar numbers of parameters, the former require multiple forward passes -- one for each modality -- through the single branch. On the other hand, the latter require only forward passes through modality-specific networks, which are comparably smaller in size.
The dynamics of including additional modality types on a trained \sibrar instance remain unknown. 
\reviewTwo{We limited our study to the beyond-accuracy evaluation in warm-start scenarios where all modalities are available. An additional in-depth analysis on the introduced biases of different modalities may provide insights on how cold-start users and items are inherently affected by the modalities used to represent them in the absence of collaborative information.}
We envision studying the practical implications and addressing the limitations as future extensions of this work.

%% file: 0_sections/7_acknowledgements.tex
This research was funded in whole or in part by the Austrian Science Fund (FWF) 
\url{https://doi.org/10.55776/P33526}, 
\url{https://doi.org/10.55776/P36413},
\url{https://doi.org/10.55776/DFH23}, \url{https://doi.org/10.55776/COE12}. 
For open access purposes, the authors have applied a CC BY public copyright license to any author-accepted manuscript version arising from this submission.

%% file: 0_sections/A_appendix.tex
\input{resources/tables/tab-hyperparams}

\input{resources/tables/tab_results_recall}
\input{resources/tables/tab_results_precision}
\input{resources/tables/tab_results_mrr}
\input{resources/tables/tab_results_map}
\input{resources/tables/tab_results_fscore}
\input{resources/tables/tab_results_coverage}

\input{resources/tables/tab_results_sibrar_mubrar_recall}
\input{resources/tables/tab_results_sibrar_mubrar_precision}
\input{resources/tables/tab_results_sibrar_mubrar_mrr}
\input{resources/tables/tab_results_sibrar_mubrar_map}
\input{resources/tables/tab_results_sibrar_mubrar_fscore}
\input{resources/tables/tab_results_sibrar_mubrar_coverage}

\input{resources/tables/tab_missing_modalities_number_random_ml}
\input{resources/tables/tab_missing_modalities_number_random_amazon}
\input{resources/tables/tab_missing_modalities_number_cs_ml}

\input{resources/figures/missing_modalities/fig-missing-modality-random-ml1m-best}
\input{resources/figures/missing_modalities/fig-missing-modality-random-amazon-best}
\input{resources/figures/missing_modalities/fig-missing-modality-random-ml1m-amazon-diff-best}

\input{resources/figures/missing_modalities/fig-missing-modality-random-ml1m-diff-samp-contr.tex}
\input{resources/figures/missing_modalities/fig-missing-modality-random-amazon-diff-samp-contr.tex}

\input{resources/figures/contrastive_loss/fig-contrastive-loss-gridsearch_appendix.tex}
\input{resources/figures/contrastive_loss/fig-contrastive-loss-ratio_appendix}

\input{resources/figures/projections/fig-embedding-space-all.tex}

\input{resources/figures/projections/fig-embedding-space-clcrec}

\input{resources/tables/tab_intra_inter_comparison_clcrec}

\input{resources/tables/tab_intra_inter_comparison_euclidean}

\input{resources/tables/tab_modality_prediction-cold}

%% file: resources/tables/tab-hyperparams.tex
\begin{center}
\begin{table*}[!htb]
    \scalebox{0.90}{
\begin{tabular}{llllrrl}
 & \multicolumn{1}{l}{parameter group}& \multicolumn{1}{l}{parameter name} & type & min & max & choices \\
 \midrule
\mf &  & embedding size & int & 32 & 512 &  \\ \hline
\multirow{8}{*}{\dmf} & \multirow{2}{0.2\linewidth}{hidden item interaction layers} & number of layers & int & 0 & 1 &  \\
 &  & layer size & int & 16 & 512 &  \\
\cdashline{2-7}[3.0pt/3.0pt]
 & \multirow{2}{0.2\linewidth}{hidden user interaction layers} & number of layers & int & 0 & 1 &  \\
 &  & layer size & int & 16 & 512 &  \\
\cdashline{2-7}[3.0pt/3.0pt]
 &  & apply ReLU activation on interactions & bool &  &  & False, True \\
 &  & apply ReLU activation on embeddings & bool &  &  & False, True \\
 &  & embedding size & int & 8 & 256 &  \\
 &  & miminum recommendation score & float & 1,00E-07 & 1,00E-02 &  \\ \hline
\multirow{5}{*}{\clcrec} & \multirow{2}{0.2\linewidth}{hidden side-information layers} & number of layers & int & 0 & 2 &  \\
 &  & layer size & int & 16 & 256 &  \\
\cdashline{2-7}[3.0pt/3.0pt]
 & \multirow{2}{0.2\linewidth}{contrastive loss} & weight & float & 1,00E-06 & 1,00E-03 &  \\
 &  & temperature & float & 1,00E-03 & 1,00E-01 &  \\
\cdashline{2-7}[3.0pt/3.0pt]
 &  & embedding size & int & 8 & 512 &  \\ \hline
\multirow{13}{*}{\dropoutnet} & \multirow{2}{0.2\linewidth}{hidden item interaction layers} & number of layers & int & 0 & 2 &  \\
 &  & layer   size & int & 16 & 512 &  \\
\cdashline{2-7}[3.0pt/3.0pt]
 & \multirow{2}{0.2\linewidth}{hidden item layers per side-information} & number of layers & int & 0 & 1 &  \\
 &  & layer size & int & 64 & 512 &  \\
\cdashline{2-7}[3.0pt/3.0pt]
 & \multirow{2}{0.2\linewidth}{hidden item shared layers} & number of layers & int & 0 & 2 &  \\
 &  & layer size & int & 32 & 512 &  \\
\cdashline{2-7}[3.0pt/3.0pt]
 & \multirow{2}{0.2\linewidth}{hidden user interaction layers} & number of layers & int & 0 & 2 &  \\
 &  & layer size & int & 16 & 512 &  \\
\cdashline{2-7}[3.0pt/3.0pt]
 & \multirow{2}{0.2\linewidth}{hidden user layers per side-information} & number of layers & int & 0 & 1 &  \\
 &  & layer size & int & 8 & 24 &  \\
\cdashline{2-7}[3.0pt/3.0pt]
 & \multirow{2}{0.2\linewidth}{hidden user shared layers} & number of layers & int & 0 & 2 &  \\
 &  & layer size & int & 32 & 512 &  \\
\cdashline{2-7}[3.0pt/3.0pt]
 &  & embedding size & int & 16 & 768 &  \\
 &  & dropout rate & float & 0.0 & 1.0 &  \\ \hline
\multirow{6}{*}{\mubrar} & \multirow{2}{0.2\linewidth}{hidden layers per modality} & number of layers & int & 1 & 3 &  \\
 &  & layer size & int & 128 & 512 &  \\
\cdashline{2-7}[3.0pt/3.0pt]
 & \multirow{2}{0.2\linewidth}{contrastive loss} & weight & float & 0 & 1,00E-03 &  \\
 &  & temperature & float & 1,00E-03 & 5,00E-01 &  \\
\cdashline{2-7}[3.0pt/3.0pt]
 &  & enable modality sampling & bool &  &  & False, True \\
 &  & embedding size & int & 128 & 768 &  \\ \hline
\multirow{9}{*}{\sibrar} & \multirow{6}{0.2\linewidth}{single-branch} & number of hidden layers & int & 0 & 5 &  \\
 &  & hidden layer size & int & 32 & 512 &  \\
 &  & input size & int & 16 & 768 &  \\
 &  & apply batch normalization & bool &  &  & False, True \\
 &  & apply ReLU output activation & bool &  &  & False, True \\
 &  & normalize input & bool &  &  & False, True \\
\cdashline{2-7}[3.0pt/3.0pt]
 & \multirow{2}{0.2\linewidth}{contrastive loss} & weight & float & 0 & 1,00E+00 &  \\
 &  & temperature & float & 1,00E-04 & 1,00E+00 &  \\
\cdashline{2-7}[3.0pt/3.0pt]
 &  & embedding size & int & 128 & 768 & \\ \midrule
\multirow{2}{0.1\linewidth}{} & \multirow{2}{0.2\linewidth}{optimizer} & learning rate & float & 1,00E-05 & 5,00E-02 &  \\
 &  & weight decay & float & 0 & 5,00E-03 &  \\ \hline
\end{tabular}}
\vspace{0.6em}
    \caption{Summary of hyperparameters evaluated for various algorithms in this study. The second and third column refer to the parameter optimized, while the fourth which type of parameter it is. Lastly, the final three columns, \textit{min}, \textit{max}, and \textit{choice}, indicate which hyperparameter ranges were considered.
    We use Sweeps module of Weights \& Biases to explore the extensive parameter space, considering various distributions, \eg \textit{int\_uniform} and \textit{log\_uniform} for integer parameters and float parameters spanning multiple orders of magnitude, respectively. Lastly, the group \textit{optimizer} refers to the parameters of the Adam optimizer, which have been equally tuned for all models.
    The best hyperparameters were selected based on \ndcgten performance on the validation set. 
    When optimal parameters were found at the search space boundary, the space was expanded to ensure true optimality. Additionally, insights gained from hyperparameter analysis of one algorithm informed the initialization of parameter searches for other algorithms.}\label{tab:hyperparams}
\end{table*}
\end{center}

%% file: resources/tables/tab_results_recall.tex
\begin{center}
\begin{table*}[!htb]
    \centering
    \begin{adjustbox}{max width=\textwidth}
    \begin{tabular}{ll|rrrcrrrcrr}
          & & \multicolumn{3}{c}{Warm-Start} & & \multicolumn{3}{c}{Item Cold-Start} & & \multicolumn{2}{c}{User Cold-Start} \\
          & & \mlonem & \onion & \amazonvid & & \mlonem & \onion & \amazonvid & & \mlonem & \onion \\
          
          \cmidrule{1-5}
          \cmidrule{7-9}
          \cmidrule{11-12}
          \multirow{2}{*}{\baserec} & \rand & 0.0027 & 0.0008 & 0.0029 & \bfseries  & 0.0309 & 0.0068 & 0.0231 & \bfseries  & 0.0027 & 0.0009 \\
           & \pop & 0.0798 & 0.0170 & 0.0515 & \bfseries  & 0.0032 & 0.0064 & 0.0149 & \bfseries  & 0.0508 & 0.0128 \\
          \cmidrule{1-5}
          \cmidrule{7-9}
          \cmidrule{11-12}
          \multirow{2}{*}{\cfbr} & \mf & 0.1806 & 0.1438 & \bfseries 0.1394 & \bfseries  & 0.0152 & 0.0056 & 0.0232 & \bfseries  & 0.0520 & 0.0116 \\
           & \dmf & 0.0774 & 0.1224 & 0.1036 & \bfseries  & 0.0032 & 0.0064 & 0.0149 & \bfseries  & 0.0510 & 0.0079 \\
          \cmidrule{1-5}
          \cmidrule{7-9}
          \cmidrule{11-12}
          \multirow{7}{*}{\hrs} & \clcrec & \bfseries 0.1851 & 0.1363 & 0.1051 & \bfseries  & 0.1153 & 0.1169 & 0.2199 & \bfseries  & 0.0527 & 0.0128 \\
           & \newresults{\dropoutnetTwo} & \newresults{0.1744} & \newresults{0.1052} & \newresults{0.0802} & \newresults{\bfseries } & \newresults{0.1488} & \newresults{0.1207} & \newresults{0.1906} & \newresults{\bfseries } & \newresults{0.0510} & \newresults{0.0137} \\
           & \mubrarTwo & 0.1813 & 0.1494 & 0.0977 & \bfseries  & 0.1616 & 0.1944 & 0.2159 & \bfseries  & 0.0504 & 0.0114 \\
           & \sibrarTwo & 0.1836 & 0.1495 & 0.1328 & \bfseries  & 0.1736 & 0.1877 & \bfseries 0.2466$\dagger$ & \bfseries  & \bfseries 0.0529 & 0.0121 \\
          \cdashline{2-5}[3.0pt/3.0pt]
          \cdashline{7-9}[3.0pt/3.0pt]
          \cdashline{11-12}[3.0pt/3.0pt]
           & \newresults{\dropoutnetBest} & \newresults{0.1744} & \newresults{0.1052} & \newresults{0.0887} & \newresults{\bfseries } & \newresults{0.1488} & \newresults{0.1618} & \newresults{0.1906} & \newresults{\bfseries } & \newresults{0.0510} & \newresults{\bfseries 0.0155} \\
           & \mubrarBest & 0.1813 & \bfseries 0.1681 & 0.0977 & \bfseries  & 0.1775 & 0.2014 & 0.2159 & \bfseries  & 0.0510 & 0.0121 \\
           & \sibrarBest & 0.1820 & 0.1624 & 0.1328 & \bfseries  & \bfseries 0.1869$\dagger$ & \bfseries 0.2114$\dagger$ & \bfseries 0.2466$\dagger$ & \bfseries  & 0.0522 & 0.0155 \\

\cmidrule{1-5}
\cmidrule{7-9}
\cmidrule{11-12}
    \end{tabular}
    \end{adjustbox}
    \vspace{0.6em}
    \caption{\recten results for different \baserecs, \cfbrs, and \hrss. Each row indicates the results of the evaluated algorithm, and the columns blocks indicate the data scenario (warm start and user/item cold start) for the datasets (\mlonem, \onion, and \amazonvid). 
    Bold indicates the best-performing RS, $\dagger$ indicates significant improvement over all others (paired $t$-tests with $p<0.05$ considering Bonferroni correction).
    }
    \label{tab:results-recall}
\end{table*}
\end{center}

%% file: resources/tables/tab_results_precision.tex
\begin{center}
\begin{table*}[!htb]
    \centering
    \begin{adjustbox}{max width=\textwidth}
    \begin{tabular}{ll|rrrcrrrcrr}
          & & \multicolumn{3}{c}{Warm-Start} & & \multicolumn{3}{c}{Item Cold-Start} & & \multicolumn{2}{c}{User Cold-Start} \\
          & & \mlonem & \onion & \amazonvid & & \mlonem & \onion & \amazonvid & & \mlonem & \onion \\
          
          \cmidrule{1-5}
          \cmidrule{7-9}
          \cmidrule{11-12}
          \multirow{2}{*}{\baserec} & \rand & 0.0042 & 0.0006 & 0.0003 & \bfseries  & 0.0459 & 0.0061 & 0.0033 & \bfseries  & 0.0435 & 0.0069 \\
           & \pop & 0.1016 & 0.0136 & 0.0059 & \bfseries  & 0.0076 & 0.0056 & 0.0021 & \bfseries  & 0.4380 & 0.0927 \\
          \cmidrule{1-5}
          \cmidrule{7-9}
          \cmidrule{11-12}
          \multirow{2}{*}{\cfbr} & \mf & 0.2011 & 0.0818 & \bfseries 0.0158 & \bfseries  & 0.0292 & 0.0054 & 0.0035 & \bfseries  & 0.4424 & 0.0858 \\
           & \dmf & 0.0989 & 0.0793 & 0.0118 & \bfseries  & 0.0076 & 0.0056 & 0.0021 & \bfseries  & 0.4253 & 0.0581 \\
          \cmidrule{1-5}
          \cmidrule{7-9}
          \cmidrule{11-12}
          \multirow{7}{*}{\hrs} & \clcrec & \bfseries 0.2069$\dagger$ & 0.0764 & 0.0119 & \bfseries  & 0.1644 & 0.0890 & 0.0314 & \bfseries  & 0.4428 & 0.0923 \\
           & \newresults{\dropoutnetTwo} & \newresults{0.1948} & \newresults{0.0663} & \newresults{0.0091} & \newresults{\bfseries } & \newresults{0.1832} & \newresults{0.0850} & \newresults{0.0276} & \newresults{\bfseries } & \newresults{0.4369} & \newresults{0.0912} \\
           & \mubrarTwo & 0.2013 & 0.0878 & 0.0111 & \bfseries  & 0.2100 & 0.1386 & 0.0314 & \bfseries  & 0.4328 & 0.0881 \\
           & \sibrarTwo & 0.1983 & 0.0905 & 0.0151 & \bfseries  & 0.2108 & 0.1348 & \bfseries 0.0355$\dagger$ & \bfseries  & 0.4497 & 0.0852 \\
          \cdashline{2-5}[3.0pt/3.0pt]
          \cdashline{7-9}[3.0pt/3.0pt]
          \cdashline{11-12}[3.0pt/3.0pt]
           & \newresults{\dropoutnetBest} & \newresults{0.1948} & \newresults{0.0663} & \newresults{0.0100} & \newresults{\bfseries } & \newresults{0.1832} & \newresults{0.1109} & \newresults{0.0276} & \newresults{\bfseries } & \newresults{0.4369} & \newresults{0.1008} \\
           & \mubrarBest & 0.2013 & \bfseries 0.1097 & 0.0111 & \bfseries  & 0.2300 & 0.1432 & 0.0314 & \bfseries  & 0.4371 & 0.0873 \\
           & \sibrarBest & 0.1977 & 0.1063 & 0.0151 & \bfseries  & \bfseries 0.2350$\dagger$ & \bfseries 0.1449 & \bfseries 0.0355$\dagger$ & \bfseries  & \bfseries 0.4534 & \bfseries 0.1044 \\

\cmidrule{1-5}
\cmidrule{7-9}
\cmidrule{11-12}
    \end{tabular}
    \end{adjustbox}
    \vspace{0.6em}
    \caption{\precten results for different \baserecs, \cfbrs, and \hrss. Each row indicates the results of the evaluated algorithm, and the columns blocks indicate the data scenario (warm start and user/item cold start) for the datasets (\mlonem, \onion, and \amazonvid). 
    Bold indicates the best-performing RS, $\dagger$ indicates significant improvement over all others (paired $t$-tests with $p<0.05$ considering Bonferroni correction).
    }
    \label{tab:results-precision}
\end{table*}
\end{center}

%% file: resources/tables/tab_results_mrr.tex
\begin{center}
\begin{table*}[!htb]
    \centering
    \begin{adjustbox}{max width=\textwidth}
    \begin{tabular}{ll|rrrcrrrcrr}
          & & \multicolumn{3}{c}{Warm-Start} & & \multicolumn{3}{c}{Item Cold-Start} & & \multicolumn{2}{c}{User Cold-Start} \\
          & & \mlonem & \onion & \amazonvid & & \mlonem & \onion & \amazonvid & & \mlonem & \onion \\
          
          \cmidrule{1-5}
          \cmidrule{7-9}
          \cmidrule{11-12}
          \multirow{2}{*}{\baserec} & \rand & 0.0118 & 0.0017 & 0.0007 & \bfseries  & 0.1105 & 0.0178 & 0.0103 & \bfseries  & 0.1077 & 0.0231 \\
           & \pop & 0.2581 & 0.0423 & 0.0189 & \bfseries  & 0.0246 & 0.0133 & 0.0076 & \bfseries  & 0.6993 & 0.2125 \\
          \cmidrule{1-5}
          \cmidrule{7-9}
          \cmidrule{11-12}
          \multirow{2}{*}{\cfbr} & \mf & 0.4552 & 0.2658 & \bfseries 0.0681$\dagger$ & \bfseries  & 0.0702 & 0.0114 & 0.0114 & \bfseries  & 0.6996 & 0.1978 \\
           & \dmf & 0.2305 & 0.2463 & 0.0438 & \bfseries  & 0.0246 & 0.0133 & 0.0076 & \bfseries  & 0.5698 & 0.1274 \\
          \cmidrule{1-5}
          \cmidrule{7-9}
          \cmidrule{11-12}
          \multirow{7}{*}{\hrs} & \clcrec & \bfseries 0.4717 & 0.2564 & 0.0448 & \bfseries  & 0.2859 & 0.2973 & 0.1272 & \bfseries  & 0.6950 & 0.2114 \\
           & \newresults{\dropoutnetTwo} & \newresults{0.4489} & \newresults{0.2087} & \newresults{0.0317} & \newresults{\bfseries } & \newresults{0.3070} & \newresults{0.2147} & \newresults{0.0897} & \newresults{\bfseries } & \newresults{\bfseries 0.7004} & \newresults{0.2011} \\
           & \mubrarTwo & 0.4600 & 0.2642 & 0.0426 & \bfseries  & \bfseries 0.5353$\dagger$ & \bfseries 0.4059 & 0.1150 & \bfseries  & 0.6976 & 0.2005 \\
           & \sibrarTwo & 0.4627 & 0.2724 & 0.0580 & \bfseries  & 0.4635 & 0.3877 & \bfseries 0.1364$\dagger$ & \bfseries  & 0.6627 & 0.1998 \\
          \cdashline{2-5}[3.0pt/3.0pt]
          \cdashline{7-9}[3.0pt/3.0pt]
          \cdashline{11-12}[3.0pt/3.0pt]
           & \newresults{\dropoutnetBest} & \newresults{0.4489} & \newresults{0.2087} & \newresults{0.0348} & \newresults{\bfseries } & \newresults{0.3070} & \newresults{0.2969} & \newresults{0.0897} & \newresults{\bfseries } & \newresults{\bfseries 0.7004} & \newresults{0.2317} \\
           & \mubrarBest & 0.4600 & \bfseries 0.3132 & 0.0426 & \bfseries  & 0.5220 & 0.3938 & 0.1150 & \bfseries  & 0.6516 & 0.2127 \\
           & \sibrarBest & 0.4627 & 0.3111 & 0.0580 & \bfseries  & 0.5181 & 0.3885 & \bfseries 0.1364$\dagger$ & \bfseries  & 0.6541 & \bfseries 0.2406 \\

\cmidrule{1-5}
\cmidrule{7-9}
\cmidrule{11-12}
    \end{tabular}
    \end{adjustbox}
    \vspace{0.6em}
    \caption{\mrrten results for different \baserecs, \cfbrs, and \hrss. Each row indicates the results of the evaluated algorithm, and the columns blocks indicate the data scenario (warm start and user/item cold start) for the datasets (\mlonem, \onion, and \amazonvid). 
    Bold indicates the best-performing RS, $\dagger$ indicates significant improvement over all others (paired $t$-tests with $p<0.05$ considering Bonferroni correction).
    }
    \label{tab:results-mrr}
\end{table*}
\end{center}

%% file: resources/tables/tab_results_map.tex
\begin{center}
\begin{table*}[!htb]
    \centering
    \begin{adjustbox}{max width=\textwidth}
    \begin{tabular}{ll|rrrcrrrcrr}
          & & \multicolumn{3}{c}{Warm-Start} & & \multicolumn{3}{c}{Item Cold-Start} & & \multicolumn{2}{c}{User Cold-Start} \\
          & & \mlonem & \onion & \amazonvid & & \mlonem & \onion & \amazonvid & & \mlonem & \onion \\
          
          \cmidrule{1-5}
          \cmidrule{7-9}
          \cmidrule{11-12}
          \multirow{2}{*}{\baserec} & \rand & 0.0007 & 0.0003 & 0.0006 & \bfseries  & 0.0098 & 0.0019 & 0.0072 & \bfseries  & 0.0010 & 0.0003 \\
           & \pop & 0.0360 & 0.0056 & 0.0169 & \bfseries  & 0.0011 & 0.0016 & 0.0055 & \bfseries  & 0.0337 & 0.0056 \\
          \cmidrule{1-5}
          \cmidrule{7-9}
          \cmidrule{11-12}
          \multirow{2}{*}{\cfbr} & \mf & 0.0924 & 0.0739 & \bfseries 0.0618$\dagger$ & \bfseries  & 0.0048 & 0.0011 & 0.0077 & \bfseries  & 0.0332 & 0.0048 \\
           & \dmf & 0.0312 & 0.0615 & 0.0392 & \bfseries  & 0.0011 & 0.0016 & 0.0055 & \bfseries  & 0.0272 & 0.0026 \\
          \cmidrule{1-5}
          \cmidrule{7-9}
          \cmidrule{11-12}
          \multirow{7}{*}{\hrs} & \clcrec & \bfseries 0.0975 & 0.0726 & 0.0405 & \bfseries  & 0.0498 & 0.0753 & 0.0974 & \bfseries  & 0.0329 & 0.0055 \\
           & \newresults{\dropoutnetTwo} & \newresults{0.0889} & \newresults{0.0499} & \newresults{0.0282} & \newresults{\bfseries } & \newresults{0.0616} & \newresults{0.0519} & \newresults{0.0675} & \newresults{\bfseries } & \newresults{0.0328} & \newresults{0.0058} \\
           & \mubrarTwo & 0.0933 & 0.0734 & 0.0384 & \bfseries  & 0.0965 & \bfseries 0.1181 & 0.0877 & \bfseries  & 0.0335 & 0.0050 \\
           & \sibrarTwo & 0.0948 & 0.0747 & 0.0523 & \bfseries  & 0.0898 & 0.1100 & \bfseries 0.1054$\dagger$ & \bfseries  & \bfseries 0.0343 & 0.0054 \\
          \cdashline{2-5}[3.0pt/3.0pt]
          \cdashline{7-9}[3.0pt/3.0pt]
          \cdashline{11-12}[3.0pt/3.0pt]
           & \newresults{\dropoutnetBest} & \newresults{0.0889} & \newresults{0.0499} & \newresults{0.0309} & \newresults{\bfseries } & \newresults{0.0616} & \newresults{0.0777} & \newresults{0.0675} & \newresults{\bfseries } & \newresults{0.0328} & \newresults{0.0072} \\
           & \mubrarBest & 0.0933 & \bfseries 0.0862 & 0.0384 & \bfseries  & 0.1098 & 0.1146 & 0.0877 & \bfseries  & 0.0318 & 0.0056 \\
           & \sibrarBest & 0.0942 & 0.0840 & 0.0523 & \bfseries  & \bfseries 0.1134$\dagger$ & 0.1148 & \bfseries 0.1054$\dagger$ & \bfseries  & 0.0330 & \bfseries 0.0073 \\

\cmidrule{1-5}
\cmidrule{7-9}
\cmidrule{11-12}
    \end{tabular}
    \end{adjustbox}
    \vspace{0.6em}
    \caption{\mapten results for different \baserecs, \cfbrs, and \hrss. Each row indicates the results of the evaluated algorithm, and the columns blocks indicate the data scenario (warm start and user/item cold start) for the datasets (\mlonem, \onion, and \amazonvid). 
    Bold indicates the best-performing RS, $\dagger$ indicates significant improvement over all others (paired $t$-tests with $p<0.05$ considering Bonferroni correction).
    }
    \label{tab:results-map}
\end{table*}
\end{center}

%% file: resources/tables/tab_results_fscore.tex
\begin{center}
\begin{table*}[!htb]
    \centering
    \begin{adjustbox}{max width=\textwidth}
    \begin{tabular}{ll|rrrcrrrcrr}
          & & \multicolumn{3}{c}{Warm-Start} & & \multicolumn{3}{c}{Item Cold-Start} & & \multicolumn{2}{c}{User Cold-Start} \\
          & & \mlonem & \onion & \amazonvid & & \mlonem & \onion & \amazonvid & & \mlonem & \onion \\
          
          \cmidrule{1-5}
          \cmidrule{7-9}
          \cmidrule{11-12}
          \multirow{2}{*}{\baserec} & \rand & 0.0027 & 0.0005 & 0.0006 & \bfseries  & 0.0299 & 0.0054 & 0.0056 & \bfseries  & 0.0049 & 0.0014 \\
          & \pop & 0.0726 & 0.0124 & 0.0104 & \bfseries  & 0.0036 & 0.0048 & 0.0036 & \bfseries  & 0.0833 & 0.0203 \\
         \cmidrule{1-5}
         \cmidrule{7-9}
         \cmidrule{11-12}
         \multirow{2}{*}{\cfbr} & \mf & 0.1564 & 0.0876 & \bfseries 0.0280 & \bfseries  & 0.0166 & 0.0045 & 0.0060 & \bfseries  & 0.0851 & 0.0184 \\
          & \dmf & 0.0711 & 0.0799 & 0.0210 & \bfseries  & 0.0036 & 0.0048 & 0.0036 & \bfseries  & 0.0830 & 0.0126 \\
         \cmidrule{1-5}
         \cmidrule{7-9}
         \cmidrule{11-12}
         \multirow{7}{*}{\hrs} & \clcrec & \bfseries 0.1610$\dagger$ & 0.0827 & 0.0212 & \bfseries  & 0.1143 & 0.0834 & 0.0534 & \bfseries  & 0.0860 & 0.0202 \\
          & \newresults{\dropoutnetTwo} & \newresults{0.1514} & \newresults{0.0679} & \newresults{0.0161} & \newresults{\bfseries } & \newresults{0.1355} & \newresults{0.0814} & \newresults{0.0469} & \newresults{\bfseries } & \newresults{0.0835} & \newresults{0.0211} \\
          & \mubrarTwo & 0.1567 & 0.0934 & 0.0197 & \bfseries  & 0.1517 & 0.1331 & 0.0531 & \bfseries  & 0.0826 & 0.0183 \\
          & \sibrarTwo & 0.1572 & 0.0947 & 0.0268 & \bfseries  & 0.1580 & 0.1291 & \bfseries 0.0602$\dagger$ & \bfseries  & \bfseries 0.0866 & 0.0189 \\
         \cdashline{2-5}[3.0pt/3.0pt]
         \cdashline{7-9}[3.0pt/3.0pt]
         \cdashline{11-12}[3.0pt/3.0pt]
          & \newresults{\dropoutnetBest} & \newresults{0.1514} & \newresults{0.0679} & \newresults{0.0178} & \newresults{\bfseries } & \newresults{0.1355} & \newresults{0.1076} & \newresults{0.0469} & \newresults{\bfseries } & \newresults{0.0835} & \newresults{0.0240} \\
          & \mubrarBest & 0.1567 & \bfseries 0.1112 & 0.0197 & \bfseries  & 0.1678 & 0.1381 & 0.0531 & \bfseries  & 0.0835 & 0.0195 \\
          & \sibrarBest & 0.1561 & 0.1076 & 0.0268 & \bfseries  & \bfseries 0.1743$\dagger$ & \bfseries 0.1413 & \bfseries 0.0602$\dagger$ & \bfseries  & 0.0857 & \bfseries 0.0241 \\

\cmidrule{1-5}
\cmidrule{7-9}
\cmidrule{11-12}
    \end{tabular}
    \end{adjustbox}
    \vspace{0.6em}
    \caption{\fscoreten results for different \baserecs, \cfbrs, and \hrss. Each row indicates the results of the evaluated algorithm, and the columns blocks indicate the data scenario (warm start and user/item cold start) for the datasets (\mlonem, \onion, and \amazonvid). 
    Bold indicates the best-performing RS, $\dagger$ indicates significant improvement over all others (paired $t$-tests with $p<0.05$ considering Bonferroni correction).
    }
    \label{tab:results-fscore}
\end{table*}
\end{center}

%% file: resources/tables/tab_results_coverage.tex
\begin{center}
\begin{table*}[!htb]
    \centering
    \begin{adjustbox}{max width=\textwidth}
    \begin{tabular}{ll|rrrcrrrcrr}
          & & \multicolumn{3}{c}{Warm-Start} & & \multicolumn{3}{c}{Item Cold-Start} & & \multicolumn{2}{c}{User Cold-Start} \\
          & & \mlonem & \onion & \amazonvid & & \mlonem & \onion & \amazonvid & & \mlonem & \onion \\
          
          \cmidrule{1-5}
          \cmidrule{7-9}
          \cmidrule{11-12}
          \multirow{2}{*}{\baserec} & \rand & \bfseries 1.0000 & \bfseries 0.9799 & \bfseries 1.0000 & \bfseries  & \bfseries 1.0000 & \bfseries 1.0000 & \bfseries 1.0000 & \bfseries  & \bfseries 0.8484 & \bfseries 0.3853 \\
           & \pop & 0.0288 & 0.0024 & 0.0041 & \bfseries  & 0.0303 & 0.0073 & 0.0239 & \bfseries  & 0.0032 & 0.0009 \\
          \cmidrule{1-5}
          \cmidrule{7-9}
          \cmidrule{11-12}
          \multirow{2}{*}{\cfbr} & \mf & 0.5156 & 0.4712 & 0.7575 & \bfseries  & 0.0303 & 0.0073 & 0.0239 & \bfseries  & 0.0032 & 0.0009 \\
           & \dmf & 0.3741 & 0.3705 & 0.8875 & \bfseries  & 0.0303 & 0.0073 & 0.0239 & \bfseries  & 0.0032 & 0.0009 \\
          \cmidrule{1-5}
          \cmidrule{7-9}
          \cmidrule{11-12}
          \multirow{7}{*}{\hrs} & \clcrec & 0.4765 & 0.4417 & 0.8667 & \bfseries  & 0.6667 & 0.9993 & \bfseries 1.0000 & \bfseries  & 0.0055 & 0.0010 \\
           & \newresults{\dropoutnetTwo} & \newresults{0.4668} & \newresults{0.4799} & \newresults{0.4044} & \newresults{\bfseries} & \newresults{0.8818} & \newresults{0.8589} & \newresults{0.9737} & \newresults{\bfseries} & \newresults{0.0032} & \newresults{0.0028} \\
           & \mubrarTwo & 0.4862 & 0.5370 & 0.8870 & \bfseries  & 0.7242 & 0.9280 & \bfseries 1.0000 & \bfseries  & 0.0032 & 0.0009 \\
           & \sibrarTwo & 0.5008 & 0.6445 & 0.7204 & \bfseries  & 0.9606 & 0.9309 & 0.9928 & \bfseries  & 0.0055 & 0.0186 \\
          \cdashline{2-5}[3.0pt/3.0pt]
          \cdashline{7-9}[3.0pt/3.0pt]
          \cdashline{11-12}[3.0pt/3.0pt]
           & \newresults{\dropoutnetBest} & \newresults{0.4668} & \newresults{0.4799} & \newresults{0.4989} & \newresults{\bfseries } & \newresults{0.8818} & \newresults{0.8905} & \newresults{0.9737} & \newresults{\bfseries } & \newresults{0.0032} & \newresults{0.0055} \\
           & \mubrarBest & 0.4862 & 0.4707 & 0.8870 & \bfseries  & 0.8061 & 0.9148 & \bfseries 1.0000 & \bfseries  & 0.0032 & 0.0010 \\
           & \sibrarBest & 0.4568 & 0.5855 & 0.7204 & \bfseries  & 0.7636 & 0.9309 & 0.9928 & \bfseries  & 0.0100 & 0.0180 \\

\cmidrule{1-5}
\cmidrule{7-9}
\cmidrule{11-12}
    \end{tabular}
    \end{adjustbox}
    \vspace{0.6em}
    \caption{\covten results for different \baserecs, \cfbrs, and \hrss. Each row indicates the results of the evaluated algorithm, and the columns blocks indicate the data scenario (warm start and user/item cold start) for the datasets (\mlonem, \onion, and \amazonvid). 
    Bold indicates the best-performing RS.
    }
    \label{tab:results-coverage}
\end{table*}
\end{center}

%% file: resources/tables/tab_results_sibrar_mubrar_recall.tex

\begin{center}
\begin{table*}[!htb]
    \centering
    \begin{adjustbox}{max width=\textwidth}
    \begin{tabular}{l|rrrcrrrcrr}
          & \multicolumn{3}{c}{Warm-Start} & & \multicolumn{3}{c}{Item Cold-Start} & & \multicolumn{2}{c}{User Cold-Start} \\
          & \mlonem & \onion & \amazonvid & & \mlonem & \onion & \amazonvid & & \mlonem & \onion \\
    
          \cmidrule{1-4}
          \cmidrule{6-8}
          \cmidrule{10-11}
          \mubrarVanilla & \bfseries 0.1823 & 0.1237 & 0.0875 & \bfseries  & 0.1431 & 0.1727 & 0.2062 & \bfseries  & 0.0510 & 0.0113 \\
          \mubrarSamp & 0.1641 & 0.1523 & 0.0918 & \bfseries  & 0.1583 & 0.2031 & 0.1920 & \bfseries  & 0.0498 & 0.0129 \\
          \mubrarSampContr & 0.1813 & \bfseries 0.1681 & 0.0977 & \bfseries  & 0.1775 & 0.2014 & 0.2159 & \bfseries  & 0.0510 & 0.0121 \\
          \cdashline{1-4}[3.0pt/3.0pt]
          \cdashline{6-8}[3.0pt/3.0pt]
          \cdashline{10-11}[3.0pt/3.0pt]
          \sibrarSamp & 0.1798 & 0.1479 & 0.1159 & \bfseries  & 0.1848 & 0.2024 & 0.1573 & \bfseries  & 0.0502 & \bfseries 0.0155 \\
          \sibrarSampContr & 0.1820 & 0.1624 & \bfseries 0.1328$\dagger$ & \bfseries  & \bfseries 0.1869 & \bfseries 0.2114$\dagger$ & \bfseries 0.2466$\dagger$ & \bfseries  & \bfseries 0.0522 & 0.0146 \\

          \cmidrule{1-4}
          \cmidrule{6-8}
          \cmidrule{10-11}
    \end{tabular}
    \end{adjustbox}
    \vspace{0.6em}
    \caption{\recten results for the best-performing \sibrar and \mubrar variants on different warm- and user and item cold-start datasets (\mlonem, \onion, and \amazonvid). Bold indicates the best-performing RSs, $\dagger$ indicates significant improvement over all other variants (paired $t$-tests with $p<0.05$ and Bonferroni correction).
    }
    \label{tab:results-recall:sibrar-mubrar}
\end{table*}
\end{center}

%% file: resources/tables/tab_results_sibrar_mubrar_precision.tex

\begin{center}
\begin{table*}[!htb]
    \centering
    \begin{adjustbox}{max width=\textwidth}
    \begin{tabular}{l|rrrcrrrcrr}
          & \multicolumn{3}{c}{Warm-Start} & & \multicolumn{3}{c}{Item Cold-Start} & & \multicolumn{2}{c}{User Cold-Start} \\
          & \mlonem & \onion & \amazonvid & & \mlonem & \onion & \amazonvid & & \mlonem & \onion \\
    
          \cmidrule{1-4}
          \cmidrule{6-8}
          \cmidrule{10-11}
          \mubrarVanilla & \bfseries 0.2014 & 0.0750 & 0.0099 & \bfseries  & 0.1952 & 0.1264 & 0.0295 & \bfseries  & 0.4381 & 0.0846 \\
          \mubrarSamp & 0.1890 & 0.0982 & 0.0104 & \bfseries  & 0.2107 & 0.1416 & 0.0277 & \bfseries  & 0.4363 & 0.0846 \\
          \mubrarSampContr & 0.2013 & \bfseries 0.1097$\dagger$ & 0.0111 & \bfseries  & 0.2300 & 0.1432 & 0.0314 & \bfseries  & 0.4371 & 0.0873 \\
          \cdashline{1-4}[3.0pt/3.0pt]
          \cdashline{6-8}[3.0pt/3.0pt]
          \cdashline{10-11}[3.0pt/3.0pt]
          \sibrarSamp & 0.1967 & 0.0979 & 0.0131 & \bfseries  & 0.2342 & 0.1389 & 0.0230 & \bfseries  & 0.4430 & \bfseries 0.1044 \\
          \sibrarSampContr & 0.1977 & 0.1063 & \bfseries 0.0151$\dagger$ & \bfseries  & \bfseries 0.2350 & \bfseries 0.1449 & \bfseries 0.0355$\dagger$ & \bfseries  & \bfseries 0.4534 & 0.1025 \\

          \cmidrule{1-4}
          \cmidrule{6-8}
          \cmidrule{10-11}
    \end{tabular}
    \end{adjustbox}
    \vspace{0.6em}
    \caption{\precten results for the best-performing \sibrar and \mubrar variants on different warm- and user and item cold-start datasets (\mlonem, \onion, and \amazonvid). Bold indicates the best-performing RSs, $\dagger$ indicates significant improvement over all other variants (paired $t$-tests with $p<0.05$ and Bonferroni correction).
    }
    \label{tab:results-precision:sibrar-mubrar}
\end{table*}
\end{center}

%% file: resources/tables/tab_results_sibrar_mubrar_mrr.tex

\begin{center}
\begin{table*}[!htb]
    \centering
    \begin{adjustbox}{max width=\textwidth}
    \begin{tabular}{l|rrrcrrrcrr}
          & \multicolumn{3}{c}{Warm-Start} & & \multicolumn{3}{c}{Item Cold-Start} & & \multicolumn{2}{c}{User Cold-Start} \\
          & \mlonem & \onion & \amazonvid & & \mlonem & \onion & \amazonvid & & \mlonem & \onion \\
    
          \cmidrule{1-4}
          \cmidrule{6-8}
          \cmidrule{10-11}
          \mubrarVanilla & 0.4627 & 0.2360 & 0.0354 & \bfseries  & 0.4171 & 0.3724 & 0.1123 & \bfseries  & \bfseries 0.6991 & 0.1866 \\
          \mubrarSamp & 0.4408 & 0.2908 & 0.0404 & \bfseries  & 0.4734 & 0.3903 & 0.0961 & \bfseries  & 0.6902 & 0.2003 \\
          \mubrarSampContr & 0.4600 & \bfseries 0.3132 & 0.0426 & \bfseries  & 0.5220 & \bfseries 0.3938 & 0.1150 & \bfseries  & 0.6516 & 0.2127 \\
          \cdashline{1-4}[3.0pt/3.0pt]
          \cdashline{6-8}[3.0pt/3.0pt]
          \cdashline{10-11}[3.0pt/3.0pt]
          \sibrarSamp & 0.4592 & 0.2797 & 0.0479 & \bfseries  & \bfseries 0.5505$\dagger$ & 0.3733 & 0.0849 & \bfseries  & 0.6953 & \bfseries 0.2406 \\
          \sibrarSampContr & \bfseries 0.4627 & 0.3111 & \bfseries 0.0580$\dagger$ & \bfseries  & 0.5181 & 0.3885 & \bfseries 0.1364$\dagger$ & \bfseries  & 0.6541 & 0.2283 \\

          \cmidrule{1-4}
          \cmidrule{6-8}
          \cmidrule{10-11}
    \end{tabular}
    \end{adjustbox}
    \vspace{0.6em}
    \caption{\mrrten results for the best-performing \sibrar and \mubrar variants on different warm- and user and item cold-start datasets (\mlonem, \onion, and \amazonvid). Bold indicates the best-performing RSs, $\dagger$ indicates significant improvement over all other variants (paired $t$-tests with $p<0.05$ and Bonferroni correction).
    }
    \label{tab:results-mrr:sibrar-mubrar}
\end{table*}
\end{center}

%% file: resources/tables/tab_results_sibrar_mubrar_map.tex

\begin{center}
\begin{table*}[!htb]
    \centering
    \begin{adjustbox}{max width=\textwidth}
    \begin{tabular}{l|rrrcrrrcrr}
          & \multicolumn{3}{c}{Warm-Start} & & \multicolumn{3}{c}{Item Cold-Start} & & \multicolumn{2}{c}{User Cold-Start} \\
          & \mlonem & \onion & \amazonvid & & \mlonem & \onion & \amazonvid & & \mlonem & \onion \\
    
          \cmidrule{1-4}
          \cmidrule{6-8}
          \cmidrule{10-11}
          \mubrarVanilla & 0.0937 & 0.0617 & 0.0319 & \bfseries  & 0.0779 & 0.1008 & 0.0868 & \bfseries  & \bfseries 0.0333 & 0.0046 \\
          \mubrarSamp & 0.0844 & 0.0761 & 0.0369 & \bfseries  & 0.0898 & 0.1132 & 0.0728 & \bfseries  & 0.0322 & 0.0053 \\
          \mubrarSampContr & 0.0933 & \bfseries 0.0862 & 0.0384 & \bfseries  & 0.1098 & 0.1146 & 0.0877 & \bfseries  & 0.0318 & 0.0056 \\
          \cdashline{1-4}[3.0pt/3.0pt]
          \cdashline{6-8}[3.0pt/3.0pt]
          \cdashline{10-11}[3.0pt/3.0pt]
          \sibrarSamp & 0.0928 & 0.0738 & 0.0432 & \bfseries  & \bfseries 0.1140 & 0.1093 & 0.0639 & \bfseries  & 0.0328 & \bfseries 0.0073 \\
          \sibrarSampContr & \bfseries 0.0942 & 0.0840 & \bfseries 0.0523$\dagger$ & \bfseries  & 0.1134 & \bfseries 0.1148 & \bfseries 0.1054$\dagger$ & \bfseries  & 0.0330 & 0.0068 \\

          \cmidrule{1-4}
          \cmidrule{6-8}
          \cmidrule{10-11}
    \end{tabular}
    \end{adjustbox}
    \vspace{0.6em}
    \caption{\mapten results for the best-performing \sibrar and \mubrar variants on different warm- and user and item cold-start datasets (\mlonem, \onion, and \amazonvid). Bold indicates the best-performing RSs, $\dagger$ indicates significant improvement over all other variants (paired $t$-tests with $p<0.05$ and Bonferroni correction).
    }
    \label{tab:results-map:sibrar-mubrar}
\end{table*}
\end{center}

%% file: resources/tables/tab_results_sibrar_mubrar_fscore.tex

\begin{center}
\begin{table*}[!htb]
    \centering
    \begin{adjustbox}{max width=\textwidth}
    \begin{tabular}{l|rrrcrrrcrr}
          & \multicolumn{3}{c}{Warm-Start} & & \multicolumn{3}{c}{Item Cold-Start} & & \multicolumn{2}{c}{User Cold-Start} \\
          & \mlonem & \onion & \amazonvid & & \mlonem & \onion & \amazonvid & & \mlonem & \onion \\
    
          \cmidrule{1-4}
          \cmidrule{6-8}
          \cmidrule{10-11}
          \mubrarVanilla & \bfseries 0.1575 & 0.0782 & 0.0176 & \bfseries  & 0.1385 & 0.1202 & 0.0502 & \bfseries  & 0.0836 & 0.0183 \\
          \mubrarSamp & 0.1446 & 0.0999 & 0.0185 & \bfseries  & 0.1512 & 0.1375 & 0.0470 & \bfseries  & 0.0819 & 0.0197 \\
          \mubrarSampContr & 0.1567 & \bfseries 0.1112$\dagger$ & 0.0197 & \bfseries  & 0.1678 & 0.1381 & 0.0531 & \bfseries  & 0.0835 & 0.0195 \\
          \cdashline{1-4}[3.0pt/3.0pt]
          \cdashline{6-8}[3.0pt/3.0pt]
          \cdashline{10-11}[3.0pt/3.0pt]
          \sibrarSamp & 0.1541 & 0.0982 & 0.0233 & \bfseries  & 0.1724 & 0.1359 & 0.0388 & \bfseries  & 0.0826 & \bfseries 0.0241 \\
          \sibrarSampContr & 0.1561 & 0.1076 & \bfseries 0.0268$\dagger$ & \bfseries  & \bfseries 0.1743 & \bfseries 0.1413 & \bfseries 0.0602$\dagger$ & \bfseries  & \bfseries 0.0857 & 0.0234 \\

          \cmidrule{1-4}
          \cmidrule{6-8}
          \cmidrule{10-11}
    \end{tabular}
    \end{adjustbox}
    \vspace{0.6em}
    \caption{\fscoreten results for the best-performing \sibrar and \mubrar variants on different warm- and user and item cold-start datasets (\mlonem, \onion, and \amazonvid). Bold indicates the best-performing RSs, $\dagger$ indicates significant improvement over all other variants (paired $t$-tests with $p<0.05$ and Bonferroni correction).
    }
    \label{tab:results-fscore:sibrar-mubrar}
\end{table*}
\end{center}

%% file: resources/tables/tab_results_sibrar_mubrar_coverage.tex

\begin{center}
\begin{table*}[!htb]
    \centering
    \begin{adjustbox}{max width=\textwidth}
    \begin{tabular}{l|rrrcrrrcrr}
          & \multicolumn{3}{c}{Warm-Start} & & \multicolumn{3}{c}{Item Cold-Start} & & \multicolumn{2}{c}{User Cold-Start} \\
          & \mlonem & \onion & \amazonvid & & \mlonem & \onion & \amazonvid & & \mlonem & \onion \\
    
          \cmidrule{1-4}
          \cmidrule{6-8}
          \cmidrule{10-11}
          \mubrarVanilla & 0.5095 & \bfseries 0.7009 & \bfseries 0.9299 & \bfseries  & 0.9424 & \bfseries 0.9890 & \bfseries 1.0000 & \bfseries  & 0.0032 & 0.0009 \\
          \mubrarSamp & 0.4274 & 0.4967 & 0.8832 & \bfseries  & \bfseries 0.9758 & 0.9530 & \bfseries 1.0000 & \bfseries  & 0.0032 & 0.0023 \\
          \mubrarSampContr & 0.4862 & 0.4707 & 0.8870 & \bfseries  & 0.8061 & 0.9148 & \bfseries 1.0000 & \bfseries  & 0.0032 & 0.0010 \\
          \cdashline{1-4}[3.0pt/3.0pt]
          \cdashline{6-8}[3.0pt/3.0pt]
          \cdashline{10-11}[3.0pt/3.0pt]
          \sibrarSamp & \bfseries 0.5362 & 0.4957 & 0.6474 & \bfseries  & 0.8182 & 0.9552 & \bfseries 1.0000 & \bfseries  & \bfseries 0.0119 & \bfseries 0.0180 \\
          \sibrarSampContr & 0.4568 & 0.5855 & 0.7204 & \bfseries  & 0.7636 & 0.9309 & 0.9928 & \bfseries  & 0.0100 & 0.0087 \\

          \cmidrule{1-4}
          \cmidrule{6-8}
          \cmidrule{10-11}
    \end{tabular}
    \end{adjustbox}
    \vspace{0.6em}
    \caption{\covten results for the best-performing \sibrar and \mubrar variants on different warm- and user and item cold-start datasets (\mlonem, \onion, and \amazonvid). Bold indicates the best-performing RSs.
    }
    \label{tab:results-coverage:sibrar-mubrar}
\end{table*}
\end{center}

%% file: resources/tables/tab_missing_modalities_number_random_ml.tex
\begin{table}[]
    \begin{tabular}{lccc}
         & \multicolumn{3}{c}{\# item modalities} \\
         & 1 & 2 & 3 \\
         \midrule
         \mubrarVanilla & 0.0866 & 0.1595 & 0.2355 \\ 
         \mubrarNoReg & 0.1260 & 0.2060 & 0.2427 \\ 
         \mubrarReg & 0.1547 & 0.2358 & 0.2560 \\ 
         \hdashline
         \sibrarNoReg & 0.1734 & 0.2332 & 0.2568 \\ 
         \sibrarReg & \bfseries 0.1851 & \bfseries 0.2388 & \bfseries 0.2581 \\ 
         \bottomrule
    \end{tabular}
    \vspace{0.6em}
    \caption{\ndcgten results of different \sibrar and \mubrar variants trained on warm-start \mlonem with the same set of item modalities (interactions, genres, and text) and evaluated on subsets of those. \ndcgten values of different subsets with the same number of modalities are averaged. Bold indicates the best-performing algorithm per the number of modalities used. 
    }
    \label{tab:missing-modality:number-of-mods:random:ml}
\end{table}

%% file: resources/tables/tab_missing_modalities_number_random_amazon.tex
\begin{table}[]
    \begin{tabular}{lccccc}
         & \multicolumn{5}{c}{\# item modalities} \\
         & 1 & 2 \\
         \midrule
        \mubrarVanilla & 0.0199 & 0.0427  \\
        \mubrarNoReg & 0.0442 & 0.0508 \\
        \mubrarReg & 0.0494 & 0.0535 \\
        \hdashline
        \sibrarNoReg & 0.0541 & 0.0615 \\
        \sibrarReg & \textbf{0.0708} & \textbf{0.0728} \\
        \bottomrule
    \end{tabular}
    \vspace{0.6em}
    \caption{\ndcgten results of different \sibrar and \mubrar variants trained on warm-start \amazonvid with the same set of item modalities (interactions, text) and evaluated on subsets of those. \ndcgten values of different subsets with the same number of modalities are averaged. Bold indicates the best-performing algorithm per the number of modalities used. 
    }
    \label{tab:missing-modality:number-of-mods:random:amazon}
\end{table}

%% file: resources/tables/tab_missing_modalities_number_cs_ml.tex
\begin{table}[]
    \begin{tabular}{lcc}
         & \multicolumn{2}{c}{\# item modalities} \\
         & 1 & 2 \\
         \midrule
        \mubrarVanilla & 0.1502 & 0.2511 \\ 
        \mubrarNoReg & 0.1748 & 0.2620 \\ 
        \mubrarReg & 0.1982 & 0.2949 \\ 
        \hdashline
        \sibrarNoReg & 0.2374 & \bfseries 0.3046 \\ 
        \sibrarReg & \bfseries 0.2413 & 0.2994 \\ 
         \bottomrule
    \end{tabular}
    \vspace{0.6em}
    \caption{\ndcgten results of different \sibrar and \mubrar variants trained on cold-start \mlonem with the same set of item modalities (interactions, genres, and text) and evaluated on subsets of those (excluding interactions). \ndcgten values of different subsets with the same number of modalities are averaged. Bold indicates the best-performing algorithm per the number of modalities used. 
    }
    \label{tab:missing-modality:number-of-mods:cs:ml}
\end{table}

%% file: resources/figures/missing_modalities/fig-missing-modality-random-ml1m-best.tex
\begin{figure*}
    \centering
    \begin{subfigure}[b]{0.48\columnwidth}
        \centering
        \includegraphics[width=\textwidth]{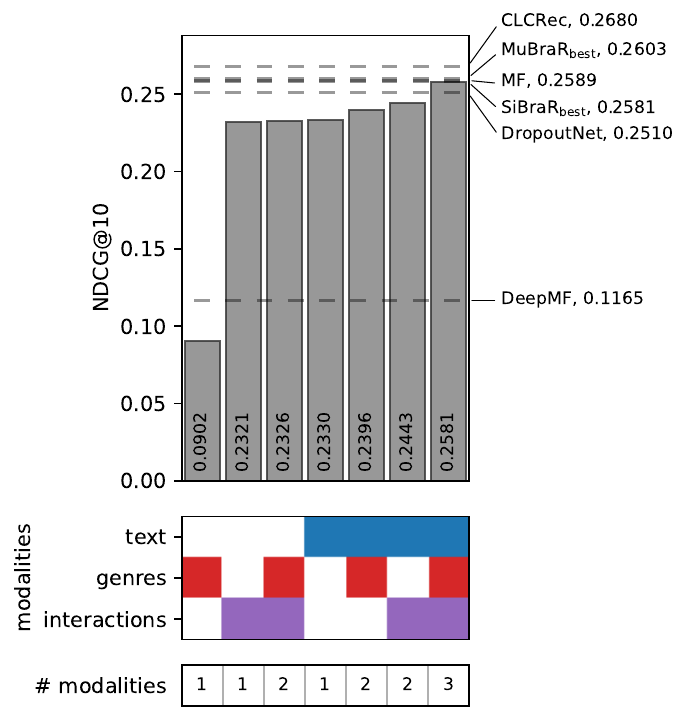}
        \caption{\sibrarBest}
        \label{fig:mm:ml:sibrar}
    \end{subfigure}
    \begin{subfigure}[b]{0.48\columnwidth}
        \centering
        \includegraphics[width=\textwidth]{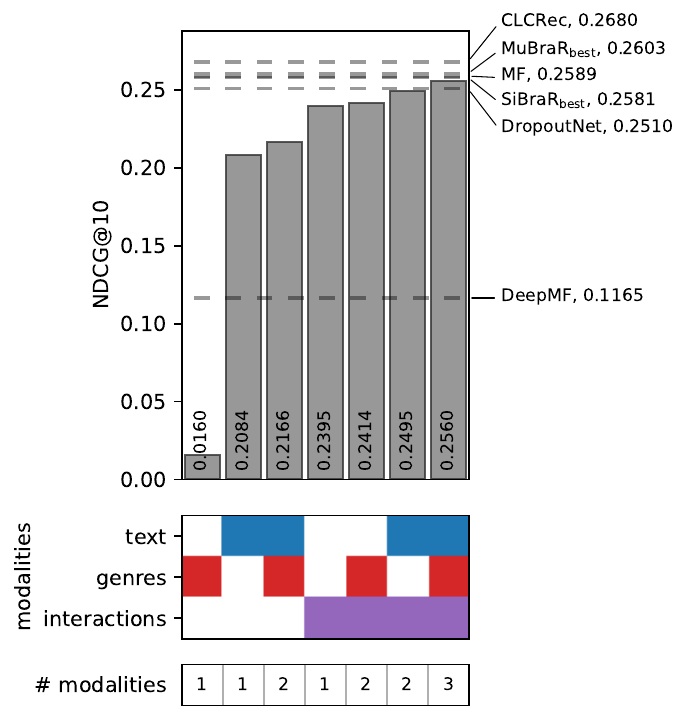}
        \caption{\mubrarBest}
        \label{fig:mm:ml:mubrar}
    \end{subfigure}    
    \caption{\ndcgten results of (a)~\sibrarBest and (b)~\mubrarBest trained on warm-start \mlonem with interactions, genres, and text modalities. The bars correspond to evaluations on the test set with varying evaluation modalities. If a modality is used, its block is filled with the corresponding color in the central plot. The bottom plot counts the number of modalities per bar. The gray dashed horizontal lines in the bar plot show the \ndcgten of the best \cfbrs and \hrss in warm-start \mlonem.
    }
    \label{fig:mm:random:ml}
\end{figure*}

%% file: resources/figures/missing_modalities/fig-missing-modality-random-amazon-best.tex
\begin{figure*}
    \centering
    \begin{subfigure}[b]{0.38\columnwidth}
        \centering
        \includegraphics[width=\textwidth]{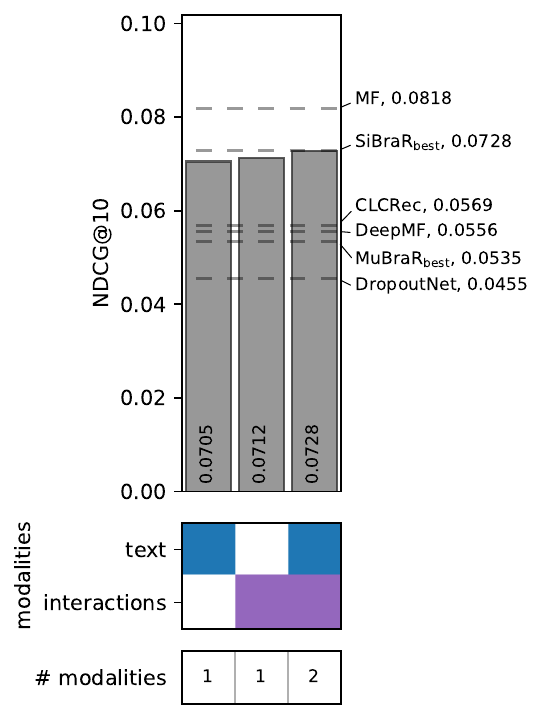}
        \caption{\sibrarBest}
        \label{fig:mm:amazon:sibrar}
    \end{subfigure}
    \hspace{1.5em}
    \begin{subfigure}[b]{0.38\columnwidth}
        \centering
        \includegraphics[width=\textwidth]{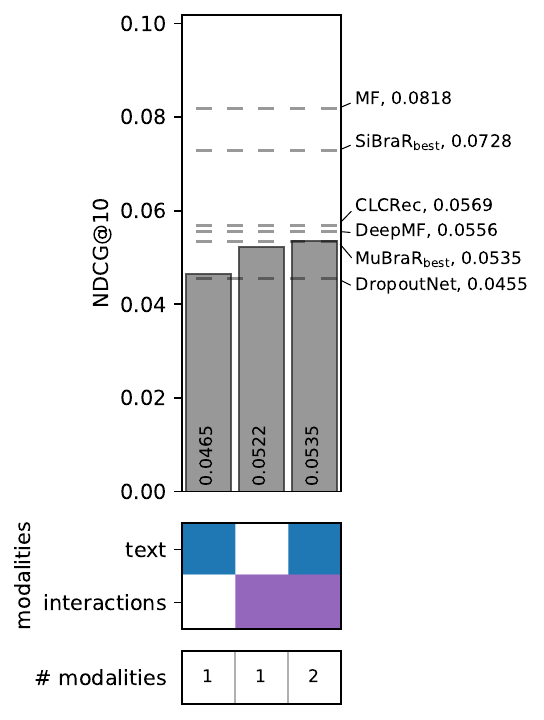}
        \caption{\mubrarBest}
        \label{fig:mm:amazon:mubrar}
    \end{subfigure}    
    \caption{\ndcgten results of (a)~\sibrarBest and (b)~\mubrarBest trained on warm-start \amazonvid with interactions, and text modalities. The bars correspond to evaluations on the test set with varying evaluation modalities. If a modality is used, its block is filled with the corresponding color in the central plot. The bottom plot counts the number of modalities per bar. The gray dashed horizontal lines in the bar plot show the \ndcgten of the best \cfbrs and \hrss in warm-start \amazonvid.
    }
    \label{fig:mm:random:amazon}
\end{figure*}

%% file: resources/figures/missing_modalities/fig-missing-modality-random-ml1m-amazon-diff-best.tex
\begin{figure*}        
    \centering
    \begin{subfigure}[b]{0.48\columnwidth}
        \centering
        \includegraphics[width=\textwidth]{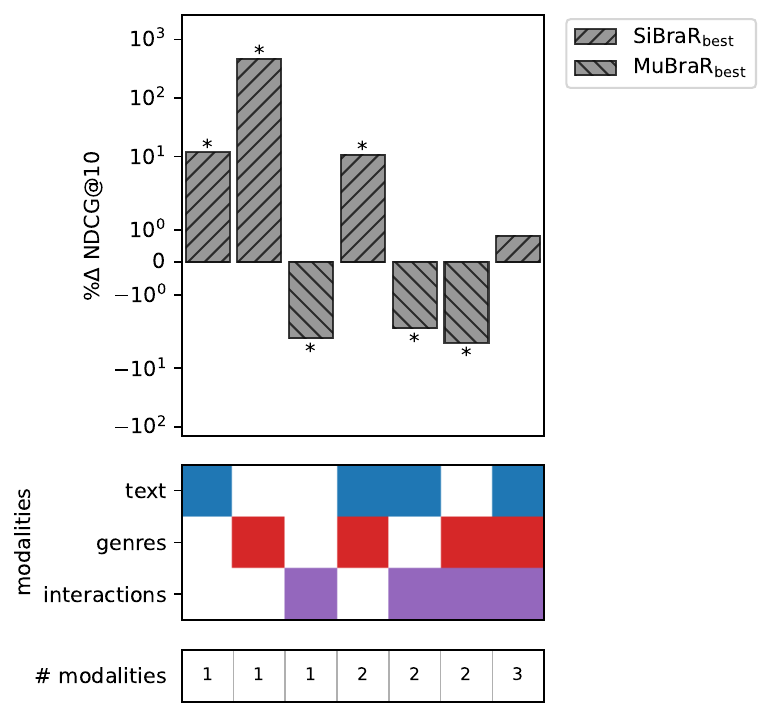}
        \caption{\mlonem}
        \label{fig:mm:random:ml:diff:best}
    \end{subfigure}
    \begin{subfigure}[b]{0.38\columnwidth}
        \centering
        \includegraphics[width=\textwidth]{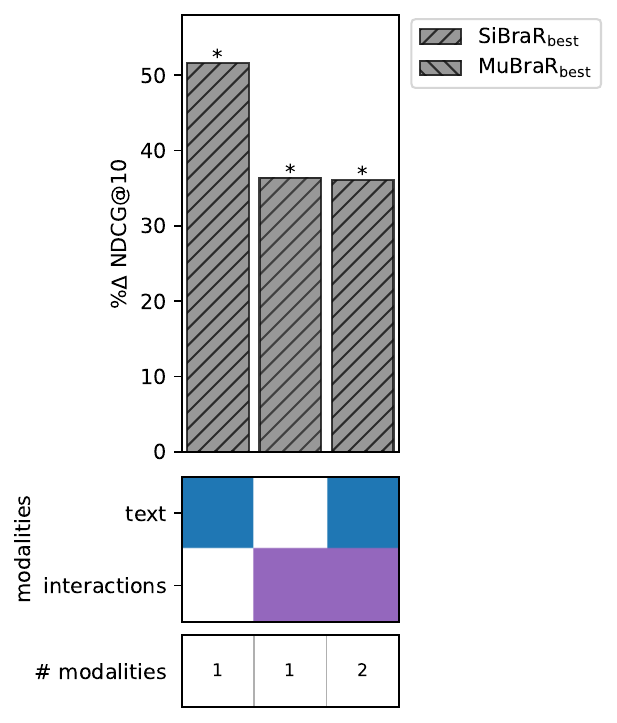}
        \caption{\amazonvid}
        \label{fig:mm:random:amazon:diff:best}
    \end{subfigure}
    \caption{\ndcgten percent difference between \sibrarBest and \mubrarBest. For (a), both models are trained on warm-start \mlonem with interactions, genres, and text as modalities. For (b), they are trained on warm-start \amazonvid with interactions and text. Positive values indicate that \sibrarBest performs better than \mubrarBest on a specific set of modalities, while negative values indicate a poorer performance.
    }
    \label{fig:mm:random:ml-amazon:diff:best}
\end{figure*}

%% file: resources/figures/missing_modalities/fig-missing-modality-random-ml1m-diff-samp-contr.tex
\begin{figure*}
    \centering
    \begin{subfigure}[b]{0.48\columnwidth}
        \centering
        \includegraphics[width=\textwidth]{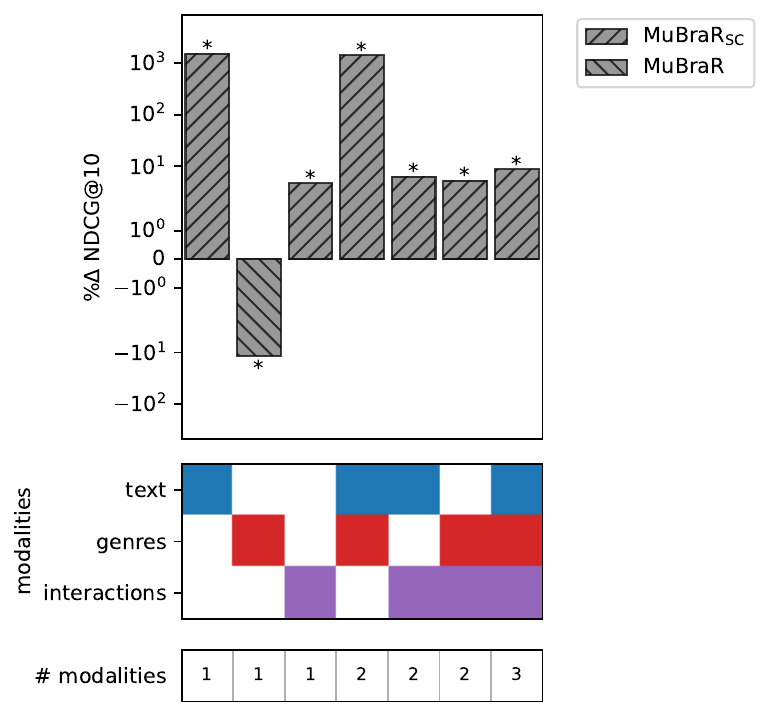}
        \caption{\mubrarSampContr vs. \mubrar}
        \label{fig:mm:random:ml:diff:samp-contr:mubrar}
    \end{subfigure}
    \begin{subfigure}[b]{0.48\columnwidth}
        \centering
        \includegraphics[width=\textwidth]{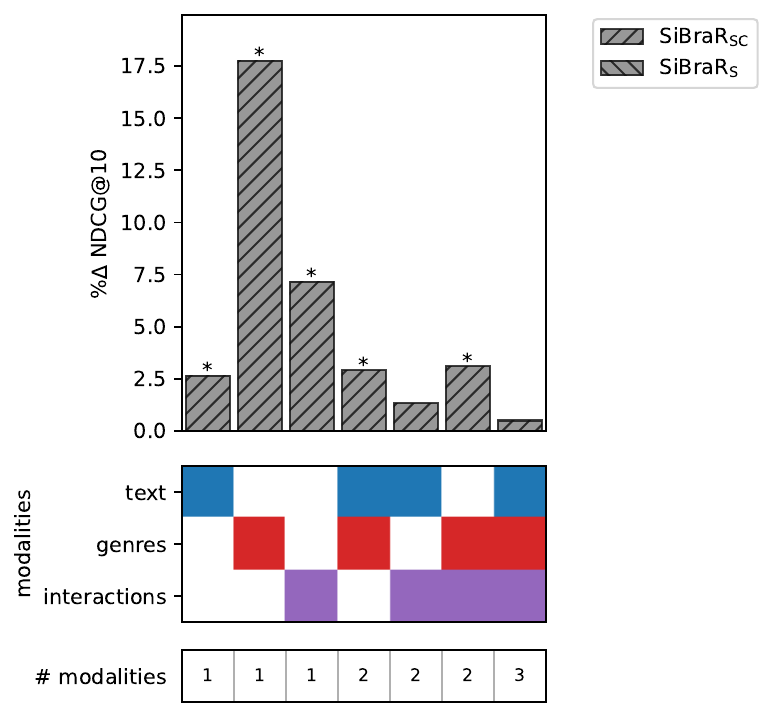}
        \caption{\sibrarSampContr vs. \sibrarSamp}
        \label{fig:mm:random:ml:diff:samp-contr:sibrar}
    \end{subfigure}
    \caption{\ndcgten comparison for single- and multi-branch architectures when trained on all modalities of warm-start \mlonem, with and without utilizing modality sampling (S) and a contrastive loss (C). (a)~shows the difference between \mubrar and \mubrarSampContr, and (b)~the difference between \sibrarSamp and \sibrarSampContr. The colored squares in the middle plot indicate the modalities used during evaluation. Moreover, the bar plot shows the percent difference of \ndcgten, where positive values indicate that a contrastive loss is beneficial for recommendation quality, and worse if negative.
    }
    \label{fig:mm:random:ml:diff:samp-contr}
\end{figure*}

%% file: resources/figures/missing_modalities/fig-missing-modality-random-amazon-diff-samp-contr.tex
\begin{figure*}
    \centering
    \begin{subfigure}[b]{0.35\columnwidth}
        \centering
        \includegraphics[width=\textwidth]{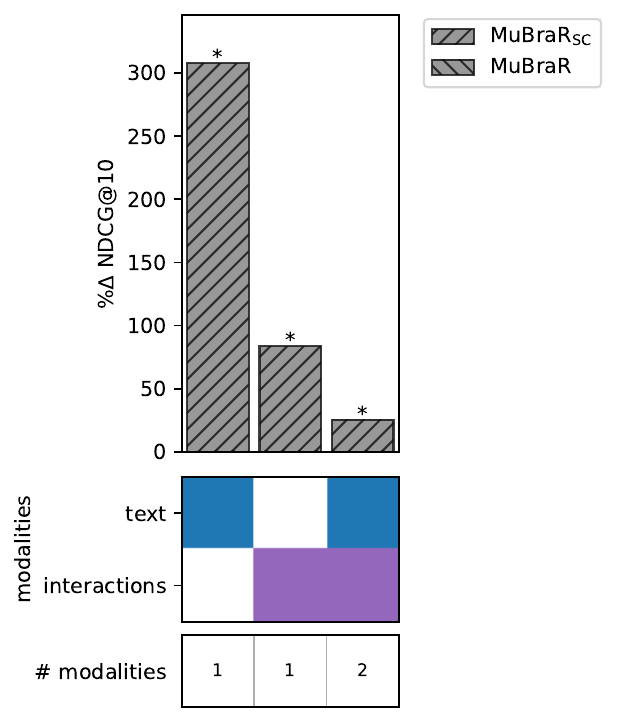}
        \caption{\mubrarSampContr vs. \mubrar}
        \label{fig:mm:random:amazon:diff:samp-contr:mubrar}
    \end{subfigure}
    \vspace{1.5em}
    \begin{subfigure}[b]{0.35\columnwidth}
        \centering
        \includegraphics[width=\textwidth]{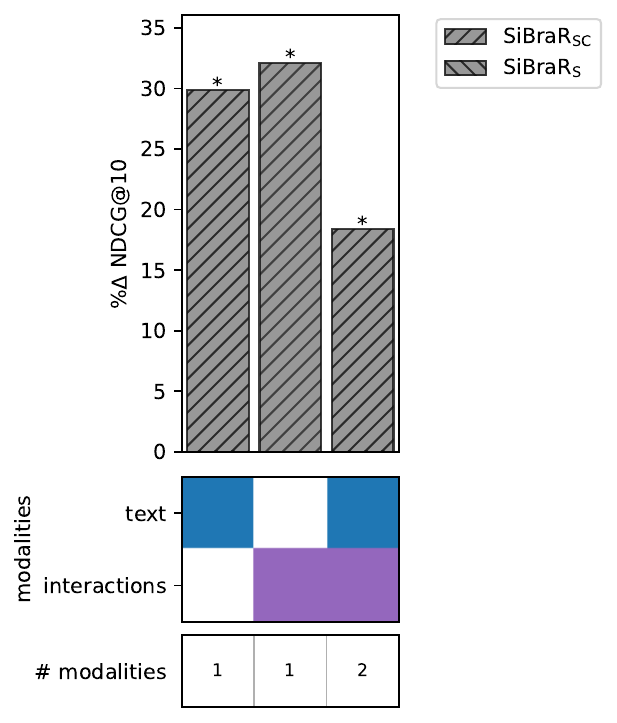}
        \caption{\sibrarSampContr vs. \sibrarSamp}
        \label{fig:mm:random:amazon:diff:samp-contr:sibrar}
    \end{subfigure}
    \vspace{-2em}
    \caption{\ndcgten comparison for single- and multi-branch architectures when trained on all modalities of warm-start \amazonvid, with and without utilizing modality sampling (S) and a contrastive loss (C). (a)~shows the difference between \mubrar and \mubrarSampContr, and (b)~the difference between \sibrarSamp and \sibrarSampContr. The colored squares in the middle plot indicate the modalities used during evaluation. Moreover, the bar plot shows the percent difference of \ndcgten, where positive values indicate that a contrastive loss is beneficial for recommendation quality, and worse if negative.
    }
    \label{fig:mm:random:amazon:diff:samp-contr}
\end{figure*}

%% file: resources/figures/contrastive_loss/fig-contrastive-loss-gridsearch_appendix.tex
\begin{figure}
    \centering
    \begin{subfigure}[b]{1\columnwidth}
        \centering
        \includegraphics[width=\textwidth]{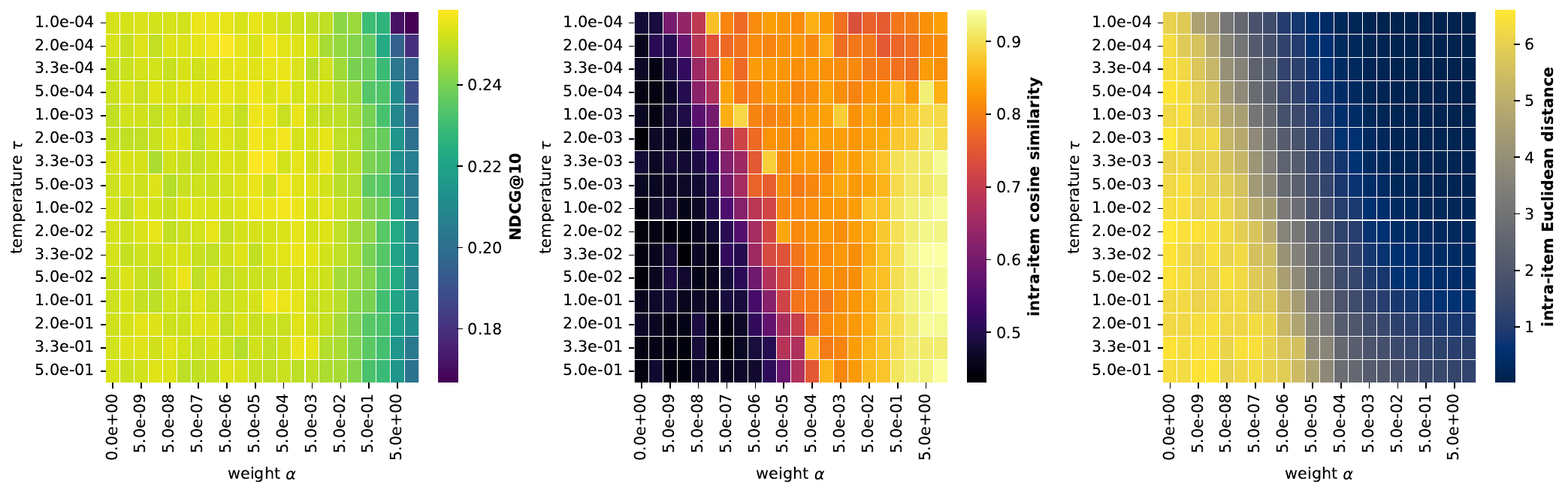}
        \caption{\mlonem}
        \label{fig:contrastive:gridsearch:ml}
    \end{subfigure}
    \begin{subfigure}[b]{1\columnwidth}
        \centering
        \includegraphics[width=\textwidth]{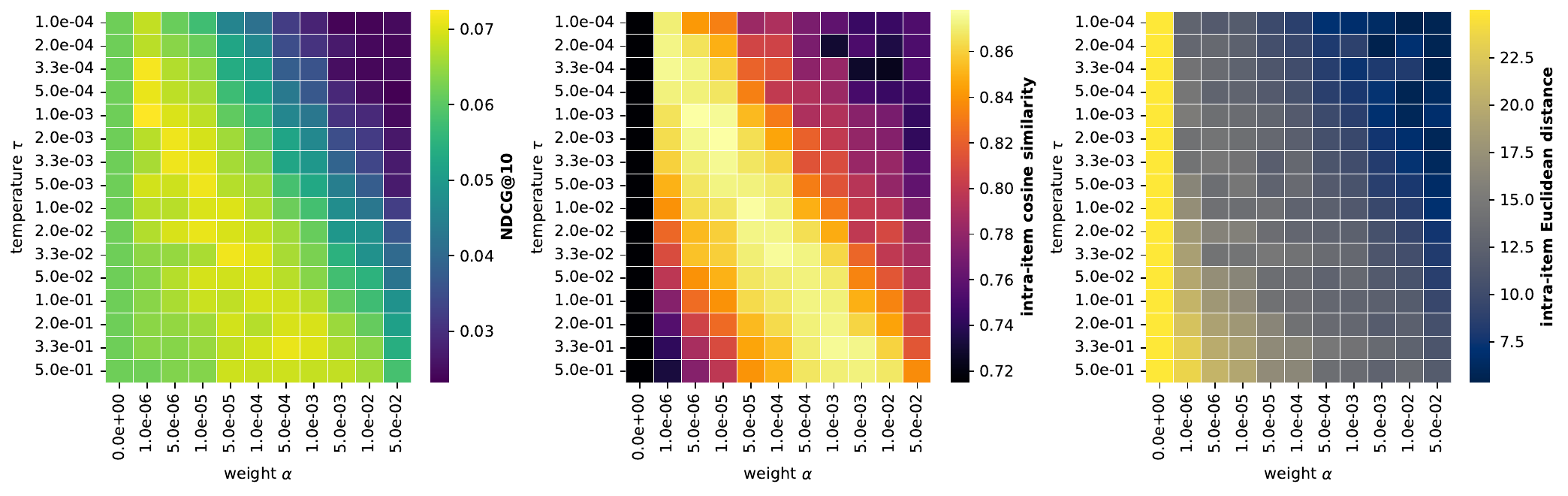}
        \caption{\amazonvid}
        \label{fig:contrastive:gridsearch:amazon}
    \end{subfigure}
    \caption{\ndcgten, intra-item cosine similarity and Euclidean distance results of \sibrarBest on different combinations of contrastive loss temperature~$\tau$ and weight~$\alpha$ on warm-start (a)~\mlonem and (b)~\amazonvid, while all other hyperparameters per dataset are fixed. The left plots show the \ndcgten results at test time, and the center and right plots show the intra-item cosine similarity and Euclidean distance, respectively. When contrastive loss $\alpha=0$, \ie when the loss does not affect training, the minor performance differences are due to randomness in the training pipeline.
    }
    \label{fig:contrastive:gridsearch:ml-amazon}
\end{figure}

%% file: resources/figures/contrastive_loss/fig-contrastive-loss-ratio_appendix.tex
\begin{figure}
    \centering

    \begin{subfigure}[b]{0.37\columnwidth}
        \centering
        \includegraphics[width=\textwidth]{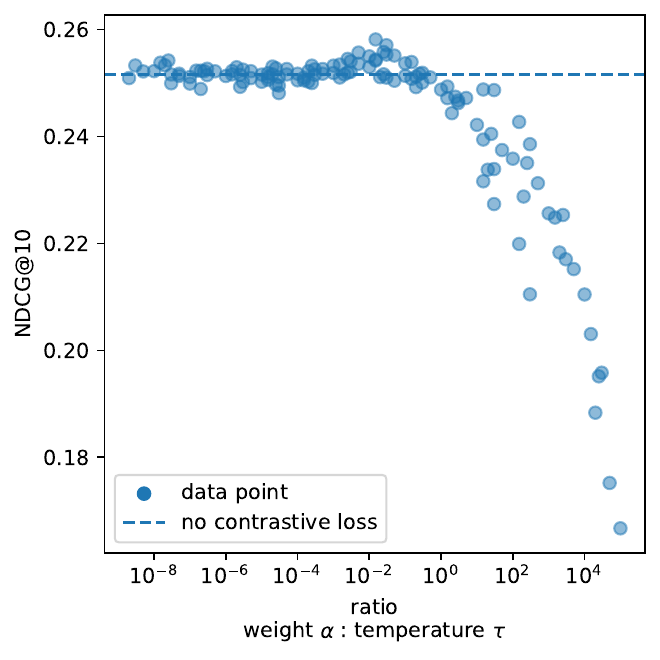}
        \caption{\mlonem}
        \label{fig:contrastive:ratio:ml}
    \end{subfigure}
    \hspace{1.5em}
    \begin{subfigure}[b]{0.37\columnwidth}
        \centering
        \includegraphics[width=\textwidth]{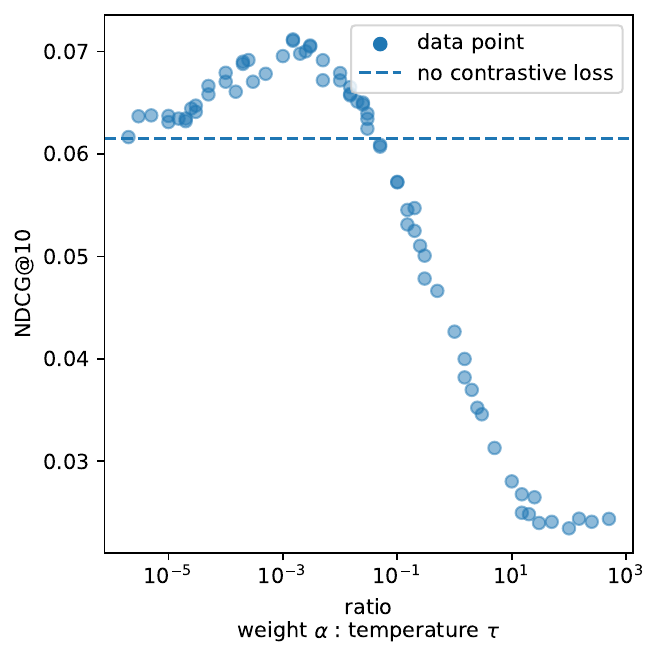}
        \caption{\amazonvid}
        \label{fig:contrastive:ratio:amazon}
    \end{subfigure}

    \caption{\ndcgten results of \sibrarBest on different combinations of contrastive loss temperature~$\tau$ and weight~$\alpha$ on warm-start (a)~\mlonem and (b)~\amazonvid, while all other hyperparameters per dataset are fixed. Each point indicates the performance of a run with a certain ratio ${\alpha}/{\tau}$ and the resulting \ndcgten. The horizontal dashed line shows the average performance for runs where the contrastive loss has no effect on training, \ie when $\alpha=0$.
    }
    \label{fig:contrastive:ratio:ml-amazon}
\end{figure}

%% file: resources/figures/projections/fig-embedding-space-all.tex
\begin{figure}[!htb]
    \centering
    
    \begin{subfigure}{1\textwidth}
        \centering
        \includegraphics[width=\textwidth]{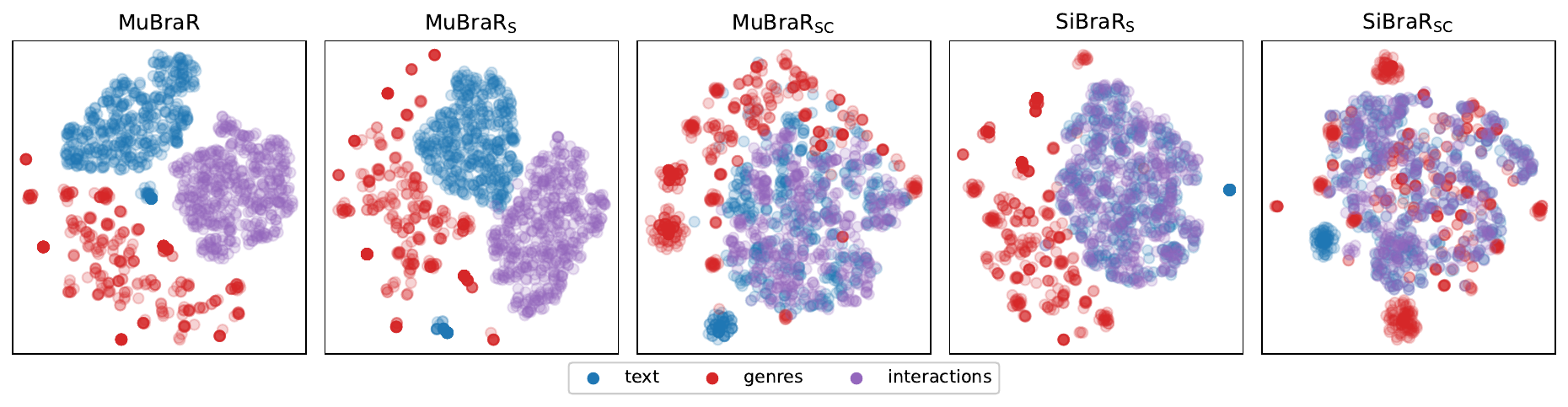}
        \caption{\mlonem}
    \end{subfigure}
    \hfill
    \begin{subfigure}{1\textwidth}
        \centering
        \includegraphics[width=\textwidth]{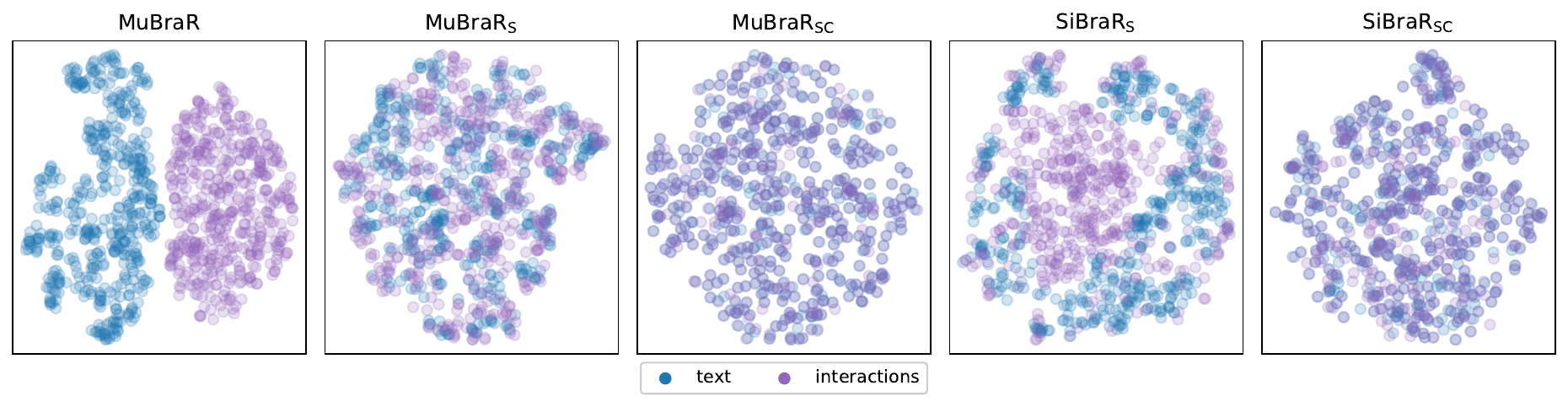}
        \caption{\amazonvid}
    \end{subfigure}    
    \caption{T-SNE projected embeddings of the modality-specific embeddings for all \mubrar and \sibrar variants on warm-start 
    \mlonem~(a) and \amazonvid~(b).
    While the different modalities form their own clusters in the embedding space of \mubrarVanilla models (1st), for models using a contrastive loss (\mubrarReg, 3rd and \sibrarReg, 5th), the modalities end up in similar regions in the projection space. 
    Genres hardly overlap with other modalities in \mlonem, which we attribute to the low number of genres and thus similar embeddings for all items. 
    }
    \label{fig:projection:all}
\end{figure}

%% file: resources/figures/projections/fig-embedding-space-clcrec.tex
\begin{figure}[!htb]
    \centering
    
    \begin{subfigure}{0.3\textwidth}
        \centering
        \includegraphics[width=\textwidth]{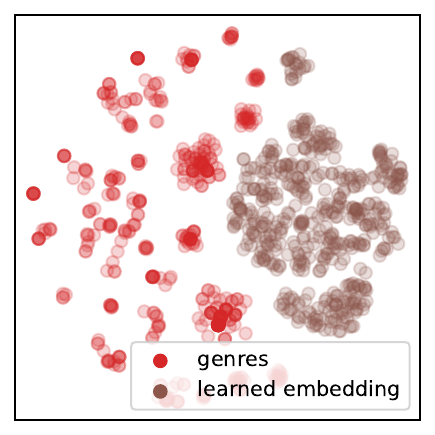}
        \caption{\mlonem}
    \end{subfigure}
    \hfill
    \begin{subfigure}{0.3\textwidth}
        \centering
        \includegraphics[width=\textwidth]{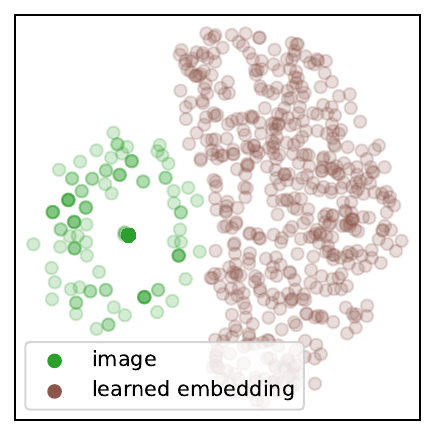}
        \caption{\onion}
    \end{subfigure}
    \hfill
    \begin{subfigure}{0.3\textwidth}
        \centering
        \includegraphics[width=\textwidth]{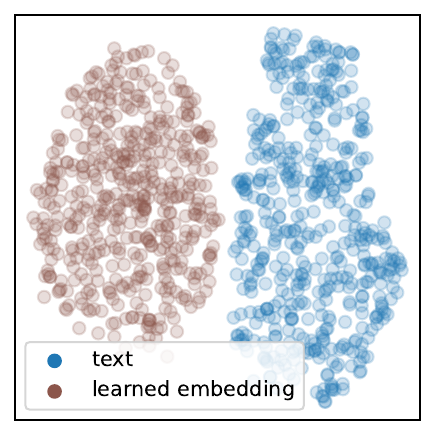}
        \caption{\amazonvid}
    \end{subfigure}    
    \caption{T-SNE projected embeddings of the modality-specific embeddings best \clcrec algorithms on warm-start 
    \mlonem~(a), \onion~(b) and \amazonvid~(c).    
    All modalities form their own clusters in the embedding space across datasets, indicating low proximity of different modalities. 
    }
    \label{fig:projection:clcrec}
\end{figure}

%% file: resources/tables/tab_intra_inter_comparison_clcrec.tex
\begin{table}[]
    \begin{adjustbox}{max width=\textwidth}
    \begin{tabular}{l|rr|rr|rr}
        & \multicolumn{2}{c|}{\mlonem} & \multicolumn{2}{c|}{\onion} & \multicolumn{2}{c}{\amazonvid} \\
        & intra & inter & intra & inter & intra & inter\\
        \midrule
\metricEuclidean 		& 2.569 & 1.811 & 7.306 & 4.890 & 6.821 & 7.403 \\
\metricCosineSim 		& 0.044 & 0.153 & 0.000 & 0.051 & 0.432 & 0.162 \\
        \bottomrule
    \end{tabular}
    \end{adjustbox}
    \vspace{0.6em}
    \caption{Average inter- and intra-item \metricEuclidean\xspace (\metricEuclideanShort) and \metricCosineSim\xspace (\metricCosineSimShort) metrics for the best \clcrec models trained each warm-start dataset. While modalities of the same type are more similar in the embedding space than between modalities of the same item for \clcrec on the \mlonem and \onion datasets (lower \metricEuclideanShort and higher \metricCosineSimShort), the modalities of individual item are more similar than to the same modalities of other items for the \amazonvid dataset.
    }
    \label{tab:intra-inter-comparison:clcrec}
\end{table}

%% file: resources/tables/tab_intra_inter_comparison_euclidean.tex
\begin{table}[]
    \begin{adjustbox}{max width=\textwidth}
    \begin{tabular}{l|rr|rr|rr}
        & \multicolumn{2}{c|}{\mlonem} & \multicolumn{2}{c|}{\onion} & \multicolumn{2}{c}{\amazonvid} \\
        & intra & inter & intra & inter & intra & inter\\
        \midrule
\mubrarVanilla 		& 188.587 & 90.606 & 420.848 & 129.887 & 30.632 & 14.551 \\
\mubrarSamp 		& 125.929 & 85.579 & 60.876 & 55.630 & 44.273 & 62.094 \\
\mubrarSampContr 	& 18.158 & 18.193 & 13.814 & 15.683 & \bfseries 4.742 & \bfseries 11.448 \\
\hdashline
\sibrarSamp 		& 2.244 & 2.463 & 9.347 & 10.218 & 24.956 & 34.775 \\
\sibrarSampContr	& \bfseries 0.756 & \bfseries 1.191 & \bfseries 4.161 & \bfseries 5.950 & 12.209 & 31.765 \\
        \bottomrule
    \end{tabular}
    \end{adjustbox}
    \vspace{0.6em}
    \caption{\newtext{Average inter- and intra-item \metricEuclidean\xspace (\metricEuclideanShort) for \sibrar and \mubrar variants. For each warm-start dataset, all variants are trained and evaluated on the same set of modalities: (interactions, genres, text) for \mlonem, (interactions, genres, audio, image, text) for \onion, and (interactions, text) for \amazonvid. Bold indicates the algorithm where modality embeddings exhibit the closest average proximity within each group.}
    }
    \label{tab:intra-inter-comparison:euclidean}
\end{table}

%% file: resources/tables/tab_modality_prediction-cold.tex
\begin{table}[]
    \begin{tabular}{lrr}
        & \mlonem & \onion \\
        \midrule
    \mubrarVanilla & 1.0000 & 1.0000 \\
    \mubrarNoReg & 1.0000 & 0.9999 \\
    \mubrarReg & 1.0000 & 0.9979 \\
    \hdashline
    \sibrarNoReg & \bfseries 0.9485$\dagger$ & 0.9008 \\
    \sibrarReg & 0.9928 & \bfseries 0.7424$\dagger$ \\
    \hdashline
    Random & 0.5000 & 0.2500 \\
        \bottomrule
    \end{tabular}
    \vspace{0.6em}
    \caption{Accuracy of random forest classifiers predicting the modality class of modality embeddings output by \mubrar and \sibrar variants trained on item cold-start \onion and \mlonem. We do not consider \amazonvid, as it works best in cold-start when trained on a single type of side-information, which makes the classification task trivial.
    Each result reports the average performance of 20 classifiers trained on different train--test splits and seeds. 80\% of each cold-start test set is used for fitting the classifiers, and 20\% for their evaluation. \newtext{Bold indicates the variant with the least separable embedding space, $\dagger$ indicates significant worse classifier accuracies in comparison to all other methods (paired $t$-tests with $p<0.05$ considering Bonferroni correction). The last row shows the expected accuracy for random guessing. }
    }
    \label{tab:modality-prediction-cold}
\end{table}